\documentclass[
 reprint,
 showkeys,
 amsmath,amssymb,
 aps,
prb,
]{revtex4-2}
\pdfoutput=1 
\usepackage{graphicx}
\usepackage{dcolumn}
\usepackage{bm}
\usepackage{multirow}
\usepackage[caption=false]{subfig}
\usepackage{xr-hyper}
\usepackage[colorlinks=true,linkcolor = blue,urlcolor=blue,citecolor=blue]{hyperref}
\usepackage[capitalize]{cleveref}
\Crefname{figure}{Fig.}{Figs.}
\Crefname{section}{Sec.}{Secs.}
\Crefname{equation}{Eq.}{Eqs.}

\newcommand\Tstrut{\rule{0pt}{2.6ex}}         
\newcommand\Bstrut{\rule[-0.9ex]{0pt}{0pt}}   
\newcommand{\nhrule}{\hline\Tstrut\Bstrut}
\newcommand{\nmidrule}[1]{\cline{#1}\Tstrut\Bstrut}

\begin{document}
\preprint{APS/123-QED}
 
\title{Accurate first-principles bandgap predictions \\in strain-engineered ternary III-V semiconductors}

\author{Badal Mondal}
\affiliation{Wilhelm-Ostwald-Institut f\"ur Physikalische und Theoretische Chemie, Universit\"at Leipzig, 04103 Leipzig, Germany}%
\affiliation{Fachbereich Physik, Philipps-Universit\"at Marburg, 35032 Marburg, Germany}

\author{Marcel Kr\"oner}%
\affiliation{Material Science Center and Department of Physics, Philipps-Universit\"at Marburg, D-35043 Marburg, Germany}%

\author{Thilo Hepp}%
\affiliation{Material Science Center and Department of Physics, Philipps-Universit\"at Marburg, D-35043 Marburg, Germany}%

\author{Kerstin Volz}%
\affiliation{Material Science Center and Department of Physics, Philipps-Universit\"at Marburg, D-35043 Marburg, Germany}%

\author{Ralf Tonner-Zech}
\email{ralf.tonner@uni-leipzig.de}
\affiliation{Wilhelm-Ostwald-Institut f\"ur Physikalische und Theoretische Chemie, Universit\"at Leipzig, 04103 Leipzig, Germany}%
\date{\today}
\begin{abstract}
Tuning the bandgap in ternary III-V semiconductors via modification of the composition or the strain in the material is a major approach for the design of optoelectronic materials. Experimental approaches screening a large range of possible target structures are hampered by the tremendous effort to optimize the material synthesis for every target structure. We present an approach based on density functional theory efficiently capable of providing the bandgap as a function of composition and strain. Using a specific density functional designed for accurate bandgap computation (TB09) together with a band unfolding procedure and special quasirandom structures, we develop a computational protocol to predict bandgaps. The approach's accuracy is validated by comparison to selected experimental data. We thus map the bandgap over the phase space of composition and strain (we call this the ``bandgap phase diagram'') for several important III-V compound semiconductors: GaAsP, GaAsN, GaPSb, GaAsSb, GaPBi, and GaAsBi. We show the application of these diagrams for identifying the most promising materials for device design. Furthermore, our computational protocol can easily be generalized to explore the vast chemical space of III-V materials with all other possible combinations of III- and V-elements.
\end{abstract}

\maketitle

\section{\label{sec1:introduction}Introduction}
Materials based on III-V semiconductor compounds are attracting much attention in science and engineering due to their diverse applications in fields such as optoelectronics \cite{Soref1993Silicon,CardanoBook}. One of the main goals of basic and applied research is to tailor materials' optical properties to a specific application \cite{Hepp2022Room,Fuchs2018,Mokkapati2009IIIV,Dimroth2016Four,Mitchell2011Four,PhilippsS2018Chapter}. One of the most critical fundamental properties in this respect is the bandgap, both in terms of size and type (direct or indirect). For example, optical telecommunication applications require materials with direct bandgaps in the range of 0.80--0.95 eV \cite{Hepp2022Room,Fuchs2018,Mokkapati2009IIIV}, while solar cell applications require a range of 0.5--2.0 eV \cite{Dimroth2016Four, Mitchell2011Four, PhilippsS2018Chapter}. Composition engineering, i.e., changing the relative composition of group 13 and 15 elements in ternary III-V compounds, is one of the most important approaches to adjusting the bandgap  \cite{Beyer2015Metastable,Vurgaftman2001Band,Beyer2017Local,Ludewig2016Movpe,Liebich2011Laser,Supplie2018Metalorganic,Stringfellow2019Fundamental,Volz2009Movpe,Feifel2017Movpe,Kunert2004Movpe,Volz2004Specific,Veletas2019Bismuth,Wegele2016Interface,Hepp2019Movpe}. Systematic application of strain such as mechanical strain (e.g., external pressure \cite{Van1990Pressure,Potter1956Indirect,Alekseev2020Effect,Katiyar2020Breaking}, mechanical bending of nanowires \cite{Lim2021Strain,Signorello2014Inducing,Signorello2013Tuning}) or strain due to lattice mismatch (e.g., core-shell mismatch in nanowires \cite{Balaghi2019Widely,Gronqvist2009Strain,Hetzl2016,Montazeri2010,Skold2005}) on a system are alternative strategies to tailor the bandgap. Combining composition and strain engineering, the bandgap can be tuned over a wide range of values, and direct or indirect semiconductors can be designed. In thin-layer heteroepitaxy, choosing the substrate-layer combination with minimum lattice mismatch is often desirable to minimize the strain effect from the substrate. However, in practice, perfect lattice matching is rarely possible. In such cases, not only the composition but the effect of inherent strain from the substrate also substantially affects the active layer's bandgap \cite{Beyer2015Metastable,Vurgaftman2001Band,Beyer2017Local,Ludewig2016Movpe,Liebich2011Laser,Supplie2018Metalorganic,Stringfellow2019Fundamental,Volz2009Movpe,Feifel2017Movpe,Kunert2004Movpe,Volz2004Specific,Veletas2019Bismuth,Wegele2016Interface,Hepp2019Movpe,Bir1974Symmetry,Sun2007Physics,Tao2020,Tsutsui2019,Fang2011}. Therefore, one requires a complete knowledge of the material-specific dependence of the bandgap on composition and strain to guide the optimal choice of materials. However, exploring the vast chemical space of all possible combinations of III- and V-elements with variation in composition and strain is experimentally not feasible. Additionally, growing a new material is often challenging because of thermodynamic or kinetic limitations, such as phase separation or surface roughening, in addition to the demanding task of optimizing the growth conditions \cite{Beyer2015Metastable,Liebich2011Laser,Volz2009Movpe,Volz2004Specific,Wegele2016Interface,Hepp2019Movpe}. This makes an experimental screening approach of vast compound and strain spaces unrealistic. We thus aim in this study to develop a reliable and predictive theoretical approach.

Two major theoretical approaches that have been used to analyze strain effects on the bandgap of III-V materials are (semi-)empirical methods and ab initio approaches. Although (semi-)empirical methods such as k$\cdot$p theory \cite{Sun2007Physics,Bannow2017} and tight-binding methods \cite{Sun2007Physics,Anderson1991Optimized,Tan2016Transferable,Jancu1998} are computationally efficient, they rely on empirical parameters which require system-specific experimental input data. This strongly limits the predictive ability of these methods for new or yet unknown materials. Additionally, in case of a large mismatch in atomic sizes of the constituting elements, the ternary material shows local strain effects, severely affecting the bandgap \cite{Rosenow2018Ab}. These local strain effects, however, can not be included in empirical approaches and, hence, are neglected. Then again, ab initio approaches such as density functional theory (DFT) \cite{Bannow2017,Mondal2022,Kim2010Towards,Tran2009Accurate,Jiang2013Band,Rehman2016Electronic,Koller2012Improving} allow for the calculation of electronic properties from first-principles and are thus predictive if accurate density functional are used. The relaxation of the atomic positions also allows to properly include and investigate the effect of the local strain on the electronic properties in these approaches. Additionally, recent advancements in the modeling strategies of alloy systems using quasirandom supercells \cite{Wang1998,Medeiros2015,Popescu2010,Rubel2014,Popescu2012} allow for electronic properties calculations in the ab initio approaches, even for diluted and disordered materials. An accurate alternative to DFT approaches is the use of GW-based methods, which are nevertheless too computationally demanding for screening approaches as intended here \cite{Rosenow2018Ab,Mondal2022,Kim2010Towards,Hinuma2014Band}. 

In a previous study, we established a computational protocol for predictive modeling based on DFT for binary III-V compounds over a wide range of strain values \cite{Mondal2022}. In this study, we are now extending this approach to ternary III-V compounds, which then allow the combination of strain and composition to fully explore a bandgap design approach. For ternary systems, only the effects of composition variations on the bandgap in unstrained materials have been studied \cite{Bannow2017,Anderson1991Optimized,Rosenow2018Ab,Wang1998,Medeiros2015,Popescu2010,Rubel2014,Popescu2012}. For strained materials, a suitable theoretical framework is still lacking. We present here a predictive first-principles protocol for a complete mapping of the mutual correlation of composition, strain, and bandgap in ternary III-V semiconductor systems. The goal is to provide guidelines for assessing and identifying the most promising target materials for experimental investigations in the future.

We start by describing the computational methods in \cref{sec2:computationaldetails}. Next, we describe the protocol for determining the nature of the bandgap from supercell calculations using GaAsP as an example in \cref{sec3:protocolfordeterminingbandgapnature}. We further present the composition-strain-bandgap correlation results for different ternary III-V semiconductors in \cref{sec4:results}. We start with GaAsP, an experimentally well-studied and promising candidate for LEDs, detectors, and Si-based multijunction solar cells \cite{Craford1973,Henning1983,Tanaka1994,Sato2002,Geisz2002,Lang2013,Hayashi1994,Grassman2016}. The results for the GaAsN compound, a promising laser-active material \cite{Rosenow2018Ab,Weyers1992,Kunert2006,Zhao2004}, are presented next. To show the general applicability of our approach, we then show selected results for (i) GaPSb, a candidate for vertical cavity emitting surface laser \cite{Loualiche1998,Shimomura1996,Nakajima2000,Russell2016,Jou1988}; (ii) GaAsSb, a material for tandem solar cell application \cite{Cherng1984,Jen1998}; (iii) GaPBi, a promising material for near-infrared photonic device application on Si \cite{Christian2016,Christian2015}; and (iv) GaAsBi, another material discussed for near- and mid-infrared photonic device application \cite{Sweeney2011,Wang2013,Cooke2006}. We then discuss the comparison of our computations with experimental data in \cref{sec5:discussion}, underlining the accuracy and predictive capability of our computational approach.

\section{\label{sec2:computationaldetails}Computational details}
The calculations were performed with DFT-based approaches as implemented in the Vienna ab initio simulation package (VASP 5.4.4) \cite{Kresse1993Ab,Kresse1994Ab,Kresse1996Efficient,Kresse1996Efficiency}, using plane wave basis sets in conjunction with the projector-augmented wave (PAW) approach \cite{Kresse1999From,Blochl1994Projector}. The ternary materials were generated using the special quasirandom structures (SQS) approach \cite{Zunger1990Special} with a supercell of size $6 \times 6 \times 6$. The SQS cells were generated using the alloy theoretic automated toolkit (ATAT) \cite{VandeWalle2002,VandeWalle2009,VanDeWalle2013}. For all the materials except GaAsN, one SQS cell was used per composition. In GaAsN, in agreement with the previous observation \cite{Rosenow2018Ab}, we found that the size of the bandgap strongly depends on the distribution of N atoms in the supercell, even in the SQS approach. We thus used 10 SQS cells for each composition in this case. 

Geometry optimization of the supercells was performed using the PBE functional \cite{Perdew1996Generalized}, including the dispersion-correction method DFT-D3 with an improved damping function \cite{Grimme2010Consistent,Grimme2011Effect}. The basis set energy cut-off was set to 450 eV. The electronic energy convergence criteria of $10^{-6}$ eV and the force convergence of $10^{-2}$ eV\AA $^{-1}$ were used. The reciprocal space was sampled at the $\Gamma$-point only, given the large supercells used \cite{Monkhorst1976}. The meta-GGA functional TB09 \cite{Tran2009Accurate} was used to calculate the electronic properties (bandgaps and band structures). The effects of spin-orbit coupling were considered in the TB09 calculations. For the meta-GGA calculations, the energy cut-off of the basis set and the convergence criterion for the electronic energy were lowered to 350 eV and $10^{-4}$ eV, respectively, to reduce the computational costs. Structure optimizations were carried out by consecutive volume and position optimization until convergence was reached. This setup was previously used to generate bandgaps in excellent agreement with experimental data \cite{Rosenow2018Ab}. 

All the materials within the composition range investigated here feature the zincblende-type structure only. Moreover, [100] crystal direction is the most common choice of substrate orientation and growth direction in epitaxy. Therefore, we modeled the strain application along [100] directions only. The isotropic strain was modeled by increasing (decreasing) all the lattice parameters of the unstrained structure by the same amount. In this case, only the atomic positions of the strained structure were optimized, keeping the volume fixed. For biaxial strain, the in-plane lattice parameters were kept fixed, and the lattice parameter in the out-of-plane direction was optimized. No structural phase transition is assumed under strain application. More details on the strain modeling can be found in Ref.~\cite{Mondal2022}. In the following, we indicate tensile strain with a positive sign and compressive strain with a negative sign.

DFT calculations were performed at discrete points in composition-strain space (Fig.~S6 \cite{Supp_info}). \nocite{Medeiros2015,Medeiros2014,Mondal2022,SciPyNMeth2020,Nielson1983,Renka1984,Www.qhull.org,Alfeld1984,Farin1986,Hunter:2007,thomas_a_caswell_2022_7084615}The calculated bandgap values were then interpolated to create the final images in \cref{fig:fig2,fig:fig3,fig:fig4,fig:fig5,fig:fig6}. Noticeably, for the systems we addressed in this article, the variations of bandgap values with concentration and strain are mostly non-monotonic (Fig.~S6 \cite{Supp_info}). This resulted in non-smooth interpolation in \cref{fig:fig2,fig:fig3,fig:fig4,fig:fig5,fig:fig6}. It is to be stressed that the origin of the non-smooth patterns is neither an interpolation artifact nor a deficiency of our DFT protocol. This solely originated because of the non-monotonic variation of the bandgap values (in the composition-strain space) of the SQS cells that we used to calculate bandgaps. A choice of positive smoothening during interpolation (e.g., bivariate B-spline, gaussian filtering) could mitigate the problem but significantly increased the deviation of the interpolated bandgap values from the calculated DFT values and was thus not chosen. Further detail of the interpolation procedures can be found in Sec.~S VI \cite{Supp_info}. Moreover, the nature of bandgaps can solely be deduced from the direct-indirect transition lines and thus requires no interpolation.

\section{\label{sec3:protocolfordeterminingbandgapnature}Protocol for determining bandgap nature}
\begin{figure}[hbtp]
\centering
  \subfloat[]{\label{fig:fig1a}\includegraphics[width=3.4in]{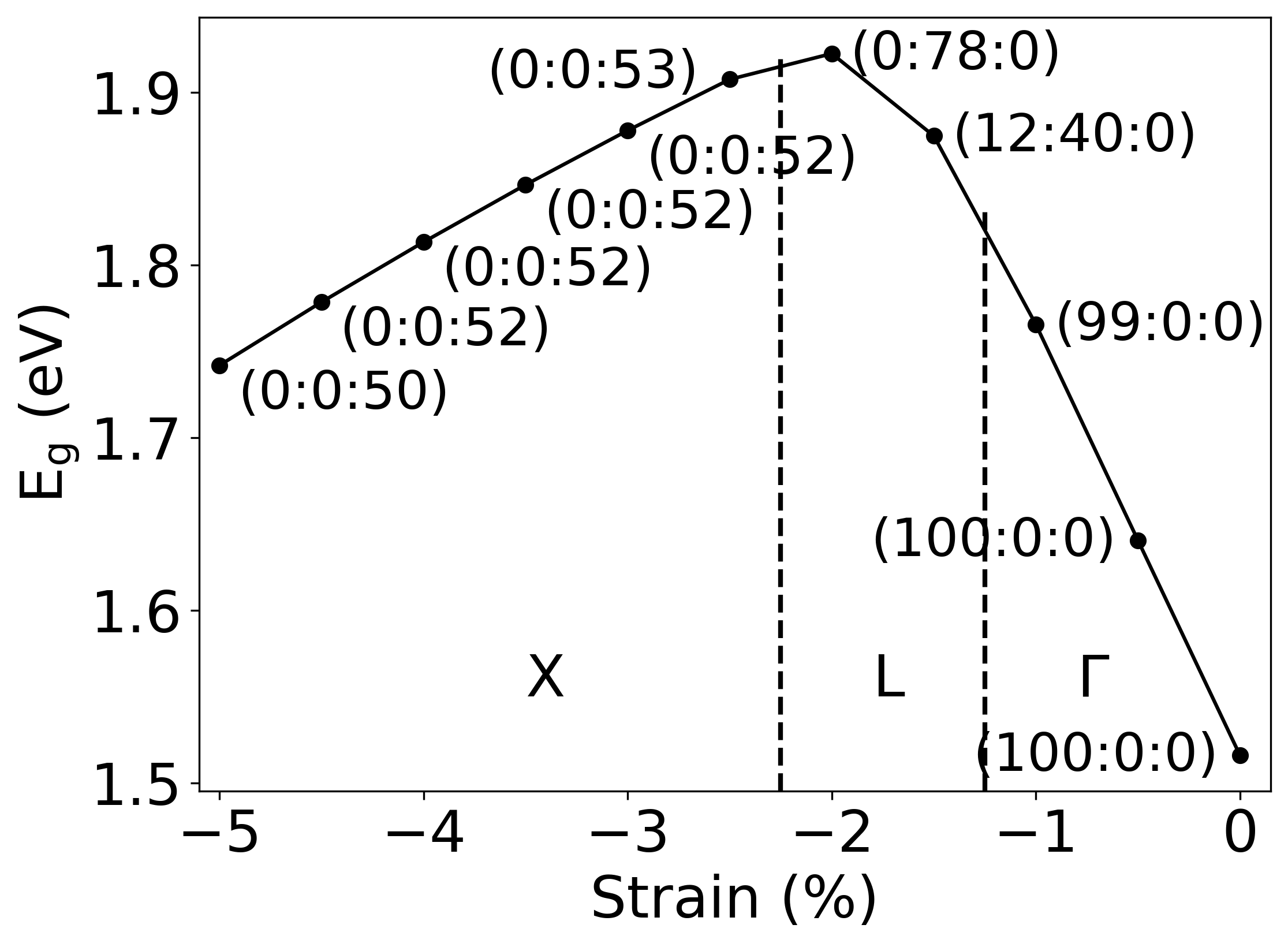}}
  \hspace{.2in}
 \subfloat[]{\label{fig:fig1b}\includegraphics[width=3.4in]{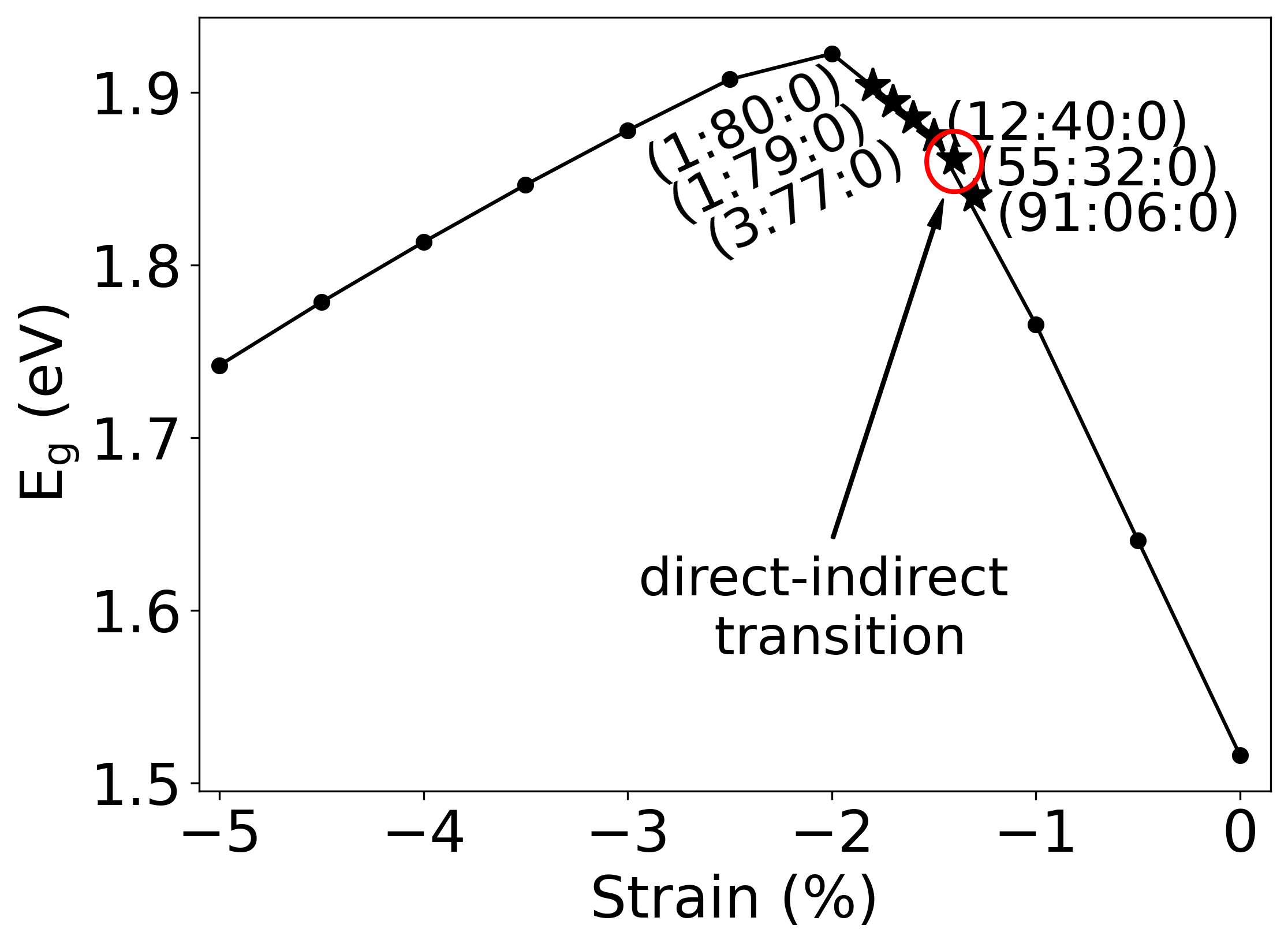}}
  \caption{\label{fig:fig1}Variation of the bandgap under isotropic compressive strain for GaAs$_{0.963}$P$_{0.037}$. The $\Gamma$-, L-, and X-BSW of the folded supercell conduction band are given in parentheses in the format ($\Gamma$:L:X). The vertical lines in (a) separate regions where the CBM changes character. In (b), the strain resolution is increased to determine the point of direct-indirect transition more accurately, indicated by the red circle (where the highest BSW changes from $\Gamma$ to L).}
\end{figure}
Supercell calculations, as required for modeling ternary semiconductors, lead to the folding of band structures \cite{Ku2010,Yang2018}. The size of the bandgap can be well extracted from the folded band structure, which represents the energy difference between the highest occupied VB and the lowest unoccupied CB obtained from supercell calculations (folded bands). However, determining the bandgap's nature requires the primitive Bloch character of the bands to be known, which gets mixed up in the supercell eigenstates. With the band unfolding method, one projects these supercell eigenstates on the eigenstates of a suitable reference primitive cell. This requires the calculation of Bloch spectral weights (BSW), which measure the fraction of the primitive Bloch character in a supercell eigenstate. The result is an effective band structure (EBS) \cite{Wang1998,Medeiros2015,Popescu2010,Rubel2014,Popescu2012}. The spectral weights, $w_{n, {\bf K}}\left( {\bf k} \right)$, can be calculated from the plane wave coefficients as described in Ref.~\cite{Popescu2012}:
\begin{equation}
    w_{n, {\bf K}}\left( {\bf k}_j \right) = \sum _{\bf g} | C_{n, {\bf K}} \left( {\bf g} + {\bf G}_j\right) |^2
\end{equation}
where $n$ represents the band index, and the reciprocal lattice vectors of the primitive and supercell are denoted by ${\bf g}$ and ${\bf G}_j$, respectively. The index $j$ accounts for the series of primitive vectors, ${\bf k}_j = {\bf K} + {\bf G}_j$. The code ``fold2Bloch'' from Ref.~\cite{Rubel2014} was used to calculate the BSW values. 

In our previous study on binary III-V systems \cite{Mondal2022}, we have shown that the valence band maxima (VBM) always remain at the $\Gamma$-point, and only the conduction band minima (CBM) change their position in reciprocal space under strain. We have also shown that the CBM occurs only at the $\Gamma$-, L-, and (near) X-point in the band structure under strain. Therefore, it is sufficient to trace the conduction band (CB) at these points to determine the nature of the bandgap. As in our previous study on binary systems, we focus here on analyzing ternary III-V compounds with zincblende structures. For these structures in the $6 \times 6 \times 6$ supercell dimensions chosen here, the $\Gamma$-, L- and X-point of the primitive band structure fold to the $\Gamma$-point in the supercell \cite{Popescu2010,Rubel2014,Popescu2012}. Therefore, it is sufficient to calculate the BSWs of solely the CB at the $\Gamma$-point in the supercell calculation to determine the nature of the bandgap. Consequently, we performed the supercell calculations by sampling the reciprocal space only at the $\Gamma$-point and unfolded the CB.

Figure~\ref{fig:fig1} shows the steps for determining the bandgap nature from supercell calculations more clearly. Figure~\ref{fig:fig1a} shows the bandgap variation for GaAsP with 3.7\% P concentration under isotropic compressive strain. The $\Gamma$-, L-, and X-BSWs of the folded supercell CB are given in parentheses. This shows 100\% $\Gamma$-BSW for the unstrained structure in line with the direct bandgap. With increasing strain, the $\Gamma$-BSW decreases (first number in brackets), and the L-BSW increases (second number in brackets). After a certain amount of strain, the L-character of the CB dominates. The bandgap becomes indirect in nature. Notably, once the strain values reach the point of direct-indirect transition (DIT) in the bandgap nature (around --1.5\% strain), the bandgap values begin to decrease further with additional strain. This trend is like what we previously observed in binary III-V semiconductor systems, where a strong dependence of the band energies ($E$) on the wavevectors (\textbf{k}) under strain was found, leading to a non-monotonic variation of bandgap values with strain \cite{Mondal2022}. Moreover, we found that such non-monotonic behaviour in bandgap values under strain points to a DIT \cite{Mondal2022}. In ternary III-V semiconductor systems, we have now found that similar non-monotonic behaviour in bandgap values under strain also indicates a DIT. Further compressing the system then leads to a transition of the CB character from L to X. 

In \cref{fig:fig1b}, we show calculations with increased resolution in strain to accurately determine the transition to L corresponding to the sought point of DIT at $-1.4$\% strain. We define the last strained structure with bandgap of direct nature before the transition to the indirect bandgap as the transition point (the red circle in \cref{fig:fig1b}). In the Supplemental Material (Fig.~S1 \cite{Supp_info}), we have given the EBSs of GaAs$_{0.963}$P$_{0.037}$ for different strain values. These confirm our analyses. 

If the difference in BSW between different points in k-space is large, the nature of the bandgap can be unanimously determined. However, close to the transition points, in some cases, the differences are more subtle (Fig.~S2 \cite{Supp_info}). We, therefore, set a cut-off criterion of 20\% BSW. If the $\Gamma$-BSW is larger than the cut-off criterion, then the direct transition has a finite probability even if the L- or X-BSW is larger than the $\Gamma$-BSW. In such cases, the bandgap is called ``partially direct''. This defines a ``region of uncertainty'' in the bandgap nature. We chose the 20\% cut-off criterion because this produces results that agree best when compared to the experiments for several systems. For the GaPBi system, however, the 10\% BSW cut-off criterion produces the best agreement.

In some systems such as GaAsN, the band originating from the added nitrogen atoms, the so-called ``defect N state'' \cite{Rubel2014,Goodrich2019,Wei1996,Wu2004,Misiewicz2005}, is strongly dispersed under strain (Figs.~S3b and S4b \cite{Supp_info}). Therefore, we set another cut-off criterion of 20\% BSW as a minimum limit for a (defect) eigenstate to be considered an eigenstate (Fig.~S5 \cite{Supp_info}). Starting from the lowest unoccupied CB, we search for eigenstates until the cut-off BSW criterion is met, at which point we consider it to be the redefined CB. If none of the CBs satisfy the cut-off criterion, we use the lowest CB for determining the bandgap nature. Accordingly, in these cases, we calculate the bandgap values as the energy difference between the highest VB and the redefined CB. When redefining, unoccupied CB states that do not satisfy the cut-off criteria are disregarded. This led to an increase in the bandgap values, as is observed in Fig.~S5 \cite{Supp_info}.

\section{\label{sec4:results}Results}
In this section, we present the bandgaps calculated for different materials and determine their nature according to the above protocol. We mapped the bandgaps in terms of their size and nature for various strained ternary III-V compounds. We start with two important ternary \mbox{III-V} semiconductor materials, GaAsP and GaAsN. Then we show selected data for the material systems GaPSb, GaAsSb, GaPBi, and GaAsBi.
\subsection{\label{subsec41:gaasp}G\lowercase{a}A\lowercase{s}P}
\begin{figure}
    \centering
    \includegraphics[width=3.4in]{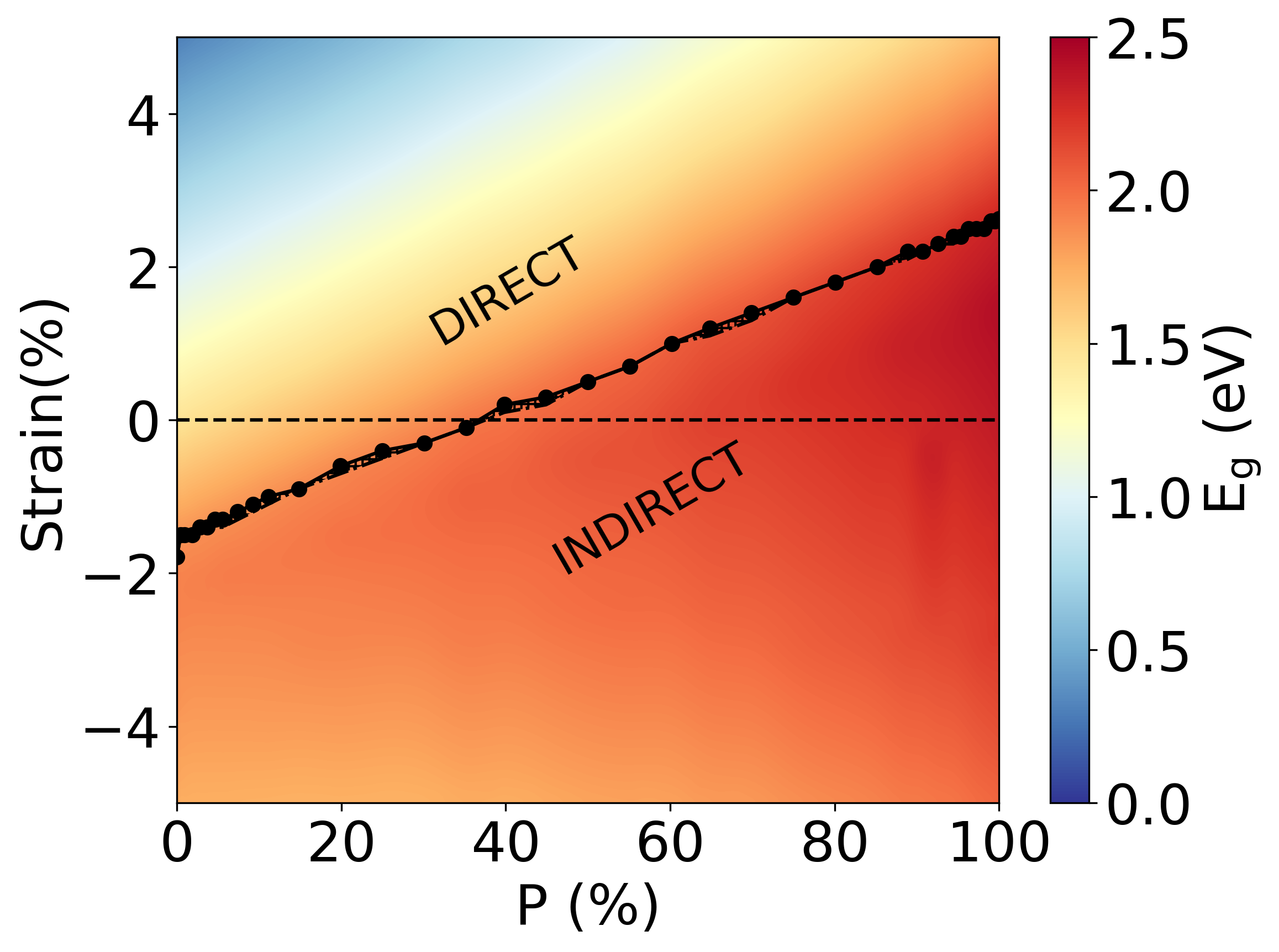}
    \caption{Isotropic strain for GaAsP. The variation of bandgap magnitudes (E$_{\text{g}}$) and type as a function of composition and strain. The dashed black horizontal line indicates unstrained GaAsP. The black circles are the calculated DIT points. The direct and indirect enclosed regions describe the nature of bandgap being direct and indirect, respectively. The hatched pattern region is the ``uncertainty region'' (see \cref{sec3:protocolfordeterminingbandgapnature}).}
    \label{fig:fig2}
\end{figure}

For the case of isotropic strain, \cref{fig:fig2} shows the bandgap as a function of composition ($x=$ 0--100\% in GaAs$_{1-x}$P$_{x}$) from 5\% tensile to 5\% compressive strain. The bandgap value varies between 0.32--2.42 eV in the strain regime investigated. For the same amount of P concentration, the bandgap primarily increases in moving from tensile to compressive. Furthermore, the figure shows that in going from compressive to tensile strain, the DIT occurs at a higher concentration of P atoms. The dashed horizontal line marks the data corresponding to the unstrained structures for different fractions of P. The intersection of this line with the DIT line shows at which percentage of phosphorous contribution the unstrained structure shows a DIT. This transition occurs at $x = 37$\%. Here, the bandgap shows a value of 1.96 eV. The terms direct and indirect in the figure correspond to the area where the bandgap is direct and indirect, respectively. Due to the similarity with commonly used phase diagrams, we call this representation a ``bandgap phase diagram". This and the following figures thus provide a 2D representation of the bandgap phase diagram for the ternary materials.   

\begin{figure}
    \centering
    \includegraphics[width=3.4in]{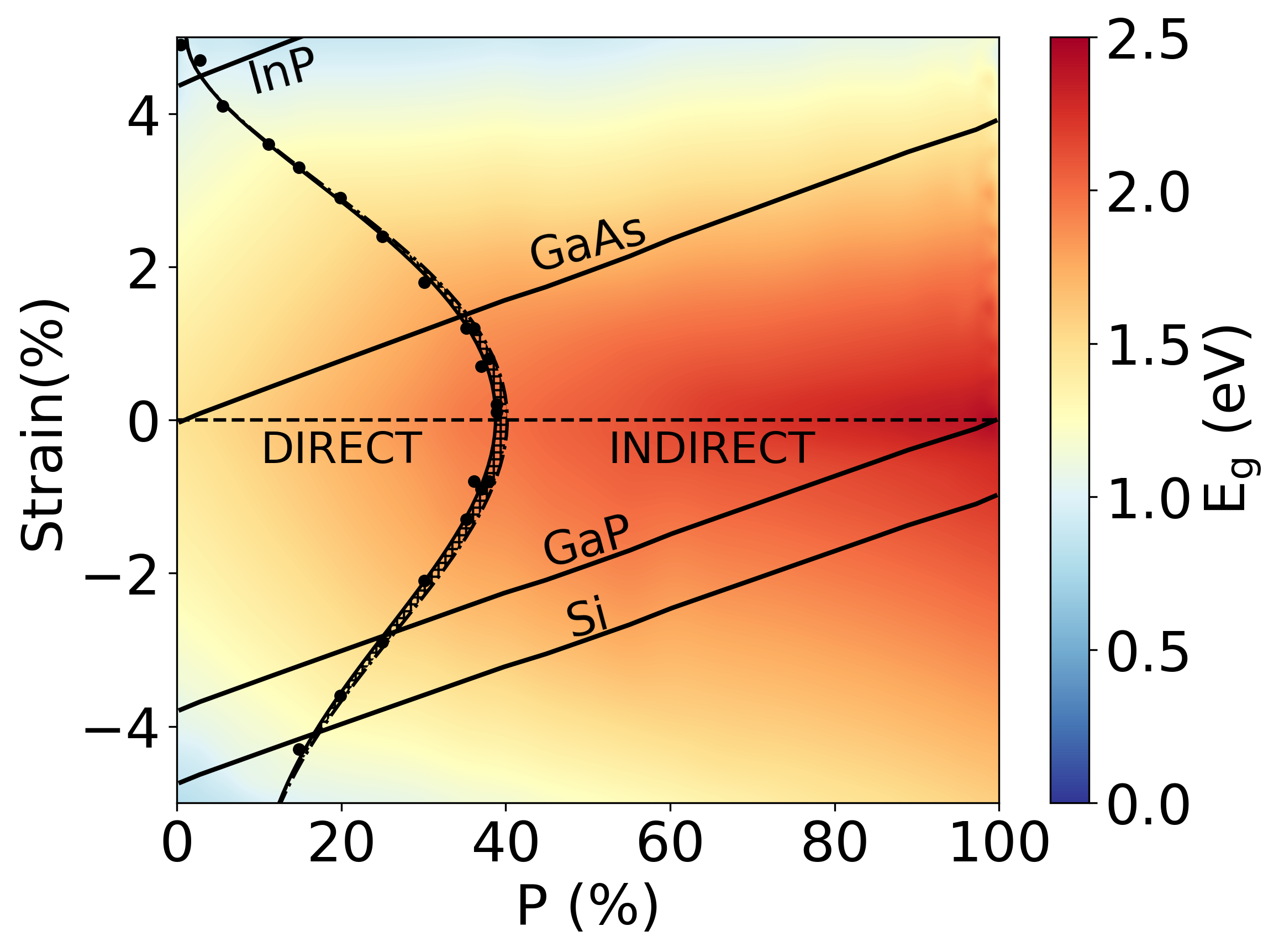}
    \caption{Biaxial strain for GaAsP. The variation of bandgap magnitudes (E$_{\text{g}}$) and type as a function of composition and strain. The dashed black horizontal line indicates unstrained GaAsP. The black circles are the calculated DIT points. The DIT points are fitted with a 5th-order polynomial. The direct and indirect enclosed regions describe the nature of bandgap being direct and indirect, respectively. The hatched pattern region is the ``uncertainty region'' (see \cref{sec3:protocolfordeterminingbandgapnature}). Solid black lines indicate the substrate lines under ``epitaxial growth'' model.}
    \label{fig:fig3}
\end{figure}

For the biaxial strain regime, \cref{fig:fig3} shows the bandgap phase diagram for GaAsP as a function of composition from 5\% tensile to 5\% compressive strain. The value of the bandgap varies in a range of 0.82--2.42 eV. For the same amount of P atoms, the bandgap reaches a maximum around the unstrained structure and gets smaller for tensile as well as compressive strain. This is different from the isotropic strain case. For unstrained GaP, the bandgap value is 2.36 eV. The nature of the bandgap also shows a different trend compared to \cref{fig:fig2}. The range of strain around the unstrained structure where a direct bandgap is found gets smaller for higher amounts of P. This is in line with GaAs (0\% P) being a direct and GaP (100\% P) being an indirect semiconductor. The largest amount of P concentration where a direct semiconductor is found is 39--40\% P in the unstrained structure. This is similar to the previous experimental result (45\% P) \cite{Vurgaftman2001Band,Capizzi1981}. 

One of the most common approaches to experimentally realize biaxial strain in III-V semiconductors is epitaxial growth. As pointed out in Ref.~\cite{Mondal2022}, biaxial strain can be used to model epitaxial growth. We thus investigate the effect of different substrates in our bandgap phase diagram (\cref{fig:fig3}), where each solid line corresponds to one substrate: GaAs, GaP, InP, or Si. These solid lines indicate how much biaxial strain would develop in the GaAsP system as the respective value of \% P when grown on the respective substrates under idealized conditions. The (substrate) strains are calculated
according to \cref{eqn:equation2}:
\begin{equation}
    \text{Substrate strain} (\%) = \frac{a_{sub}-a}{a}\times 100 \label{eqn:equation2}
\end{equation}
where $a_{sub}$ is the equilibrium lattice parameters of the substrates, and $a$ is the lattice parameters of unstrained GaAsP systems at their respective P concentrations. E.g., for 100\% P, the strain on the GaP substrate is zero, while growing GaAs (0\% P) on GaP would result in 3.8\% in-plane compressive strain. This, of course, neglects defect formation and strain relaxations and assumes perfect epitaxial growth. Clearly, by choosing different substrates, the nature can be changed, and the size of the bandgap can be tuned over a wide range. We refer to the next section for a comparison of our calculations to experimental data.

\subsection{\label{subsec42:gaasn}G\lowercase{a}A\lowercase{s}N}
\begin{figure}
    \centering
    \includegraphics[width=3.4in]{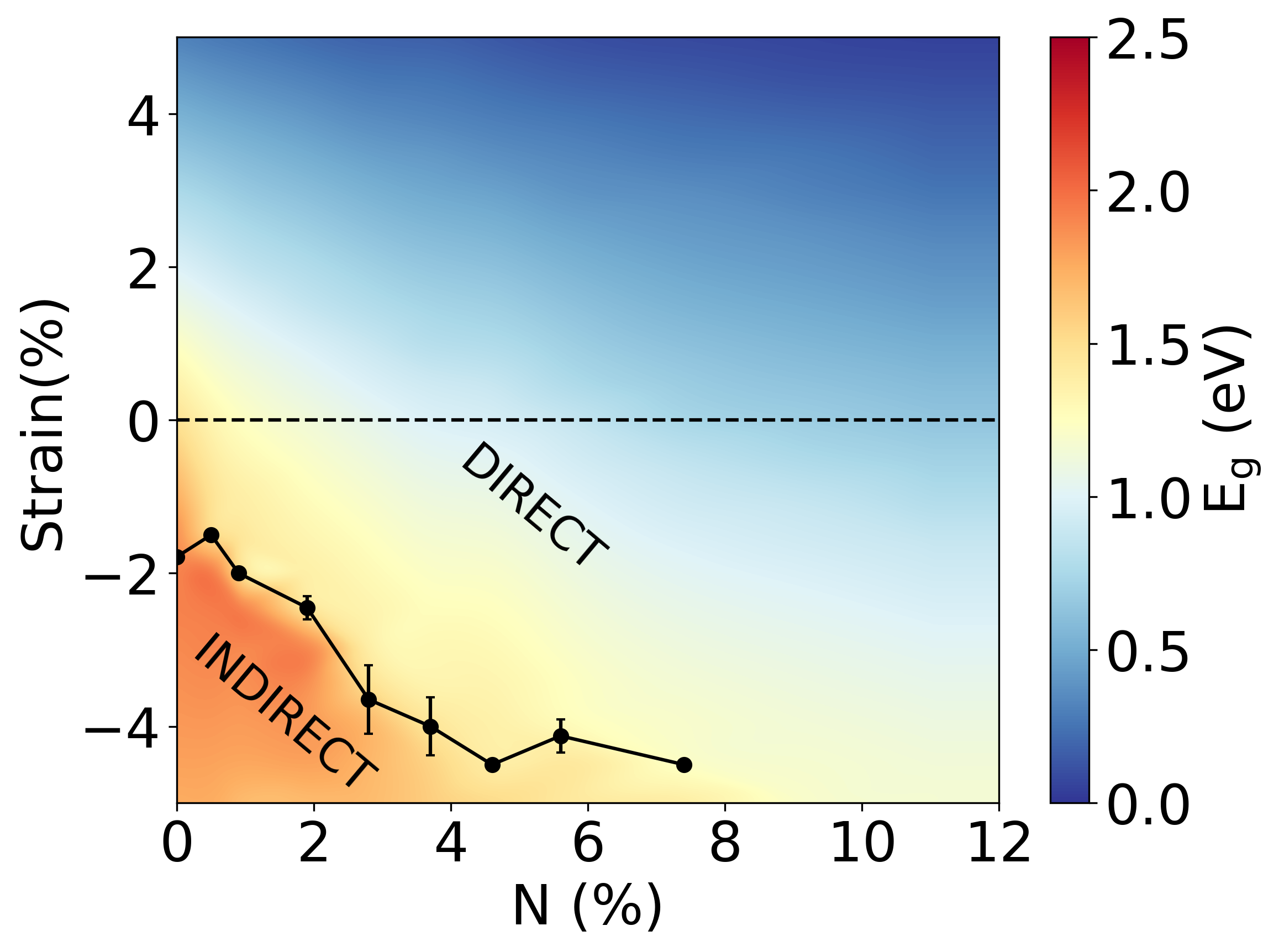}
    \caption{Isotropic strain for GaAsN (up to 12\% N). The variation of bandgap magnitudes (E$_{\text{g}}$) and type as a function of composition and strain. The dashed black horizontal line indicates unstrained GaAsN. The black circles are the calculated DIT points. Beyond 7\% N, the DIT is outside the investigated strain regime. 10 SQS cells are used for each configuration and strain point. The bandgaps plotted are the average bandgaps. The error bars indicate the standard deviation in DIT points estimation. The direct and indirect enclosed regions describe the nature of bandgap being direct and indirect, respectively.}
    \label{fig:fig4}
\end{figure}
\begin{figure}
    \centering
    \includegraphics[width=3.4in]{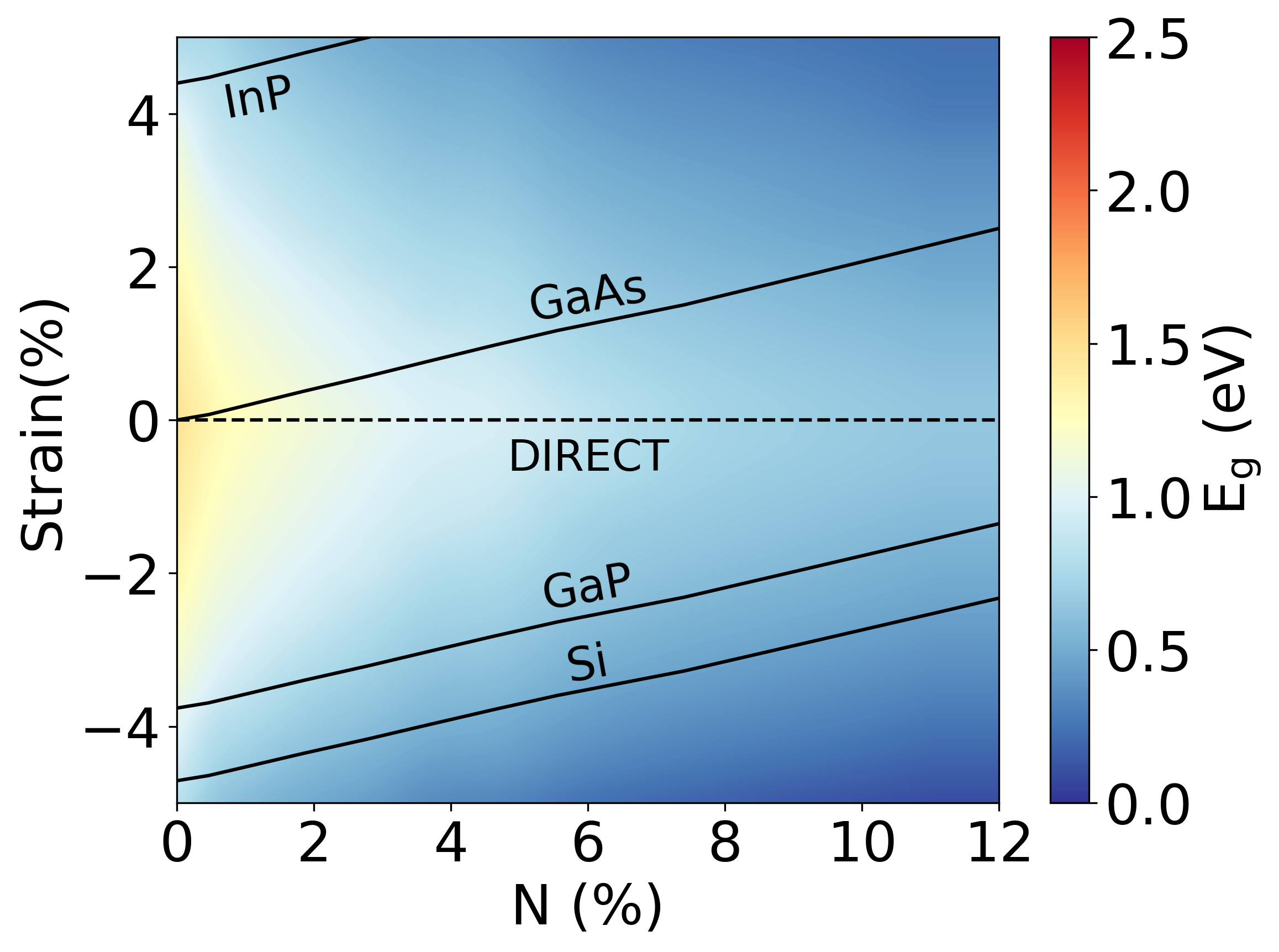}
    \caption{Biaxial strain for GaAsN (up to 12\% N). The variation of bandgap magnitudes (E$_{\text{g}}$) and type as a function of composition and strain. The dashed black horizontal line indicates unstrained GaAsN. 10 SQS cells are used for each configuration and strain point. The bandgaps plotted are the average bandgaps. All the bandgaps are direct in nature. Solid black lines indicate the substrate lines under the ``epitaxial growth'' model.}
    \label{fig:fig5}
\end{figure}
As the next material, we investigate GaAsN. First, we show results for isotropic strain, which results in the bandgap phase diagram shown in \cref{fig:fig4}. The results are markedly different from GaAsP, and the data set is much more limited. In this case, we found a strong dependency of the bandgap on the N atoms distribution in the supercell \cite{Rosenow2018Ab}. We thus used 10 SQS cells for each data point in the figure and averaged the resulting bandgaps. This results in an error bar for the DIT points, which is rather large for medium amounts of nitrogen atoms due to the formation of small clusters and chains. Calculations were only possible for up to 12\% N. For higher concentration and/or high compressive strain, our chosen supercell is not large enough to avoid the unphysical electronic interaction of N atoms with their images in the periodic boundary condition approach. This effect has already been discussed in Ref.~\cite{Rosenow2018Ab}. For the strain and composition regions where computation was possible, an indirect gap is only found for low values of \% N and rather large compressive strain values. The EBSs for selected \% N and strain values are shown in Figs.~S3 and S4 \cite{Supp_info}. 

For biaxial strain in GaAsN, the data are shown in \cref{fig:fig5}. In contrast to GaAsP, the bandgap gets smaller with the increasing amount of nitrogen in the system, from 1.47 eV for the unstrained case of GaAs to 0.10 eV for the highly strained systems with a large number of N atoms. All bandgaps computed are direct. Epitaxial growth on GaAs is reasonably possible for moderate strain values and results in a variation of bandgap from 1.47 eV to 0.45 eV. For GaP and Si substrates, a large strain would be exerted on the system, and mostly lower bandgap values are found. 

\subsection{\label{subsec43:gapsbgaassbgapbigaasbi}G\lowercase{a}PS\lowercase{b}, G\lowercase{a}A\lowercase{s}S\lowercase{b}, G\lowercase{a}PB\lowercase{i}, G\lowercase{a}A\lowercase{s}B\lowercase{i}}
\begin{figure*}
\centering
    \subfloat[GaPSb]{\label{fig:fig6a}\includegraphics[width=3.4in]{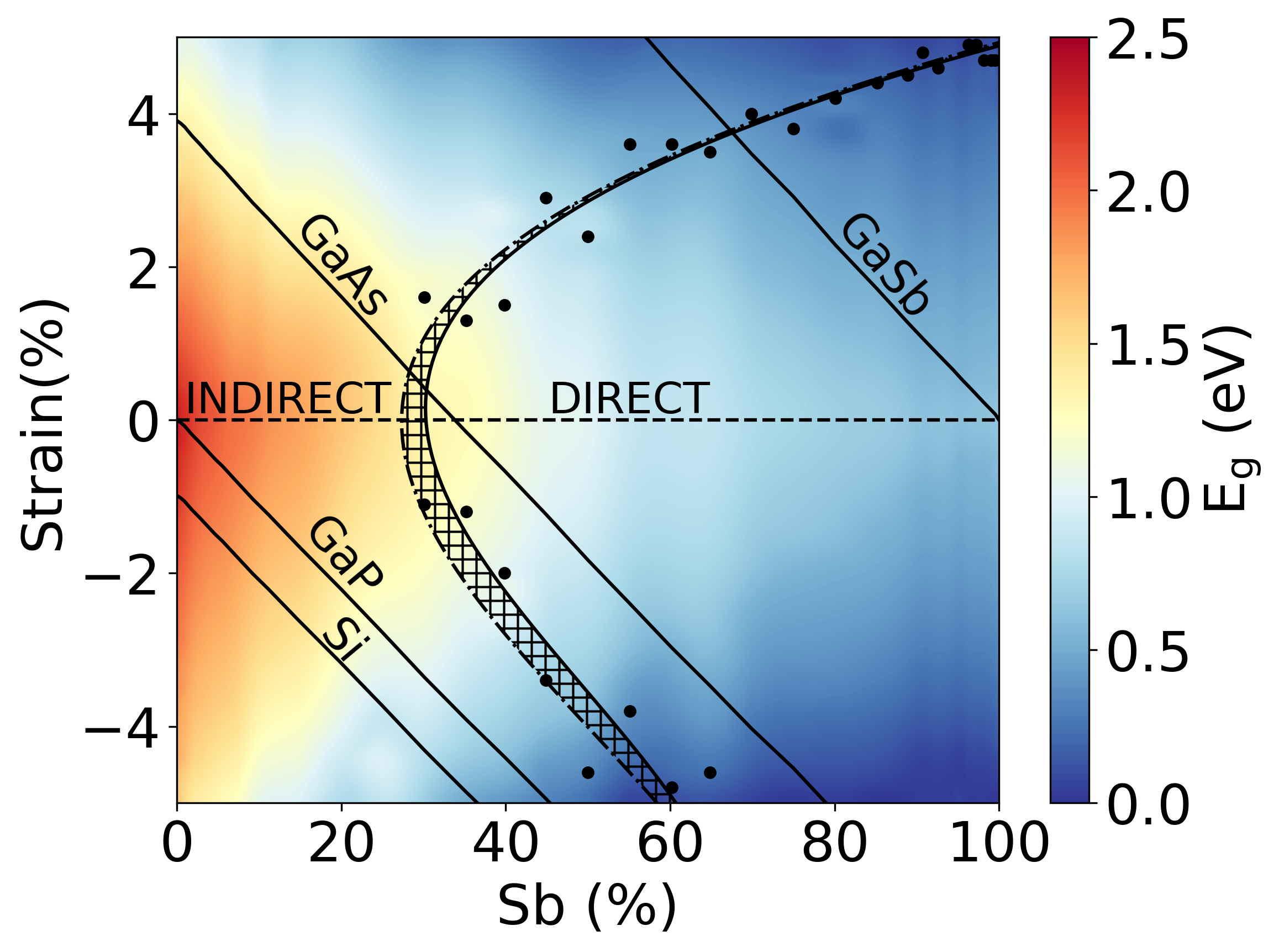}}
    \hspace{0.2in}
    \subfloat[GaAsSb]{\label{fig:fig6b}\includegraphics[width=3.4in]{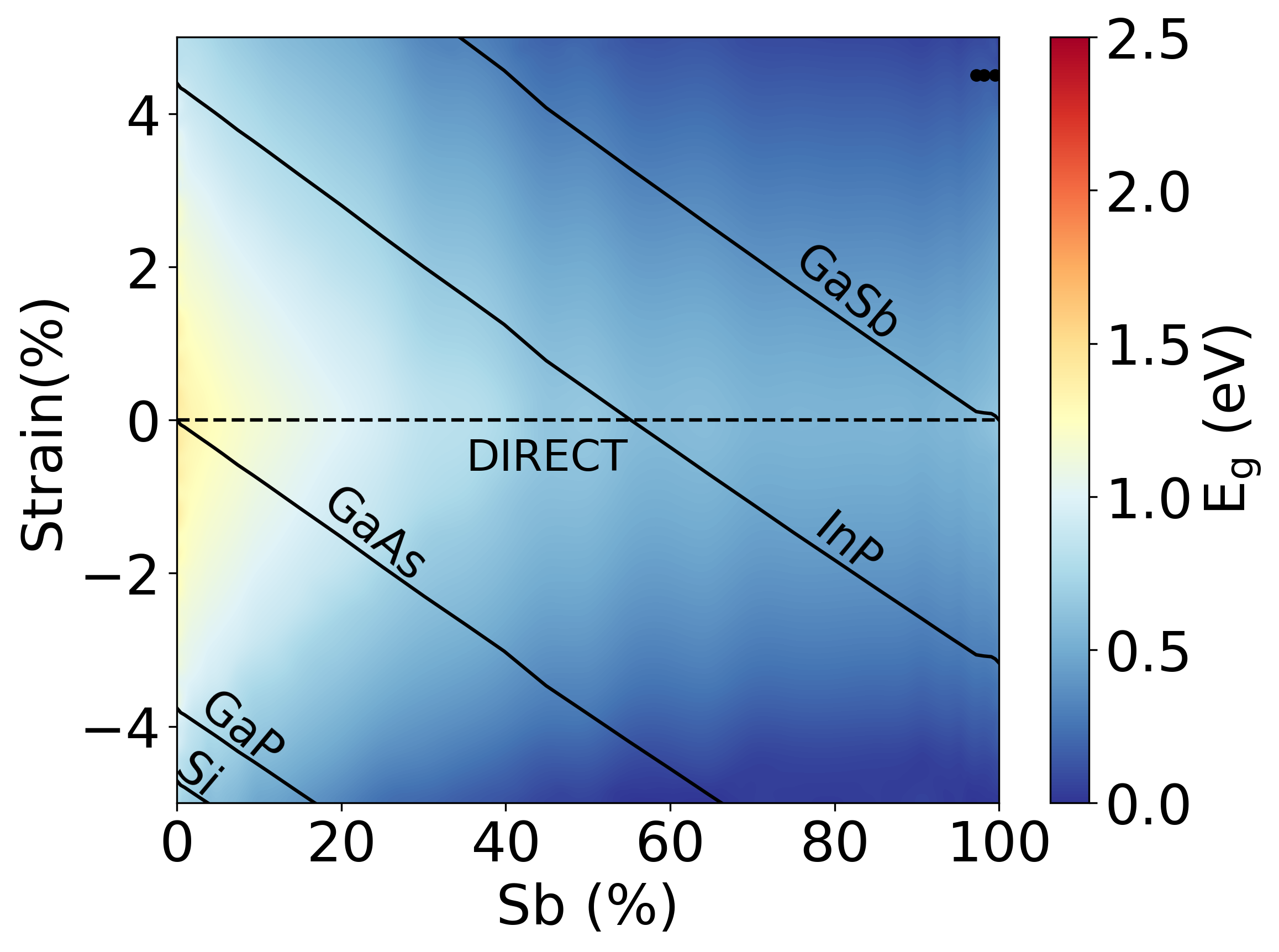}} \\
    \subfloat[GaPBi]{\label{fig:fig6c}\includegraphics[width=3.4in]{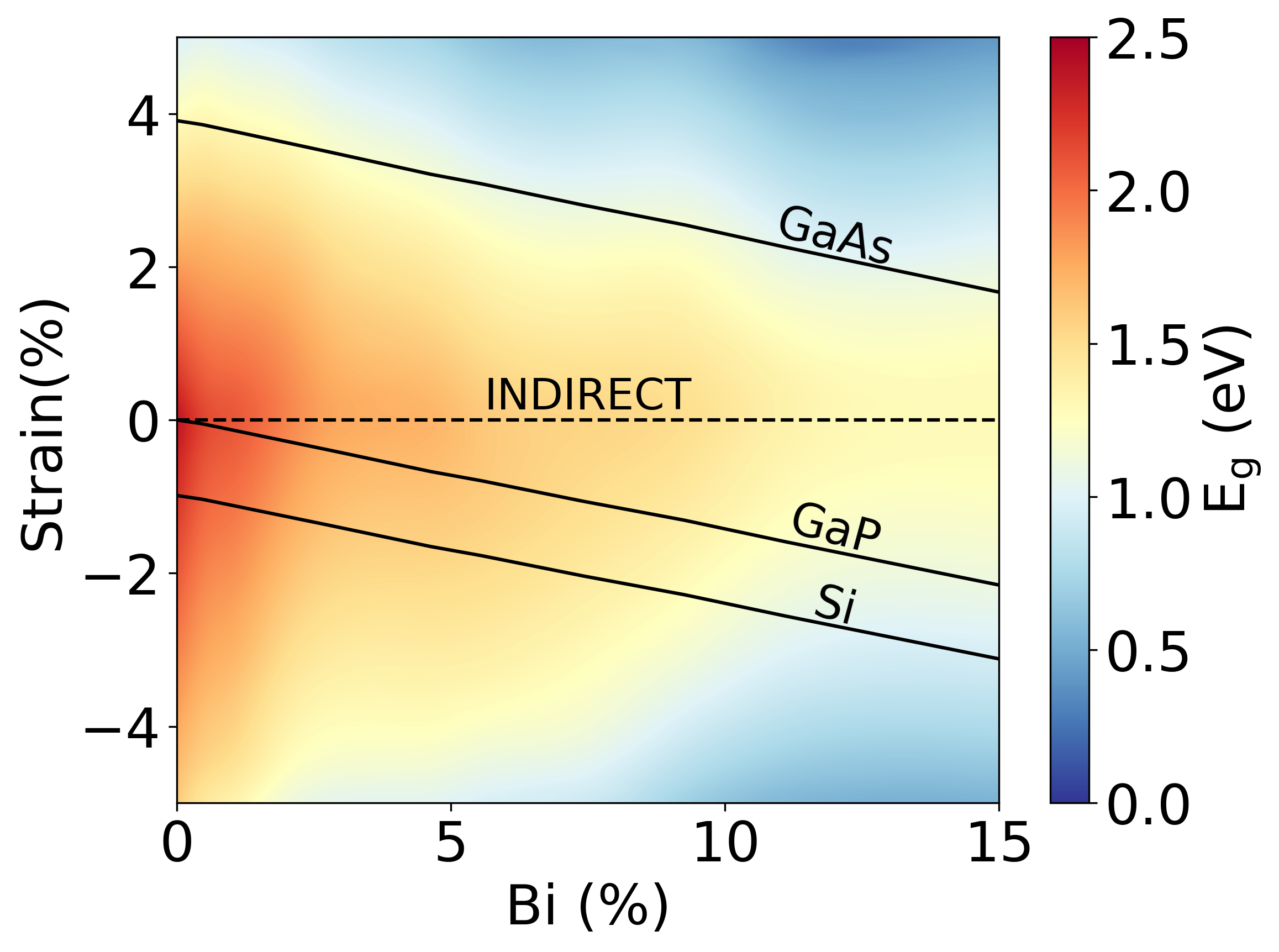}}
    \hspace{0.2in}
    \subfloat[GaAsBi]{\label{fig:fig6d}\includegraphics[width=3.4in]{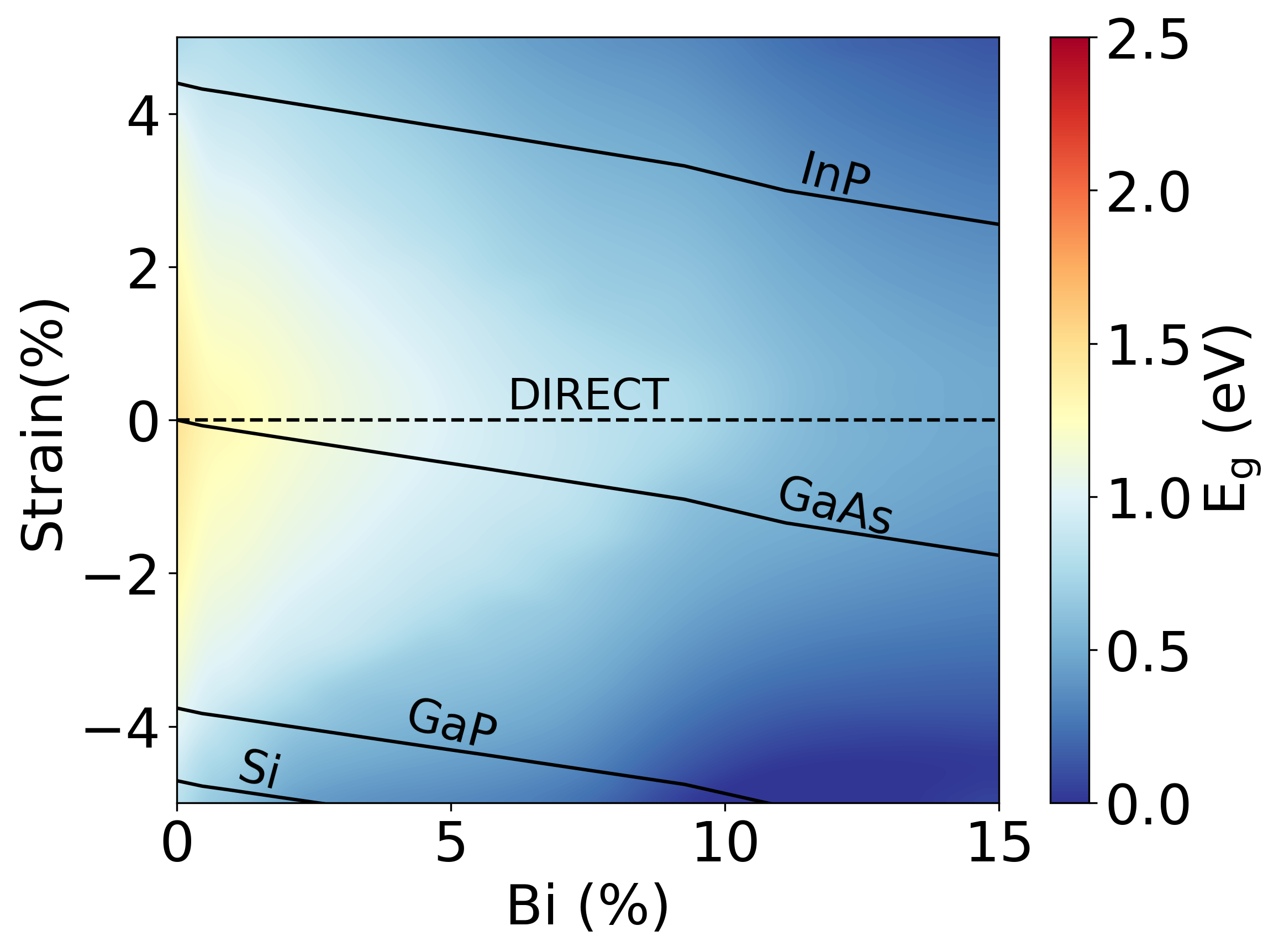}}
  \caption{\label{fig:fig6}Bandgap phase diagram for ternary III-V semiconductors GaEY (E= P, As; Y = Sb, Bi) under biaxial strain. The bandgap magnitudes (E$_{\text{g}}$) are shown in the color bar. The dashed black horizontal line indicates unstrained structures. The black circles are the calculated DIT points. The direct and indirect enclosed regions describe the nature of bandgap being direct and indirect, respectively. The hatched pattern region is the ``uncertainty region'' (see \cref{sec3:protocolfordeterminingbandgapnature}). Solid black lines indicate the substrate lines under the ``epitaxial growth'' model.}
\end{figure*}
The approach outlined here can be extended to other combinations of elements in III-V semiconductor materials. Exemplarily, we present the bandgap phase diagrams for four other important ternary compounds in \cref{fig:fig6}. Since epitaxial growth is the most interesting experimental realization method for these compounds, we only present the data for biaxial strain.

For all compounds investigated, we find a bandgap range of 0.00--2.36 eV, with the largest values found for the host materials GaP and GaAs in the unstrained case. The alloys with Sb could be investigated over the full range of 0--100\% Sb in GaPSb and GaAsSb (\cref{fig:fig6a,fig:fig6b}). We find a DIT for unstrained GaPSb at 30\% Sb concentration (\cref{fig:fig6a}). For this compound, the DIT is shown as a region, including the uncertainty in determining the nature of the bandgap, as outlined in \cref{sec3:protocolfordeterminingbandgapnature}. With the increase in the Sb fraction, the strain at which the DITs take place increases. For GaAsSb and GaAsBi, the bandgap is direct throughout the range investigated (\cref{fig:fig6b,fig:fig6d}), while it is indirect for GaPBi (\cref{fig:fig6c}). Notably, we only investigate the bismides up to a fraction of 15\% Bi. The reason is that, similar to GaAsN (Fig.~S4 \cite{Supp_info}), for structures with large Bi content, the strongly dispersed bands decrease the reliability in the determination of bandgap nature.  Additionally, GaPBi and GaAsBi become metallic for higher Bi fractions. Although we find no transition within 15\% Bi, it can not be excluded that the DIT appears at higher percentages of bismuth.

Again, we indicate the strain values associated with different typical substrates for epitaxial growth by solid black lines in the figures. The data show that deviating from the substrate-layer lattice-matching condition quickly leads to high strain, and defects are highly likely to occur during growth. Also, Si can be used as a substrate for GaPSb and GaPBi epitaxial growth if the Sb or Bi content is not too large. The epitaxial growth of the respective GaAs-based materials (GaAsP, GaAsN, GaAsSb, GaAsBi) will give rise to high strain on Si-substrate throughout the whole composition region. A noticeable change in the slope in substrate lines is found close to 10\% Bi and 40\% Sb concentration in GaAsBi and GaAsSb, respectively. Although we did not find any structural phase transition in those regions, the origin of the change in the slope is not clear to us yet.

From the above discussions, it becomes clear that bandgap phase diagrams can be a valuable aid in deciding which substrates are good choices for targeting a specific bandgap size and nature for a given ternary material. And vice versa, which material to grow for a specific application and a given substrate? We will discuss this further in the next section.

\section{\label{sec5:discussion}Discussion}
All data were derived from DFT computations to this point. In \cref{tab:table1}, we now compare our calculated bandgaps with experimental data from measurements on heteroepitaxial layer structures. The GaAsP/GaAs samples were grown by low-pressure hydride vapor phase epitaxy (LP-HVPE). Further details can be found in Ref.~\cite{Stromberg2020}. The remaining samples were grown by metalorganic vapor phase epitaxy (MOVPE). The details of the growth characteristics of the MOVPE samples can be found in Refs.~\cite{Hepp2019Movpe,Jou1988,Nattermann2017,Ludewig2013,Ludewig2017,Lewis2012,Masnadi-Shirazi2014,Ludewig2014GrowthMOVPE}.  Experimentally, the layer thickness and bandgaps of the MOVPE samples were determined using X-ray diffraction and room-temperature photoluminescence (RT-PL), respectively. Except for GaPSb samples from Ref.~\cite{Jou1988}, in which cases, the PL were measured at 10 K. 
\begin{table*}
\caption{\label{tab:table1}Comparison of the calculated bandgaps for investigated ternary III-V semiconductors under biaxial strain with experiments. The experimental data are for the heteroepitaxial layer structures, and the bandgaps are determined from photoluminescence (PL) measurements. The ``I'' in the brackets indicate the indirect nature of the bandgap. The remaining bandgaps are direct. (RT-PL: room temperature PL)}
\begin{ruledtabular}
\begin{tabular}{ccccccccc}
 & & & Layer &\multicolumn{2}{c}{Bandgap (eV)} & Deviation & Percentage & RMSD \\ \nmidrule{5-6}
 System & Substrate & $x(\%)$ & thickness (nm) & Calculated & Experiment & (eV) & deviation (\%) & (eV) \\ \nhrule 
 \multirow{3}{*}{GaAs$_{1-x}$P$_x$} & \multirow{3}{*}{GaAs} & 18.0 & 9000 & 1.65 & 1.66 & 0.01 & 0.6 & \multirow{3}{*}{0.01} \\
 & & 25.0 & 12250 & 1.72 & 1.72 & 0.00 & 0.0 & \\
 \cite{Stromberg2020}& & 28.0 & 13000 & 1.75 & 1.76 & 0.01 & 0.6 &  \\ \nhrule

 \multirow{5}{*}{GaP$_{1-x}$Sb$_x$\footnote{For GaPSb samples, no specific thicknesses were reported in the reference.}} & GaP & 14.0 & $-$ & 1.66(I) & 1.61(I) & $-0.05$ & $-3.1$ & \multirow{5}{*}{0.10} \\ \nmidrule{2-8}
 & \multirow{3}{*}{GaAs} & 29.0 & $-$ & 1.33(I) & 1.39(I) & 0.06 & 4.3 & \\
 & & 32.0 & $-$ & 1.30 & 1.31 & 0.01 & 0.8 & \\
 \cite{Jou1988} & & 37.0 & $-$ & 1.24 & 1.33 & 0.09 & 6.8 & \\ \nmidrule{2-8}
 & GaSb & 93.0 & $-$ & 0.56 & 0.74 & 0.18 & 24.3 & \\ \nhrule

\multirow{2}{*}{GaAs$_{1-x}$Sb$_x$} & \multirow{2}{*}{GaAs} & 5.5 & 46.3 & 1.22 & 1.34 & 0.12 & 9.0 & \multirow{2}{*}{0.13} \\
 & & 7.0 & 51.2 & 1.17 & 1.31 & 0.14 & 10.7 & \\ \nhrule

 \multirow{8}{*}{GaAs$_{1-x}$Bi$_x$} & \multirow{8}{*}{GaAs} & 0.9 & 75.0 & 1.28 & 1.33 & 0.05 & 3.8 & \multirow{8}{*}{0.10} \\
 & & 1.9 & 67.0 & 1.19 & 1.26 & 0.07 & 5.6 & \\
 & & 2.9 & 60.0 & 1.10 & 1.20 & 0.10 & 8.3 & \\ 
 & & 3.2 & 59.0 & 1.08 & 1.18 & 0.10 & 8.5 & \\ 
 & & 3.8 & 54.0 & 1.04 & 1.14 & 0.10 & 8.8 & \\ 
 \cite{Nattermann2017,Ludewig2013,Lewis2012} & & 4.8 & 25.0 & 0.98 & 1.11 & 0.13 & 11.7 & \\ 
 \cite{Masnadi-Shirazi2014,Ludewig2014GrowthMOVPE,Mohmad2011TheGaAs1xBix}& & 5.3 & 50.0 & 0.95 & 1.07 & 0.12 & 11.2 & \\ 
 & & 6.0 & 25.0 & 0.91 & 1.04 & 0.13 & 12.5 & \\ \nhrule

 \multirow{5}{*}{GaAs$_{1-x}$N$_x$} & \multirow{5}{*}{GaAs} & 1.2 & 6.3 & 1.20 & 1.25 & 0.05 & 4.0 & \multirow{5}{*}{0.12} \\
 & & 2.0 & 17.0 & 1.10 & 1.16 & 0.06 & 5.2 & \\
 & & 2.3 & 7.0 & 1.06 & 1.17 & 0.11 & 9.4 & \\ 
 \cite{Vurgaftman2001Band,Rosenow2018Ab,Ludewig2017}& & 2.9 & 7.0 & 1.00 & 1.11 & 0.11 & 9.9 & \\ 
 & & 5.0 & 4.0 & 0.82 & 1.01 & 0.19 & 18.8 & \\ \nhrule
 \multirow{2}{*}{GaAs$_{1-x}$Sb$_x$} & \multirow{2}{*}{GaAs} & 27.8 & 3.7 & 0.67 & 1.10 & 0.43 & 39.1 & \multirow{2}{*}{0.42} \\
 & & 28.0 & 4.1 & 0.66 & 1.07 & 0.41 & 38.3 & \\ \nhrule
 \multirow{5}{*}{GaAs$_{1-x}$N$_x$} & GaP & 4.9 & 6.0 & 0.66 & 1.18 & 0.52 & 44.1 & \multirow{5}{*}{0.66}  \\ \nmidrule{2-8}
 & \multirow{4}{*}{Si} & 6.9 & 5.5 & 0.47 & 1.21 & 0.74 & 61.2 &\\
 & & 8.9 & 5.5 & 0.46 & 1.17 & 0.71 & 60.7 & \\
 \cite{Vurgaftman2001Band,Rosenow2018Ab,Ludewig2017}& & 9.5 & 6.0 & 0.45 & 1.11 & 0.66 & 59.5 & \\ 
 & & 10.9 & 5.4 & 0.45 & 1.11 & 0.66 & 59.5 & \\ \nhrule
 GaP$_{1-x}$Sb$_x$ & Si & $\leq 17.5$ & 7--9 & 1.91--1.36(I) & \multicolumn{3}{c}{\textrm{No RT-PL observed}} \\ \nhrule
 GaP$_{1-x}$Bi$_x$ & \multirow{2}{*}{GaP} & \multirow{2}{*}{$\leq 12.0$} & \multirow{2}{*}{17--73} & \multirow{2}{*}{2.03--1.18(I)} & \multicolumn{3}{c}{\multirow{2}{*}{No RT-PL observed}} \\
 \cite{Nattermann2017} & & & & &\multicolumn{3}{c}{}  \\
\end{tabular}
\end{ruledtabular}
\end{table*}

The comparison of the experimental bandgaps with our computed results shows good agreement. The deviation is determined with respect to the root-mean-square deviation (RMSD) from all available experimental samples. For most structures, the RMSD is around 0.1 eV. Most computed values deviate by less than 10\% from the experimental values (exceptions are discussed separately); in the case of GaAsP, the deviation is even more accurate ($<$1\%). This confirms our previous findings on unstrained structures that the DFT protocol we developed gives excellent agreement to experimental bandgaps \cite{Beyer2017Local,Bannow2017,Rosenow2018Ab,Mondal2022}. In this study, we show that it is also applicable to compound semiconductors under strain. For samples with very small layer thickness, the matching of experiment and computation is less good. This can be observed for GaAsSb/GaAs thin samples with RMSD of ca. 0.4 eV. We attribute this to the 2D quantum confinement effect, which is found for thin samples. This confinement effect leads to an increase in the bandgap with respect to thicker samples \cite{Cipriano2020}. This effect is not captured in our computational model as the calculations were performed for 3D periodic strained structures. The large deviation observed for GaAsN/GaP and GaAsN/Si samples can not be explained by this effect alone, though. An additional effect here is the strong dependency of the bandgap on the distribution of N atoms which has been found for unstrained GaAsN before \cite{Rosenow2018Ab}. The dependency is further amplified under large strain (around 3\%, see \cref{fig:fig5}) in those samples. In the case of GaAsN/GaAs samples, where the N concentration investigated was around 1--5\%, the strain is relatively small ($<$1\%), resulting in better agreement with the experiment as compared to the GaAsN/GaP and GaAsN/Si samples. No RT-PL was observed for GaPSb/Si and GaPBi/GaP samples. 
\begin{figure*}
\centering
  \subfloat[]{\label{fig:fig7a}\includegraphics[width=4.9in]{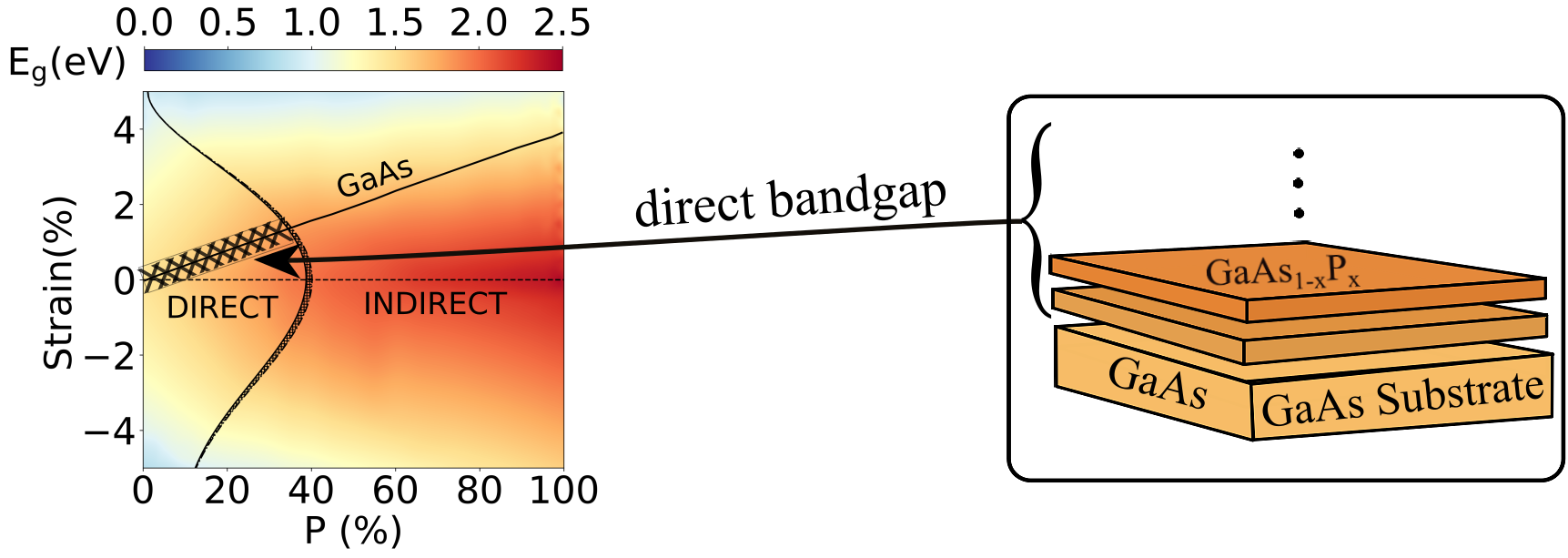}}
  \hfill
  \subfloat[]{\label{fig:fig7b}\includegraphics[width=4.9in]{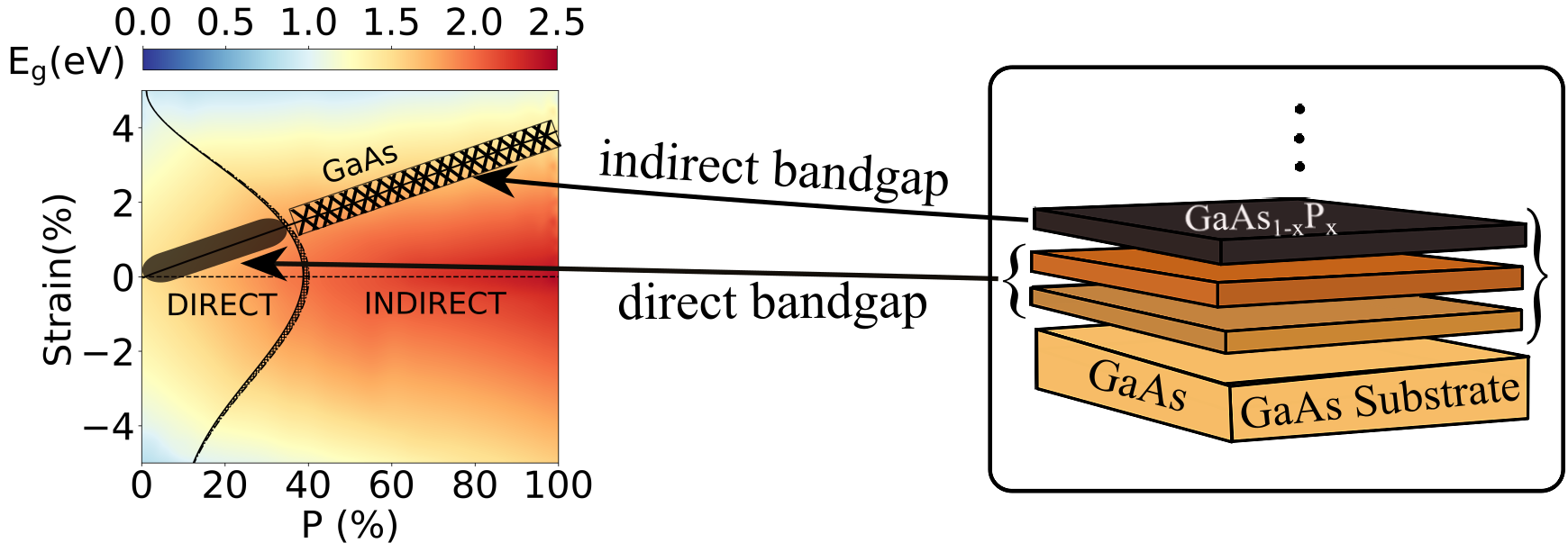}}
  \hfill
  \subfloat[]{\label{fig:fig7c}\includegraphics[width=4.9in]{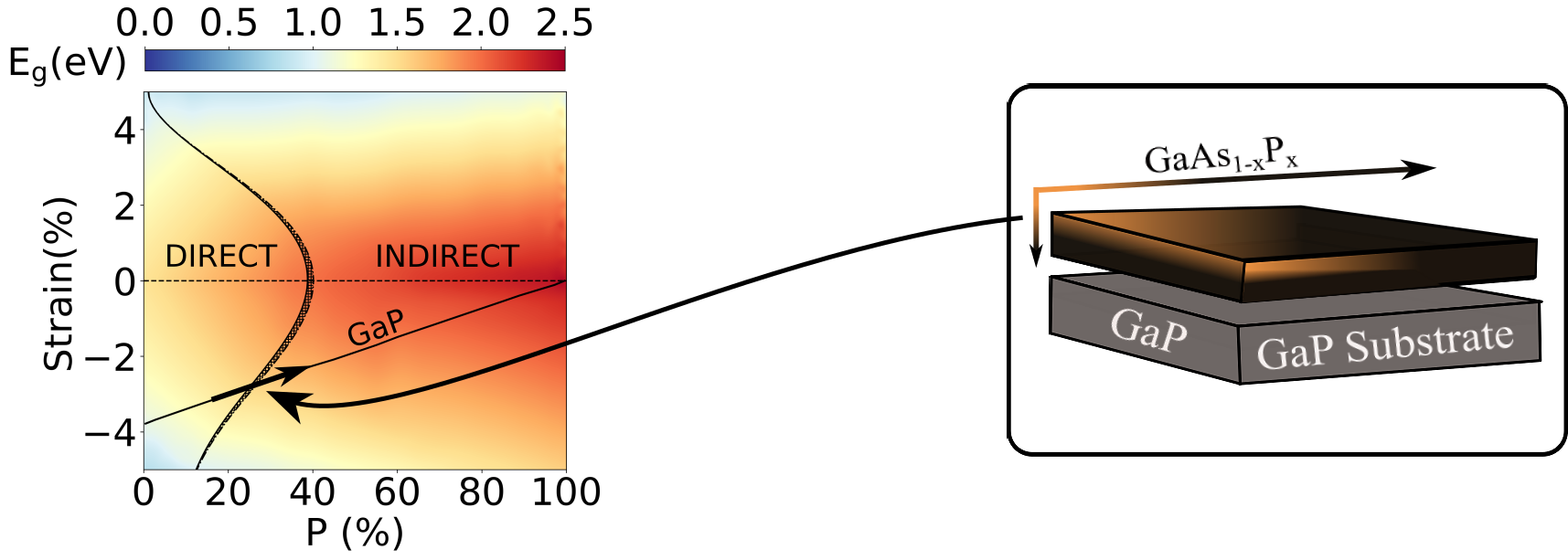}}
  \hfill
  \subfloat[]{\label{fig:fig7d}\includegraphics[width=4.9in]{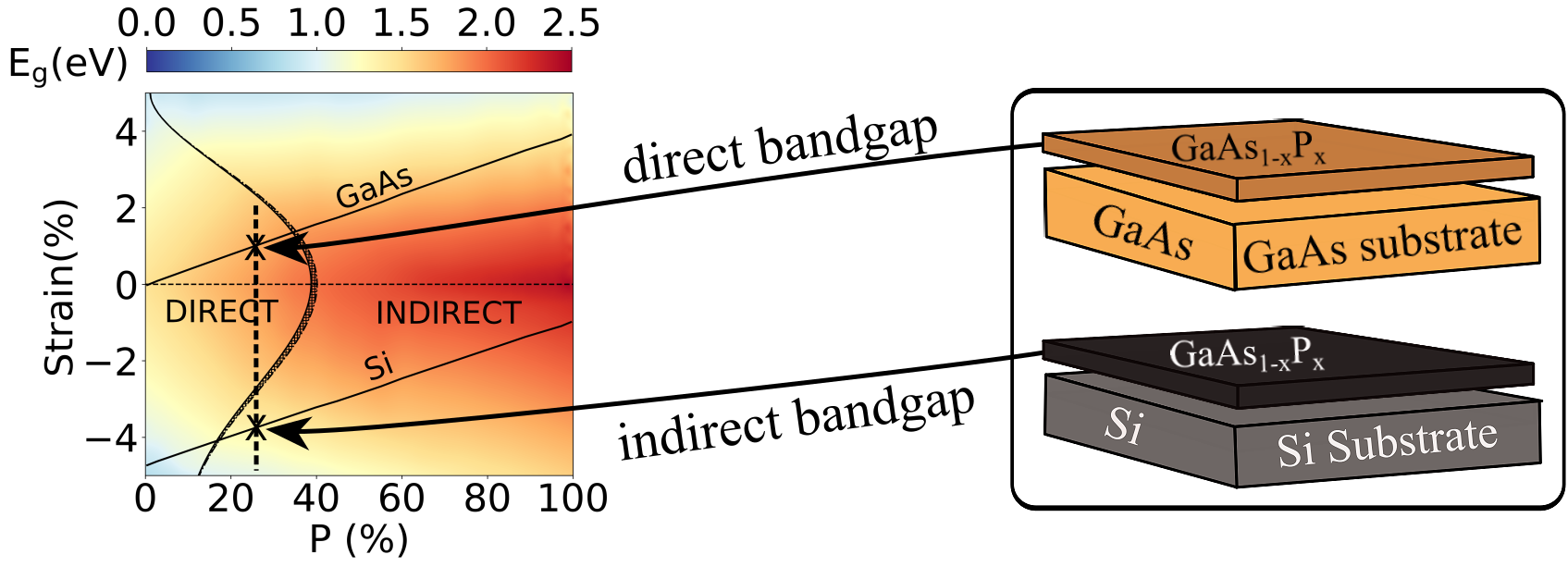}} 
  \caption{Proposals on how the bandgap phase diagram of biaxially strained GaAsP can be used in designing optoelectronic devices. (a) Defines the bound of composition region for creating a QWH with direct bandgap GaAsP on GaAs substrate. (b) Choosing the different composition regions appropriately to make a multijunction photovoltaic with successive direct and indirect cells on the GaAs substrate. (c) In the vicinity of the transition point, the bandgap properties of the GaAsP epilayer on the GaP substrate can be changed by appropriately varying the composition. (d) Depending on the choice of substrate, GaAs or Si, the particular composition indicated by the vertical line can be made direct or indirect bandgap, respectively.}
  \label{fig:fig7}
\end{figure*}
This is consistent with our findings that those materials show indirect bandgaps (\cref{fig:fig6a,fig:fig6c}). Experimental measurements of the magnitude of the indirect bandgaps are not available yet.

The consistent agreement between the experiment and calculated bandgaps (both in magnitude and in nature) suggests that we are able to quantitatively predict the bandgap over a wide range of compounds, compositions, and strain regions. However, as discussed above, the effect of 2D confinement is also crucial for relatively thin quantum well heterostructures and, hence, needs further investigation. 

Finally, based on the bandgap phase diagram, we propose several design strategies to optimize the selection of material combinations for achieving specific optical applications and new design principles for devices (\cref{fig:fig7}). 

In \cref{fig:fig7a}, we propose a quantum-well heterostructure (QWH) composed of biaxially strained GaAsP on GaAs substrate. As the QW layers are made out of a single material with varied composition only, the epitaxial growth could be performed efficiently. The bandgap phase diagram shows the areas in compositional phase space where a direct bandgap in GaAs$_{1-x}$P$_x$ can be achieved ($x <$ 34\%). For $x >$ 35\%, the bandgaps are indirect and hence, are inappropriate for the heterostructure.

Figure~\ref{fig:fig7b} shows an efficient approach for the monolithic integration of multiple QWH to construct multijunction photovoltaics. In this case, the QWHs are separated by thin indirect bandgap layers of the same material as QWH but only with a different composition. This would make the integration approach efficient, as no sample transfer is required during growth.


In \cref{fig:fig7c}, we propose a device with a gradual change in the bandgap properties. The concept utilizes the continuous transition in the nature of bandgap with alloy concentration in the vicinity of the DIT region. At the amount of P chosen here ($x =$ 15--35\%), we propose to grow the GaAsP epitaxial layer on GaP with P concentration continuously changing from the direct to indirect bandgap region or vice versa. This way, changes in the bandgap magnitude, as well as the nature of the bandgap, are possible. Note that the concentration gradient can be implemented both in the horizontal and vertical directions. 

Figure~\ref{fig:fig7d} shows another application of this concept. By appropriately choosing the substrate, we can tune the epitaxial layer (here: GaAsP) to show either a direct or indirect bandgap. Depending on the substrate, GaAs, or Si, the particular composition indicated by the vertical line will show direct or indirect bandgap, respectively.

\section{\label{sec6:summary}Summary}
Using density functional theory and the concept of band unfolding, we developed a first-principles computational protocol for the comprehensive mapping of the bandgap magnitude and type over a wide range of composition and strain values for several ternary III-V semiconductors. We constructed the composition-strain-bandgap relationship, the bandgap phase diagram, for several ternary III-V semiconductors: GaAsP, GaAsN, GaPSb, GaAsSb, GaPBi, and GaAsBi. We showed that this way of mapping the effect of strain could be used to choose application-specific best-suited material systems and hence, is highly beneficial to device design. In addition, we developed an efficient approach based on Bloch spectral density for determining the nature of bandgap from supercell calculation. Notably, our computational protocol can be generalized to explore the vast chemical space of III-V materials with all other possible combinations of III- and V-elements. The comparison to experimental bandgap data underlines the accuracy of the computational approach chosen. This approach will be extended to more complex materials in the future. 

\begin{acknowledgments}
We thank the German Research Foundation (DFG) in the framework of the Research Training Group ``Functionalization of Semiconductors" (GRK 1782) for funding this project and to HRZ Marburg, GOETHE-CSC Frankfurt, ZIH Dresden, and HLR Stuttgart for providing the computational resources.
\end{acknowledgments}

\section*{\label{datavailabilitystatement}DATA AVAILABILITY STATEMENT}
The density functional theory calculations data are openly available in the NOMAD repository (\url{https://doi.org/10.17172/NOMAD/2023.02.27-1}). The interactive bandgap phase diagrams (in HTML format) are available in the Supplemental Material \cite{Supp_info}. To view the diagrams, open the HTML files in a web browser. Alternatively, the diagrams can be viewed directly on GitHub (\url{https://bmondal94.github.io/Bandgap-Phase-Diagram/}, last accessed 10.05.2023).\\

\section*{\label{ordisids}ORCID \lowercase{i}D\lowercase{s}}
\noindent Badal Mondal \\ \hspace*{10pt} \url{https://orcid.org/0000-0002-0522-1254}\\
Ralf Tonner-Zech \\ \hspace*{10pt} \url{https://orcid.org/0000-0002-6759-8559}\\
Thilo Hepp \\ \hspace*{10pt} \url{https://orcid.org/0000-0002-0782-987X}\\
Kerstin Volz \\ \hspace*{10pt} \url{https://orcid.org/0000-0002-4456-5439}

\pagebreak
\section*{\label{supportinginformation}Supplemental Material}
The Supplemental Material contains the effective bandstructures of GaAsP, GaAsN, and GaAsN under selected isotropic strain values; a visual representation of the assessment of uncertainty in the bandgap nature near the direct-indirect transition region; the DFT calculated bandgap values for all the systems described in the manuscript; the details of the interpolation procedure; and the bandgap phase diagrams.

\twocolumngrid

\bibliographystyle{apsrev4-2}
\bibliography{ms}

\providecommand{\noopsort}[1]{}\providecommand{\singleletter}[1]{#1}
\begin{thebibliography}{119}%
\makeatletter
\providecommand \@ifxundefined [1]{%
 \@ifx{#1\undefined}
}%
\providecommand \@ifnum [1]{%
 \ifnum #1\expandafter \@firstoftwo
 \else \expandafter \@secondoftwo
 \fi
}%
\providecommand \@ifx [1]{%
 \ifx #1\expandafter \@firstoftwo
 \else \expandafter \@secondoftwo
 \fi
}%
\providecommand \natexlab [1]{#1}%
\providecommand \enquote  [1]{``#1''}%
\providecommand \bibnamefont  [1]{#1}%
\providecommand \bibfnamefont [1]{#1}%
\providecommand \citenamefont [1]{#1}%
\providecommand \href@noop [0]{\@secondoftwo}%
\providecommand \href [0]{\begingroup \@sanitize@url \@href}%
\providecommand \@href[1]{\@@startlink{#1}\@@href}%
\providecommand \@@href[1]{\endgroup#1\@@endlink}%
\providecommand \@sanitize@url [0]{\catcode `\\12\catcode `\$12\catcode
  `\&12\catcode `\#12\catcode `\^12\catcode `\_12\catcode `\%12\relax}%
\providecommand \@@startlink[1]{}%
\providecommand \@@endlink[0]{}%
\providecommand \url  [0]{\begingroup\@sanitize@url \@url }%
\providecommand \@url [1]{\endgroup\@href {#1}{\urlprefix }}%
\providecommand \urlprefix  [0]{URL }%
\providecommand \Eprint [0]{\href }%
\providecommand \doibase [0]{https://doi.org/}%
\providecommand \selectlanguage [0]{\@gobble}%
\providecommand \bibinfo  [0]{\@secondoftwo}%
\providecommand \bibfield  [0]{\@secondoftwo}%
\providecommand \translation [1]{[#1]}%
\providecommand \BibitemOpen [0]{}%
\providecommand \bibitemStop [0]{}%
\providecommand \bibitemNoStop [0]{.\EOS\space}%
\providecommand \EOS [0]{\spacefactor3000\relax}%
\providecommand \BibitemShut  [1]{\csname bibitem#1\endcsname}%
\let\auto@bib@innerbib\@empty
\bibitem [{\citenamefont {Soref}(1993)}]{Soref1993Silicon}%
  \BibitemOpen
  \bibfield  {author} {\bibinfo {author} {\bibfnamefont {R.}~\bibnamefont
  {Soref}},\ }\href {https://doi.org/10.1109/5.248958} {\bibfield  {journal}
  {\bibinfo  {journal} {Proc. IEEE}\ }\textbf {\bibinfo {volume} {81}},\
  \bibinfo {pages} {1687} (\bibinfo {year} {1993})}\BibitemShut {NoStop}%
\bibitem [{\citenamefont {Yu}\ and\ \citenamefont
  {Cardona}(2010)}]{CardanoBook}%
  \BibitemOpen
  \bibfield  {author} {\bibinfo {author} {\bibfnamefont {P.~Y.}\ \bibnamefont
  {Yu}}\ and\ \bibinfo {author} {\bibfnamefont {M.}~\bibnamefont {Cardona}},\
  }\href {https://books.google.de/books?id=4VCxDAEACAAJ} {\emph {\bibinfo
  {title} {Fundamentals of Semiconductors: Physics and Materials
  Properties}}},\ Graduate Texts in Physics\ (\bibinfo  {publisher} {Springer
  Berlin Heidelberg},\ \bibinfo {year} {2010})\BibitemShut {NoStop}%
\bibitem [{\citenamefont {Hepp}\ \emph {et~al.}(2022)\citenamefont {Hepp},
  \citenamefont {Lehr}, \citenamefont {G{\"{u}}nkel}, \citenamefont
  {Ma{\ss}meyer}, \citenamefont {Glowatzki}, \citenamefont {Ruiz~Perez},
  \citenamefont {Reinhard}, \citenamefont {Stolz},\ and\ \citenamefont
  {Volz}}]{Hepp2022Room}%
  \BibitemOpen
  \bibfield  {author} {\bibinfo {author} {\bibfnamefont {T.}~\bibnamefont
  {Hepp}}, \bibinfo {author} {\bibfnamefont {J.}~\bibnamefont {Lehr}}, \bibinfo
  {author} {\bibfnamefont {R.}~\bibnamefont {G{\"{u}}nkel}}, \bibinfo {author}
  {\bibfnamefont {O.}~\bibnamefont {Ma{\ss}meyer}}, \bibinfo {author}
  {\bibfnamefont {J.}~\bibnamefont {Glowatzki}}, \bibinfo {author}
  {\bibfnamefont {A.}~\bibnamefont {Ruiz~Perez}}, \bibinfo {author}
  {\bibfnamefont {S.}~\bibnamefont {Reinhard}}, \bibinfo {author}
  {\bibfnamefont {W.}~\bibnamefont {Stolz}},\ and\ \bibinfo {author}
  {\bibfnamefont {K.}~\bibnamefont {Volz}},\ }\href
  {https://doi.org/10.1049/ell2.12353} {\bibfield  {journal} {\bibinfo
  {journal} {Electron. Lett.}\ }\textbf {\bibinfo {volume} {58}},\ \bibinfo
  {pages} {70} (\bibinfo {year} {2022})}\BibitemShut {NoStop}%
\bibitem [{\citenamefont {Fuchs}\ \emph {et~al.}(2018)\citenamefont {Fuchs},
  \citenamefont {Br{\"{u}}ggemann}, \citenamefont {Weseloh}, \citenamefont
  {Berger}, \citenamefont {M{\"{o}}ller}, \citenamefont {Reinhard},
  \citenamefont {Hader}, \citenamefont {Moloney}, \citenamefont
  {B{\"{a}}umner}, \citenamefont {Koch},\ and\ \citenamefont
  {Stolz}}]{Fuchs2018}%
  \BibitemOpen
  \bibfield  {author} {\bibinfo {author} {\bibfnamefont {C.}~\bibnamefont
  {Fuchs}}, \bibinfo {author} {\bibfnamefont {A.}~\bibnamefont
  {Br{\"{u}}ggemann}}, \bibinfo {author} {\bibfnamefont {M.~J.}\ \bibnamefont
  {Weseloh}}, \bibinfo {author} {\bibfnamefont {C.}~\bibnamefont {Berger}},
  \bibinfo {author} {\bibfnamefont {C.}~\bibnamefont {M{\"{o}}ller}}, \bibinfo
  {author} {\bibfnamefont {S.}~\bibnamefont {Reinhard}}, \bibinfo {author}
  {\bibfnamefont {J.}~\bibnamefont {Hader}}, \bibinfo {author} {\bibfnamefont
  {J.~V.}\ \bibnamefont {Moloney}}, \bibinfo {author} {\bibfnamefont
  {A.}~\bibnamefont {B{\"{a}}umner}}, \bibinfo {author} {\bibfnamefont {S.~W.}\
  \bibnamefont {Koch}},\ and\ \bibinfo {author} {\bibfnamefont
  {W.}~\bibnamefont {Stolz}},\ }\href
  {https://doi.org/10.1038/s41598-018-19189-1} {\bibfield  {journal} {\bibinfo
  {journal} {Sci. Rep.}\ }\textbf {\bibinfo {volume} {8}},\ \bibinfo {pages}
  {1422} (\bibinfo {year} {2018})}\BibitemShut {NoStop}%
\bibitem [{\citenamefont {Mokkapati}\ and\ \citenamefont
  {Jagadish}(2009)}]{Mokkapati2009IIIV}%
  \BibitemOpen
  \bibfield  {author} {\bibinfo {author} {\bibfnamefont {S.}~\bibnamefont
  {Mokkapati}}\ and\ \bibinfo {author} {\bibfnamefont {C.}~\bibnamefont
  {Jagadish}},\ }\href {https://doi.org/10.1016/S1369-7021(09)70110-5}
  {\bibfield  {journal} {\bibinfo  {journal} {Mater. Today}\ }\textbf {\bibinfo
  {volume} {12}},\ \bibinfo {pages} {22} (\bibinfo {year} {2009})}\BibitemShut
  {NoStop}%
\bibitem [{\citenamefont {Dimroth}\ \emph {et~al.}(2016)\citenamefont
  {Dimroth}, \citenamefont {Tibbits}, \citenamefont {Niemeyer}, \citenamefont
  {Predan}, \citenamefont {Beutel}, \citenamefont {Karcher}, \citenamefont
  {Oliva}, \citenamefont {Siefer}, \citenamefont {Lackner}, \citenamefont
  {Fu{\ss}-Kailuweit}, \citenamefont {Bett}, \citenamefont {Krause},
  \citenamefont {Drazek}, \citenamefont {Guiot}, \citenamefont {Wasselin},
  \citenamefont {Tauzin},\ and\ \citenamefont
  {Signamarcheix}}]{Dimroth2016Four}%
  \BibitemOpen
  \bibfield  {author} {\bibinfo {author} {\bibfnamefont {F.}~\bibnamefont
  {Dimroth}}, \bibinfo {author} {\bibfnamefont {T.~N.~D.}\ \bibnamefont
  {Tibbits}}, \bibinfo {author} {\bibfnamefont {M.}~\bibnamefont {Niemeyer}},
  \bibinfo {author} {\bibfnamefont {F.}~\bibnamefont {Predan}}, \bibinfo
  {author} {\bibfnamefont {P.}~\bibnamefont {Beutel}}, \bibinfo {author}
  {\bibfnamefont {C.}~\bibnamefont {Karcher}}, \bibinfo {author} {\bibfnamefont
  {E.}~\bibnamefont {Oliva}}, \bibinfo {author} {\bibfnamefont
  {G.}~\bibnamefont {Siefer}}, \bibinfo {author} {\bibfnamefont
  {D.}~\bibnamefont {Lackner}}, \bibinfo {author} {\bibfnamefont
  {P.}~\bibnamefont {Fu{\ss}-Kailuweit}}, \bibinfo {author} {\bibfnamefont
  {A.~W.}\ \bibnamefont {Bett}}, \bibinfo {author} {\bibfnamefont
  {R.}~\bibnamefont {Krause}}, \bibinfo {author} {\bibfnamefont
  {C.}~\bibnamefont {Drazek}}, \bibinfo {author} {\bibfnamefont
  {E.}~\bibnamefont {Guiot}}, \bibinfo {author} {\bibfnamefont
  {J.}~\bibnamefont {Wasselin}}, \bibinfo {author} {\bibfnamefont
  {A.}~\bibnamefont {Tauzin}},\ and\ \bibinfo {author} {\bibfnamefont
  {T.}~\bibnamefont {Signamarcheix}},\ }\href
  {https://doi.org/10.1109/JPHOTOV.2015.2501729} {\bibfield  {journal}
  {\bibinfo  {journal} {IEEE J. Photovolt.}\ }\textbf {\bibinfo {volume} {6}},\
  \bibinfo {pages} {343} (\bibinfo {year} {2016})}\BibitemShut {NoStop}%
\bibitem [{\citenamefont {Mitchell}\ \emph {et~al.}(2011)\citenamefont
  {Mitchell}, \citenamefont {Peharz}, \citenamefont {Siefer}, \citenamefont
  {Peters}, \citenamefont {Gandy}, \citenamefont {Goldschmidt}, \citenamefont
  {Benick}, \citenamefont {Glunz}, \citenamefont {Bett},\ and\ \citenamefont
  {Dimroth}}]{Mitchell2011Four}%
  \BibitemOpen
  \bibfield  {author} {\bibinfo {author} {\bibfnamefont {B.}~\bibnamefont
  {Mitchell}}, \bibinfo {author} {\bibfnamefont {G.}~\bibnamefont {Peharz}},
  \bibinfo {author} {\bibfnamefont {G.}~\bibnamefont {Siefer}}, \bibinfo
  {author} {\bibfnamefont {M.}~\bibnamefont {Peters}}, \bibinfo {author}
  {\bibfnamefont {T.}~\bibnamefont {Gandy}}, \bibinfo {author} {\bibfnamefont
  {J.~C.}\ \bibnamefont {Goldschmidt}}, \bibinfo {author} {\bibfnamefont
  {J.}~\bibnamefont {Benick}}, \bibinfo {author} {\bibfnamefont {S.~W.}\
  \bibnamefont {Glunz}}, \bibinfo {author} {\bibfnamefont {A.~W.}\ \bibnamefont
  {Bett}},\ and\ \bibinfo {author} {\bibfnamefont {F.}~\bibnamefont
  {Dimroth}},\ }\href {https://doi.org/10.1002/pip.988} {\bibfield  {journal}
  {\bibinfo  {journal} {Prog. Photovolt: Res. Appl.}\ }\textbf {\bibinfo
  {volume} {19}},\ \bibinfo {pages} {61} (\bibinfo {year} {2011})}\BibitemShut
  {NoStop}%
\bibitem [{\citenamefont {Philipps}\ \emph {et~al.}(2018)\citenamefont
  {Philipps}, \citenamefont {Dimroth},\ and\ \citenamefont
  {Bett}}]{PhilippsS2018Chapter}%
  \BibitemOpen
  \bibfield  {author} {\bibinfo {author} {\bibfnamefont {S.~P.}\ \bibnamefont
  {Philipps}}, \bibinfo {author} {\bibfnamefont {F.}~\bibnamefont {Dimroth}},\
  and\ \bibinfo {author} {\bibfnamefont {A.~W.}\ \bibnamefont {Bett}},\ }in\
  \href {https://doi.org/https://doi.org/10.1016/B978-0-12-809921-6.00012-4}
  {\emph {\bibinfo {booktitle} {McEvoy's Handbook of Photovoltaics (Third
  Edition)}}},\ \bibinfo {editor} {edited by\ \bibinfo {editor} {\bibfnamefont
  {S.~A.}\ \bibnamefont {Kalogirou}}}\ (\bibinfo  {publisher} {Academic
  Press},\ \bibinfo {year} {2018})\ pp.\ \bibinfo {pages}
  {439--472}\BibitemShut {NoStop}%
\bibitem [{\citenamefont {Beyer}\ \emph {et~al.}(2015)\citenamefont {Beyer},
  \citenamefont {Stolz},\ and\ \citenamefont {Volz}}]{Beyer2015Metastable}%
  \BibitemOpen
  \bibfield  {author} {\bibinfo {author} {\bibfnamefont {A.}~\bibnamefont
  {Beyer}}, \bibinfo {author} {\bibfnamefont {W.}~\bibnamefont {Stolz}},\ and\
  \bibinfo {author} {\bibfnamefont {K.}~\bibnamefont {Volz}},\ }\href
  {https://doi.org/10.1016/j.pcrysgrow.2015.10.002} {\bibfield  {journal}
  {\bibinfo  {journal} {Prog. Cryst. Growth Charact. Mater.}\ }\textbf
  {\bibinfo {volume} {61}},\ \bibinfo {pages} {46} (\bibinfo {year}
  {2015})}\BibitemShut {NoStop}%
\bibitem [{\citenamefont {Vurgaftman}\ \emph {et~al.}(2001)\citenamefont
  {Vurgaftman}, \citenamefont {Meyer},\ and\ \citenamefont
  {Ram-Mohan}}]{Vurgaftman2001Band}%
  \BibitemOpen
  \bibfield  {author} {\bibinfo {author} {\bibfnamefont {I.}~\bibnamefont
  {Vurgaftman}}, \bibinfo {author} {\bibfnamefont {J.~R.}\ \bibnamefont
  {Meyer}},\ and\ \bibinfo {author} {\bibfnamefont {L.~R.}\ \bibnamefont
  {Ram-Mohan}},\ }\href {https://doi.org/10.1063/1.1368156} {\bibfield
  {journal} {\bibinfo  {journal} {J. Appl. Phys.}\ }\textbf {\bibinfo {volume}
  {89}},\ \bibinfo {pages} {5815} (\bibinfo {year} {2001})}\BibitemShut
  {NoStop}%
\bibitem [{\citenamefont {Beyer}\ \emph {et~al.}(2017)\citenamefont {Beyer},
  \citenamefont {Knaub}, \citenamefont {Rosenow}, \citenamefont {Jandieri},
  \citenamefont {Ludewig}, \citenamefont {Bannow}, \citenamefont {Koch},
  \citenamefont {Tonner},\ and\ \citenamefont {Volz}}]{Beyer2017Local}%
  \BibitemOpen
  \bibfield  {author} {\bibinfo {author} {\bibfnamefont {A.}~\bibnamefont
  {Beyer}}, \bibinfo {author} {\bibfnamefont {N.}~\bibnamefont {Knaub}},
  \bibinfo {author} {\bibfnamefont {P.}~\bibnamefont {Rosenow}}, \bibinfo
  {author} {\bibfnamefont {K.}~\bibnamefont {Jandieri}}, \bibinfo {author}
  {\bibfnamefont {P.}~\bibnamefont {Ludewig}}, \bibinfo {author} {\bibfnamefont
  {L.}~\bibnamefont {Bannow}}, \bibinfo {author} {\bibfnamefont {S.~W.}\
  \bibnamefont {Koch}}, \bibinfo {author} {\bibfnamefont {R.}~\bibnamefont
  {Tonner}},\ and\ \bibinfo {author} {\bibfnamefont {K.}~\bibnamefont {Volz}},\
  }\href {https://doi.org/10.1016/j.apmt.2016.11.007} {\bibfield  {journal}
  {\bibinfo  {journal} {Appl. Mater. Today}\ }\textbf {\bibinfo {volume} {6}},\
  \bibinfo {pages} {22} (\bibinfo {year} {2017})}\BibitemShut {NoStop}%
\bibitem [{\citenamefont {Ludewig}\ \emph {et~al.}(2016)\citenamefont
  {Ludewig}, \citenamefont {Reinhard}, \citenamefont {Jandieri}, \citenamefont
  {Wegele}, \citenamefont {Beyer}, \citenamefont {Tapfer}, \citenamefont
  {Volz},\ and\ \citenamefont {Stolz}}]{Ludewig2016Movpe}%
  \BibitemOpen
  \bibfield  {author} {\bibinfo {author} {\bibfnamefont {P.}~\bibnamefont
  {Ludewig}}, \bibinfo {author} {\bibfnamefont {S.}~\bibnamefont {Reinhard}},
  \bibinfo {author} {\bibfnamefont {K.}~\bibnamefont {Jandieri}}, \bibinfo
  {author} {\bibfnamefont {T.}~\bibnamefont {Wegele}}, \bibinfo {author}
  {\bibfnamefont {A.}~\bibnamefont {Beyer}}, \bibinfo {author} {\bibfnamefont
  {L.}~\bibnamefont {Tapfer}}, \bibinfo {author} {\bibfnamefont
  {K.}~\bibnamefont {Volz}},\ and\ \bibinfo {author} {\bibfnamefont
  {W.}~\bibnamefont {Stolz}},\ }\href
  {https://doi.org/10.1016/j.jcrysgro.2015.12.024} {\bibfield  {journal}
  {\bibinfo  {journal} {J. Cryst. Growth}\ }\textbf {\bibinfo {volume} {438}},\
  \bibinfo {pages} {63} (\bibinfo {year} {2016})}\BibitemShut {NoStop}%
\bibitem [{\citenamefont {Liebich}\ \emph {et~al.}(2011)\citenamefont
  {Liebich}, \citenamefont {Zimprich}, \citenamefont {Beyer}, \citenamefont
  {Lange}, \citenamefont {Franzbach}, \citenamefont {Chatterjee}, \citenamefont
  {Hossain}, \citenamefont {Sweeney}, \citenamefont {Volz}, \citenamefont
  {Kunert},\ and\ \citenamefont {Stolz}}]{Liebich2011Laser}%
  \BibitemOpen
  \bibfield  {author} {\bibinfo {author} {\bibfnamefont {S.}~\bibnamefont
  {Liebich}}, \bibinfo {author} {\bibfnamefont {M.}~\bibnamefont {Zimprich}},
  \bibinfo {author} {\bibfnamefont {A.}~\bibnamefont {Beyer}}, \bibinfo
  {author} {\bibfnamefont {C.}~\bibnamefont {Lange}}, \bibinfo {author}
  {\bibfnamefont {D.~J.}\ \bibnamefont {Franzbach}}, \bibinfo {author}
  {\bibfnamefont {S.}~\bibnamefont {Chatterjee}}, \bibinfo {author}
  {\bibfnamefont {N.}~\bibnamefont {Hossain}}, \bibinfo {author} {\bibfnamefont
  {S.~J.}\ \bibnamefont {Sweeney}}, \bibinfo {author} {\bibfnamefont
  {K.}~\bibnamefont {Volz}}, \bibinfo {author} {\bibfnamefont {B.}~\bibnamefont
  {Kunert}},\ and\ \bibinfo {author} {\bibfnamefont {W.}~\bibnamefont
  {Stolz}},\ }\href {https://doi.org/10.1063/1.3624927} {\bibfield  {journal}
  {\bibinfo  {journal} {Appl. Phys. Lett.}\ }\textbf {\bibinfo {volume} {99}},\
  \bibinfo {pages} {071109} (\bibinfo {year} {2011})}\BibitemShut {NoStop}%
\bibitem [{\citenamefont {Supplie}\ \emph {et~al.}(2018)\citenamefont
  {Supplie}, \citenamefont {Romanyuk}, \citenamefont {Koppka}, \citenamefont
  {Steidl}, \citenamefont {N{\"{a}}gelein}, \citenamefont {Paszuk},
  \citenamefont {Winterfeld}, \citenamefont {Dobrich}, \citenamefont
  {Kleinschmidt}, \citenamefont {Runge},\ and\ \citenamefont
  {Hannappel}}]{Supplie2018Metalorganic}%
  \BibitemOpen
  \bibfield  {author} {\bibinfo {author} {\bibfnamefont {O.}~\bibnamefont
  {Supplie}}, \bibinfo {author} {\bibfnamefont {O.}~\bibnamefont {Romanyuk}},
  \bibinfo {author} {\bibfnamefont {C.}~\bibnamefont {Koppka}}, \bibinfo
  {author} {\bibfnamefont {M.}~\bibnamefont {Steidl}}, \bibinfo {author}
  {\bibfnamefont {A.}~\bibnamefont {N{\"{a}}gelein}}, \bibinfo {author}
  {\bibfnamefont {A.}~\bibnamefont {Paszuk}}, \bibinfo {author} {\bibfnamefont
  {L.}~\bibnamefont {Winterfeld}}, \bibinfo {author} {\bibfnamefont
  {A.}~\bibnamefont {Dobrich}}, \bibinfo {author} {\bibfnamefont
  {P.}~\bibnamefont {Kleinschmidt}}, \bibinfo {author} {\bibfnamefont
  {E.}~\bibnamefont {Runge}},\ and\ \bibinfo {author} {\bibfnamefont
  {T.}~\bibnamefont {Hannappel}},\ }\href
  {https://doi.org/10.1016/j.pcrysgrow.2018.07.002} {\bibfield  {journal}
  {\bibinfo  {journal} {Prog. Cryst. Growth Charact. Mater.}\ }\textbf
  {\bibinfo {volume} {64}},\ \bibinfo {pages} {103} (\bibinfo {year}
  {2018})}\BibitemShut {NoStop}%
\bibitem [{\citenamefont {Stringfellow}(2019)}]{Stringfellow2019Fundamental}%
  \BibitemOpen
  \bibfield  {author} {\bibinfo {author} {\bibfnamefont {G.}~\bibnamefont
  {Stringfellow}},\ }in\ \href {https://doi.org/10.1002/9781119313021.ch2}
  {\emph {\bibinfo {booktitle} {Met. Vap. Phase Ep.}}}\ (\bibinfo  {publisher}
  {Wiley},\ \bibinfo {year} {2019})\ pp.\ \bibinfo {pages} {19--69}\BibitemShut
  {NoStop}%
\bibitem [{\citenamefont {Volz}\ \emph {et~al.}(2009)\citenamefont {Volz},
  \citenamefont {Koch}, \citenamefont {H{\"{o}}hnsdorf}, \citenamefont
  {Kunert},\ and\ \citenamefont {Stolz}}]{Volz2009Movpe}%
  \BibitemOpen
  \bibfield  {author} {\bibinfo {author} {\bibfnamefont {K.}~\bibnamefont
  {Volz}}, \bibinfo {author} {\bibfnamefont {J.}~\bibnamefont {Koch}}, \bibinfo
  {author} {\bibfnamefont {F.}~\bibnamefont {H{\"{o}}hnsdorf}}, \bibinfo
  {author} {\bibfnamefont {B.}~\bibnamefont {Kunert}},\ and\ \bibinfo {author}
  {\bibfnamefont {W.}~\bibnamefont {Stolz}},\ }\href
  {https://doi.org/10.1016/j.jcrysgro.2008.09.210} {\bibfield  {journal}
  {\bibinfo  {journal} {J. Cryst. Growth}\ }\textbf {\bibinfo {volume} {311}},\
  \bibinfo {pages} {2418} (\bibinfo {year} {2009})}\BibitemShut {NoStop}%
\bibitem [{\citenamefont {Feifel}\ \emph {et~al.}(2017)\citenamefont {Feifel},
  \citenamefont {Ohlmann}, \citenamefont {Benick}, \citenamefont {Rachow},
  \citenamefont {Janz}, \citenamefont {Hermle}, \citenamefont {Dimroth},
  \citenamefont {Belz}, \citenamefont {Beyer}, \citenamefont {Volz},\ and\
  \citenamefont {Lackner}}]{Feifel2017Movpe}%
  \BibitemOpen
  \bibfield  {author} {\bibinfo {author} {\bibfnamefont {M.}~\bibnamefont
  {Feifel}}, \bibinfo {author} {\bibfnamefont {J.}~\bibnamefont {Ohlmann}},
  \bibinfo {author} {\bibfnamefont {J.}~\bibnamefont {Benick}}, \bibinfo
  {author} {\bibfnamefont {T.}~\bibnamefont {Rachow}}, \bibinfo {author}
  {\bibfnamefont {S.}~\bibnamefont {Janz}}, \bibinfo {author} {\bibfnamefont
  {M.}~\bibnamefont {Hermle}}, \bibinfo {author} {\bibfnamefont
  {F.}~\bibnamefont {Dimroth}}, \bibinfo {author} {\bibfnamefont
  {J.}~\bibnamefont {Belz}}, \bibinfo {author} {\bibfnamefont {A.}~\bibnamefont
  {Beyer}}, \bibinfo {author} {\bibfnamefont {K.}~\bibnamefont {Volz}},\ and\
  \bibinfo {author} {\bibfnamefont {D.}~\bibnamefont {Lackner}},\ }\href
  {https://doi.org/10.1109/JPHOTOV.2016.2642645} {\bibfield  {journal}
  {\bibinfo  {journal} {IEEE J. Photovolt.}\ }\textbf {\bibinfo {volume} {7}},\
  \bibinfo {pages} {502} (\bibinfo {year} {2017})}\BibitemShut {NoStop}%
\bibitem [{\citenamefont {Kunert}\ \emph {et~al.}(2004)\citenamefont {Kunert},
  \citenamefont {Koch}, \citenamefont {Torunski}, \citenamefont {Volz},\ and\
  \citenamefont {Stolz}}]{Kunert2004Movpe}%
  \BibitemOpen
  \bibfield  {author} {\bibinfo {author} {\bibfnamefont {B.}~\bibnamefont
  {Kunert}}, \bibinfo {author} {\bibfnamefont {J.}~\bibnamefont {Koch}},
  \bibinfo {author} {\bibfnamefont {T.}~\bibnamefont {Torunski}}, \bibinfo
  {author} {\bibfnamefont {K.}~\bibnamefont {Volz}},\ and\ \bibinfo {author}
  {\bibfnamefont {W.}~\bibnamefont {Stolz}},\ }\href
  {https://doi.org/10.1016/j.jcrysgro.2004.08.091} {\bibfield  {journal}
  {\bibinfo  {journal} {J. Cryst. Growth}\ }\textbf {\bibinfo {volume} {272}},\
  \bibinfo {pages} {753} (\bibinfo {year} {2004})}\BibitemShut {NoStop}%
\bibitem [{\citenamefont {Volz}\ \emph {et~al.}(2004)\citenamefont {Volz},
  \citenamefont {Torunski}, \citenamefont {Kunert}, \citenamefont {Rubel},
  \citenamefont {Nau}, \citenamefont {Reinhard},\ and\ \citenamefont
  {Stolz}}]{Volz2004Specific}%
  \BibitemOpen
  \bibfield  {author} {\bibinfo {author} {\bibfnamefont {K.}~\bibnamefont
  {Volz}}, \bibinfo {author} {\bibfnamefont {T.}~\bibnamefont {Torunski}},
  \bibinfo {author} {\bibfnamefont {B.}~\bibnamefont {Kunert}}, \bibinfo
  {author} {\bibfnamefont {O.}~\bibnamefont {Rubel}}, \bibinfo {author}
  {\bibfnamefont {S.}~\bibnamefont {Nau}}, \bibinfo {author} {\bibfnamefont
  {S.}~\bibnamefont {Reinhard}},\ and\ \bibinfo {author} {\bibfnamefont
  {W.}~\bibnamefont {Stolz}},\ }\href
  {https://doi.org/10.1016/j.jcrysgro.2004.09.012} {\bibfield  {journal}
  {\bibinfo  {journal} {J. Cryst. Growth}\ }\textbf {\bibinfo {volume} {272}},\
  \bibinfo {pages} {739} (\bibinfo {year} {2004})}\BibitemShut {NoStop}%
\bibitem [{\citenamefont {Veletas}\ \emph {et~al.}(2019)\citenamefont
  {Veletas}, \citenamefont {Hepp}, \citenamefont {Volz},\ and\ \citenamefont
  {Chatterjee}}]{Veletas2019Bismuth}%
  \BibitemOpen
  \bibfield  {author} {\bibinfo {author} {\bibfnamefont {J.}~\bibnamefont
  {Veletas}}, \bibinfo {author} {\bibfnamefont {T.}~\bibnamefont {Hepp}},
  \bibinfo {author} {\bibfnamefont {K.}~\bibnamefont {Volz}},\ and\ \bibinfo
  {author} {\bibfnamefont {S.}~\bibnamefont {Chatterjee}},\ }\href
  {https://doi.org/10.1063/1.5111913} {\bibfield  {journal} {\bibinfo
  {journal} {J. Appl. Phys.}\ }\textbf {\bibinfo {volume} {126}},\ \bibinfo
  {pages} {135705} (\bibinfo {year} {2019})}\BibitemShut {NoStop}%
\bibitem [{\citenamefont {Wegele}\ \emph {et~al.}(2016)\citenamefont {Wegele},
  \citenamefont {Beyer}, \citenamefont {Ludewig}, \citenamefont {Rosenow},
  \citenamefont {Duschek}, \citenamefont {Jandieri}, \citenamefont {Tonner},
  \citenamefont {Stolz},\ and\ \citenamefont {Volz}}]{Wegele2016Interface}%
  \BibitemOpen
  \bibfield  {author} {\bibinfo {author} {\bibfnamefont {T.}~\bibnamefont
  {Wegele}}, \bibinfo {author} {\bibfnamefont {A.}~\bibnamefont {Beyer}},
  \bibinfo {author} {\bibfnamefont {P.}~\bibnamefont {Ludewig}}, \bibinfo
  {author} {\bibfnamefont {P.}~\bibnamefont {Rosenow}}, \bibinfo {author}
  {\bibfnamefont {L.}~\bibnamefont {Duschek}}, \bibinfo {author} {\bibfnamefont
  {K.}~\bibnamefont {Jandieri}}, \bibinfo {author} {\bibfnamefont
  {R.}~\bibnamefont {Tonner}}, \bibinfo {author} {\bibfnamefont
  {W.}~\bibnamefont {Stolz}},\ and\ \bibinfo {author} {\bibfnamefont
  {K.}~\bibnamefont {Volz}},\ }\href
  {https://doi.org/10.1088/0022-3727/49/7/075108} {\bibfield  {journal}
  {\bibinfo  {journal} {J. Phys. D. Appl. Phys.}\ }\textbf {\bibinfo {volume}
  {49}},\ \bibinfo {pages} {075108} (\bibinfo {year} {2016})}\BibitemShut
  {NoStop}%
\bibitem [{\citenamefont {Hepp}\ \emph {et~al.}(2019)\citenamefont {Hepp},
  \citenamefont {Nattermann},\ and\ \citenamefont {Volz}}]{Hepp2019Movpe}%
  \BibitemOpen
  \bibfield  {author} {\bibinfo {author} {\bibfnamefont {T.}~\bibnamefont
  {Hepp}}, \bibinfo {author} {\bibfnamefont {L.}~\bibnamefont {Nattermann}},\
  and\ \bibinfo {author} {\bibfnamefont {K.}~\bibnamefont {Volz}},\ }in\ \href
  {https://doi.org/10.1007/978-981-13-8078-5_3} {\emph {\bibinfo {booktitle}
  {Bismuth-Containing Alloys and Nanostructures}}},\ \bibinfo {editor} {edited
  by\ \bibinfo {editor} {\bibfnamefont {S.}~\bibnamefont {Wang}}\ and\ \bibinfo
  {editor} {\bibfnamefont {P.}~\bibnamefont {Lu}}}\ (\bibinfo  {publisher}
  {Springer},\ \bibinfo {address} {Singapur},\ \bibinfo {year} {2019})\ pp.\
  \bibinfo {pages} {37--58}\BibitemShut {NoStop}%
\bibitem [{\citenamefont {Van~Camp}\ \emph {et~al.}(1990)\citenamefont
  {Van~Camp}, \citenamefont {Van~Doren},\ and\ \citenamefont
  {Devreese}}]{Van1990Pressure}%
  \BibitemOpen
  \bibfield  {author} {\bibinfo {author} {\bibfnamefont {P.~E.}\ \bibnamefont
  {Van~Camp}}, \bibinfo {author} {\bibfnamefont {V.~E.}\ \bibnamefont
  {Van~Doren}},\ and\ \bibinfo {author} {\bibfnamefont {J.~T.}\ \bibnamefont
  {Devreese}},\ }\href {https://doi.org/10.1103/PhysRevB.41.1598} {\bibfield
  {journal} {\bibinfo  {journal} {Phys. Rev. B}\ }\textbf {\bibinfo {volume}
  {41}},\ \bibinfo {pages} {1598} (\bibinfo {year} {1990})}\BibitemShut
  {NoStop}%
\bibitem [{\citenamefont {Potter}(1956)}]{Potter1956Indirect}%
  \BibitemOpen
  \bibfield  {author} {\bibinfo {author} {\bibfnamefont {R.~F.}\ \bibnamefont
  {Potter}},\ }\href {https://doi.org/10.1103/PhysRev.103.861} {\bibfield
  {journal} {\bibinfo  {journal} {Phys. Rev.}\ }\textbf {\bibinfo {volume}
  {103}},\ \bibinfo {pages} {861} (\bibinfo {year} {1956})}\BibitemShut
  {NoStop}%
\bibitem [{\citenamefont {Alekseev}\ \emph {et~al.}(2020)\citenamefont
  {Alekseev}, \citenamefont {Sharov}, \citenamefont {Borodin}, \citenamefont
  {Dunaevskiy}, \citenamefont {Reznik},\ and\ \citenamefont
  {Cirlin}}]{Alekseev2020Effect}%
  \BibitemOpen
  \bibfield  {author} {\bibinfo {author} {\bibfnamefont {P.~A.}\ \bibnamefont
  {Alekseev}}, \bibinfo {author} {\bibfnamefont {V.~A.}\ \bibnamefont
  {Sharov}}, \bibinfo {author} {\bibfnamefont {B.~R.}\ \bibnamefont {Borodin}},
  \bibinfo {author} {\bibfnamefont {M.~S.}\ \bibnamefont {Dunaevskiy}},
  \bibinfo {author} {\bibfnamefont {R.~R.}\ \bibnamefont {Reznik}},\ and\
  \bibinfo {author} {\bibfnamefont {G.~E.}\ \bibnamefont {Cirlin}},\ }\href
  {https://doi.org/10.3390/mi11060581} {\bibfield  {journal} {\bibinfo
  {journal} {Micromachines}\ }\textbf {\bibinfo {volume} {11}},\ \bibinfo
  {pages} {581} (\bibinfo {year} {2020})}\BibitemShut {NoStop}%
\bibitem [{\citenamefont {Katiyar}\ \emph {et~al.}(2020)\citenamefont
  {Katiyar}, \citenamefont {Thai}, \citenamefont {Yun}, \citenamefont {Lee},\
  and\ \citenamefont {Ahn}}]{Katiyar2020Breaking}%
  \BibitemOpen
  \bibfield  {author} {\bibinfo {author} {\bibfnamefont {A.~K.}\ \bibnamefont
  {Katiyar}}, \bibinfo {author} {\bibfnamefont {K.~Y.}\ \bibnamefont {Thai}},
  \bibinfo {author} {\bibfnamefont {W.~S.}\ \bibnamefont {Yun}}, \bibinfo
  {author} {\bibfnamefont {J.}~\bibnamefont {Lee}},\ and\ \bibinfo {author}
  {\bibfnamefont {J.-h.}\ \bibnamefont {Ahn}},\ }\href
  {https://doi.org/10.1126/sciadv.abb0576} {\bibfield  {journal} {\bibinfo
  {journal} {Sci. Adv.}\ }\textbf {\bibinfo {volume} {6}},\ \bibinfo {pages}
  {eabb0576} (\bibinfo {year} {2020})}\BibitemShut {NoStop}%
\bibitem [{\citenamefont {Lim}\ \emph {et~al.}(2021)\citenamefont {Lim},
  \citenamefont {Cui},\ and\ \citenamefont {Ringer}}]{Lim2021Strain}%
  \BibitemOpen
  \bibfield  {author} {\bibinfo {author} {\bibfnamefont {B.}~\bibnamefont
  {Lim}}, \bibinfo {author} {\bibfnamefont {X.~Y.}\ \bibnamefont {Cui}},\ and\
  \bibinfo {author} {\bibfnamefont {S.~P.}\ \bibnamefont {Ringer}},\ }\href
  {https://doi.org/10.1039/D1CP00457C} {\bibfield  {journal} {\bibinfo
  {journal} {Phys. Chem. Chem. Phys.}\ }\textbf {\bibinfo {volume} {23}},\
  \bibinfo {pages} {5407} (\bibinfo {year} {2021})}\BibitemShut {NoStop}%
\bibitem [{\citenamefont {Signorello}\ \emph {et~al.}(2014)\citenamefont
  {Signorello}, \citenamefont {L{\"{o}}rtscher}, \citenamefont {Khomyakov},
  \citenamefont {Karg}, \citenamefont {Dheeraj}, \citenamefont {Gotsmann},
  \citenamefont {Weman},\ and\ \citenamefont {Riel}}]{Signorello2014Inducing}%
  \BibitemOpen
  \bibfield  {author} {\bibinfo {author} {\bibfnamefont {G.}~\bibnamefont
  {Signorello}}, \bibinfo {author} {\bibfnamefont {E.}~\bibnamefont
  {L{\"{o}}rtscher}}, \bibinfo {author} {\bibfnamefont {P.}~\bibnamefont
  {Khomyakov}}, \bibinfo {author} {\bibfnamefont {S.}~\bibnamefont {Karg}},
  \bibinfo {author} {\bibfnamefont {D.}~\bibnamefont {Dheeraj}}, \bibinfo
  {author} {\bibfnamefont {B.}~\bibnamefont {Gotsmann}}, \bibinfo {author}
  {\bibfnamefont {H.}~\bibnamefont {Weman}},\ and\ \bibinfo {author}
  {\bibfnamefont {H.}~\bibnamefont {Riel}},\ }\href
  {https://doi.org/10.1038/ncomms4655} {\bibfield  {journal} {\bibinfo
  {journal} {Nat. Commun.}\ }\textbf {\bibinfo {volume} {5}},\ \bibinfo {pages}
  {3655} (\bibinfo {year} {2014})}\BibitemShut {NoStop}%
\bibitem [{\citenamefont {Signorello}\ \emph {et~al.}(2013)\citenamefont
  {Signorello}, \citenamefont {Karg}, \citenamefont {Bj{\"{o}}rk},
  \citenamefont {Gotsmann},\ and\ \citenamefont {Riel}}]{Signorello2013Tuning}%
  \BibitemOpen
  \bibfield  {author} {\bibinfo {author} {\bibfnamefont {G.}~\bibnamefont
  {Signorello}}, \bibinfo {author} {\bibfnamefont {S.}~\bibnamefont {Karg}},
  \bibinfo {author} {\bibfnamefont {M.~T.}\ \bibnamefont {Bj{\"{o}}rk}},
  \bibinfo {author} {\bibfnamefont {B.}~\bibnamefont {Gotsmann}},\ and\
  \bibinfo {author} {\bibfnamefont {H.}~\bibnamefont {Riel}},\ }\href
  {https://doi.org/10.1021/nl303694c} {\bibfield  {journal} {\bibinfo
  {journal} {Nano Lett.}\ }\textbf {\bibinfo {volume} {13}},\ \bibinfo {pages}
  {917} (\bibinfo {year} {2013})}\BibitemShut {NoStop}%
\bibitem [{\citenamefont {Balaghi}\ \emph {et~al.}(2019)\citenamefont
  {Balaghi}, \citenamefont {Bussone}, \citenamefont {Grifone}, \citenamefont
  {H{\"{u}}bner}, \citenamefont {Grenzer}, \citenamefont {Ghorbani-Asl},
  \citenamefont {Krasheninnikov}, \citenamefont {Schneider}, \citenamefont
  {Helm},\ and\ \citenamefont {Dimakis}}]{Balaghi2019Widely}%
  \BibitemOpen
  \bibfield  {author} {\bibinfo {author} {\bibfnamefont {L.}~\bibnamefont
  {Balaghi}}, \bibinfo {author} {\bibfnamefont {G.}~\bibnamefont {Bussone}},
  \bibinfo {author} {\bibfnamefont {R.}~\bibnamefont {Grifone}}, \bibinfo
  {author} {\bibfnamefont {R.}~\bibnamefont {H{\"{u}}bner}}, \bibinfo {author}
  {\bibfnamefont {J.}~\bibnamefont {Grenzer}}, \bibinfo {author} {\bibfnamefont
  {M.}~\bibnamefont {Ghorbani-Asl}}, \bibinfo {author} {\bibfnamefont {A.~V.}\
  \bibnamefont {Krasheninnikov}}, \bibinfo {author} {\bibfnamefont
  {H.}~\bibnamefont {Schneider}}, \bibinfo {author} {\bibfnamefont
  {M.}~\bibnamefont {Helm}},\ and\ \bibinfo {author} {\bibfnamefont
  {E.}~\bibnamefont {Dimakis}},\ }\href
  {https://doi.org/10.1038/s41467-019-10654-7} {\bibfield  {journal} {\bibinfo
  {journal} {Nat. Commun.}\ }\textbf {\bibinfo {volume} {10}},\ \bibinfo
  {pages} {2793} (\bibinfo {year} {2019})}\BibitemShut {NoStop}%
\bibitem [{\citenamefont {Gr{\"{o}}nqvist}\ \emph {et~al.}(2009)\citenamefont
  {Gr{\"{o}}nqvist}, \citenamefont {S{\o}ndergaard}, \citenamefont {Boxberg},
  \citenamefont {Guhr}, \citenamefont {{\AA}berg},\ and\ \citenamefont
  {Xu}}]{Gronqvist2009Strain}%
  \BibitemOpen
  \bibfield  {author} {\bibinfo {author} {\bibfnamefont {J.}~\bibnamefont
  {Gr{\"{o}}nqvist}}, \bibinfo {author} {\bibfnamefont {N.}~\bibnamefont
  {S{\o}ndergaard}}, \bibinfo {author} {\bibfnamefont {F.}~\bibnamefont
  {Boxberg}}, \bibinfo {author} {\bibfnamefont {T.}~\bibnamefont {Guhr}},
  \bibinfo {author} {\bibfnamefont {S.}~\bibnamefont {{\AA}berg}},\ and\
  \bibinfo {author} {\bibfnamefont {H.~Q.}\ \bibnamefont {Xu}},\ }\href
  {https://doi.org/10.1063/1.3207838} {\bibfield  {journal} {\bibinfo
  {journal} {J. Appl. Phys.}\ }\textbf {\bibinfo {volume} {106}},\ \bibinfo
  {pages} {053508} (\bibinfo {year} {2009})}\BibitemShut {NoStop}%
\bibitem [{\citenamefont {Hetzl}\ \emph {et~al.}(2016)\citenamefont {Hetzl},
  \citenamefont {Kraut}, \citenamefont {Winnerl}, \citenamefont {Francaviglia},
  \citenamefont {D{\"{o}}blinger}, \citenamefont {Matich}, \citenamefont
  {FontcubertaMorral},\ and\ \citenamefont {Stutzmann}}]{Hetzl2016}%
  \BibitemOpen
  \bibfield  {author} {\bibinfo {author} {\bibfnamefont {M.}~\bibnamefont
  {Hetzl}}, \bibinfo {author} {\bibfnamefont {M.}~\bibnamefont {Kraut}},
  \bibinfo {author} {\bibfnamefont {J.}~\bibnamefont {Winnerl}}, \bibinfo
  {author} {\bibfnamefont {L.}~\bibnamefont {Francaviglia}}, \bibinfo {author}
  {\bibfnamefont {M.}~\bibnamefont {D{\"{o}}blinger}}, \bibinfo {author}
  {\bibfnamefont {S.}~\bibnamefont {Matich}}, \bibinfo {author} {\bibfnamefont
  {A.}~\bibnamefont {FontcubertaMorral}},\ and\ \bibinfo {author}
  {\bibfnamefont {M.}~\bibnamefont {Stutzmann}},\ }\href
  {https://doi.org/10.1021/acs.nanolett.6b03354} {\bibfield  {journal}
  {\bibinfo  {journal} {Nano Lett.}\ }\textbf {\bibinfo {volume} {16}},\
  \bibinfo {pages} {7098} (\bibinfo {year} {2016})}\BibitemShut {NoStop}%
\bibitem [{\citenamefont {Montazeri}\ \emph {et~al.}(2010)\citenamefont
  {Montazeri}, \citenamefont {Fickenscher}, \citenamefont {Smith},
  \citenamefont {Jackson}, \citenamefont {Yarrison-Rice}, \citenamefont {Kang},
  \citenamefont {Gao}, \citenamefont {Hoe~Tan}, \citenamefont {Jagadish},
  \citenamefont {Guo}, \citenamefont {Zou}, \citenamefont {Pistol},\ and\
  \citenamefont {Pryor}}]{Montazeri2010}%
  \BibitemOpen
  \bibfield  {author} {\bibinfo {author} {\bibfnamefont {M.}~\bibnamefont
  {Montazeri}}, \bibinfo {author} {\bibfnamefont {M.}~\bibnamefont
  {Fickenscher}}, \bibinfo {author} {\bibfnamefont {L.~M.}\ \bibnamefont
  {Smith}}, \bibinfo {author} {\bibfnamefont {H.~E.}\ \bibnamefont {Jackson}},
  \bibinfo {author} {\bibfnamefont {J.}~\bibnamefont {Yarrison-Rice}}, \bibinfo
  {author} {\bibfnamefont {J.~H.}\ \bibnamefont {Kang}}, \bibinfo {author}
  {\bibfnamefont {Q.}~\bibnamefont {Gao}}, \bibinfo {author} {\bibfnamefont
  {H.}~\bibnamefont {Hoe~Tan}}, \bibinfo {author} {\bibfnamefont
  {C.}~\bibnamefont {Jagadish}}, \bibinfo {author} {\bibfnamefont
  {Y.}~\bibnamefont {Guo}}, \bibinfo {author} {\bibfnamefont {J.}~\bibnamefont
  {Zou}}, \bibinfo {author} {\bibfnamefont {M.~E.}\ \bibnamefont {Pistol}},\
  and\ \bibinfo {author} {\bibfnamefont {C.~E.}\ \bibnamefont {Pryor}},\ }\href
  {https://doi.org/10.1021/NL903547R} {\bibfield  {journal} {\bibinfo
  {journal} {Nano Lett.}\ }\textbf {\bibinfo {volume} {10}},\ \bibinfo {pages}
  {880} (\bibinfo {year} {2010})}\BibitemShut {NoStop}%
\bibitem [{\citenamefont {Sk{\"{o}}ld}\ \emph {et~al.}(2005)\citenamefont
  {Sk{\"{o}}ld}, \citenamefont {Karlsson}, \citenamefont {Larsson},
  \citenamefont {Pistol}, \citenamefont {Seifert}, \citenamefont
  {Tr{\"{a}}g{\aa}rdh},\ and\ \citenamefont {Samuelson}}]{Skold2005}%
  \BibitemOpen
  \bibfield  {author} {\bibinfo {author} {\bibfnamefont {N.}~\bibnamefont
  {Sk{\"{o}}ld}}, \bibinfo {author} {\bibfnamefont {L.~S.}\ \bibnamefont
  {Karlsson}}, \bibinfo {author} {\bibfnamefont {M.~W.}\ \bibnamefont
  {Larsson}}, \bibinfo {author} {\bibfnamefont {M.~E.}\ \bibnamefont {Pistol}},
  \bibinfo {author} {\bibfnamefont {W.}~\bibnamefont {Seifert}}, \bibinfo
  {author} {\bibfnamefont {J.}~\bibnamefont {Tr{\"{a}}g{\aa}rdh}},\ and\
  \bibinfo {author} {\bibfnamefont {L.}~\bibnamefont {Samuelson}},\ }\href
  {https://doi.org/10.1021/nl051304s} {\bibfield  {journal} {\bibinfo
  {journal} {Nano Lett.}\ }\textbf {\bibinfo {volume} {5}},\ \bibinfo {pages}
  {1943} (\bibinfo {year} {2005})}\BibitemShut {NoStop}%
\bibitem [{\citenamefont {Bir}\ and\ \citenamefont
  {Pikus}(1974)}]{Bir1974Symmetry}%
  \BibitemOpen
  \bibfield  {author} {\bibinfo {author} {\bibfnamefont {G.}~\bibnamefont
  {Bir}}\ and\ \bibinfo {author} {\bibfnamefont {G.}~\bibnamefont {Pikus}},\
  }\href {https://books.google.de/books?id=38m2QgAACAAJ} {\emph {\bibinfo
  {title} {Symmetry and Strain-induced Effects in Semiconductors}}},\ A Halsted
  Press book\ (\bibinfo  {publisher} {Wiley},\ \bibinfo {year}
  {1974})\BibitemShut {NoStop}%
\bibitem [{\citenamefont {Sun}\ \emph {et~al.}(2007)\citenamefont {Sun},
  \citenamefont {Thompson},\ and\ \citenamefont {Nishida}}]{Sun2007Physics}%
  \BibitemOpen
  \bibfield  {author} {\bibinfo {author} {\bibfnamefont {Y.}~\bibnamefont
  {Sun}}, \bibinfo {author} {\bibfnamefont {S.~E.}\ \bibnamefont {Thompson}},\
  and\ \bibinfo {author} {\bibfnamefont {T.}~\bibnamefont {Nishida}},\ }\href
  {https://doi.org/10.1063/1.2730561} {\bibfield  {journal} {\bibinfo
  {journal} {J. Appl. Phys.}\ }\textbf {\bibinfo {volume} {101}},\ \bibinfo
  {pages} {104503} (\bibinfo {year} {2007})}\BibitemShut {NoStop}%
\bibitem [{\citenamefont {Tao}\ \emph {et~al.}(2020)\citenamefont {Tao},
  \citenamefont {Ou}, \citenamefont {Li}, \citenamefont {Liao}, \citenamefont
  {Zhang}, \citenamefont {Gan},\ and\ \citenamefont {Ou}}]{Tao2020}%
  \BibitemOpen
  \bibfield  {author} {\bibinfo {author} {\bibfnamefont {L.}~\bibnamefont
  {Tao}}, \bibinfo {author} {\bibfnamefont {W.}~\bibnamefont {Ou}}, \bibinfo
  {author} {\bibfnamefont {Y.}~\bibnamefont {Li}}, \bibinfo {author}
  {\bibfnamefont {H.}~\bibnamefont {Liao}}, \bibinfo {author} {\bibfnamefont
  {J.}~\bibnamefont {Zhang}}, \bibinfo {author} {\bibfnamefont
  {F.}~\bibnamefont {Gan}},\ and\ \bibinfo {author} {\bibfnamefont
  {X.}~\bibnamefont {Ou}},\ }\href {https://doi.org/10.1088/1361-6641/ab8e0b}
  {\bibfield  {journal} {\bibinfo  {journal} {Semicond. Sci. Technol.}\
  }\textbf {\bibinfo {volume} {35}},\ \bibinfo {pages} {103002} (\bibinfo
  {year} {2020})}\BibitemShut {NoStop}%
\bibitem [{\citenamefont {Tsutsui}\ \emph {et~al.}(2019)\citenamefont
  {Tsutsui}, \citenamefont {Mochizuki}, \citenamefont {Loubet}, \citenamefont
  {Bedell},\ and\ \citenamefont {Sadana}}]{Tsutsui2019}%
  \BibitemOpen
  \bibfield  {author} {\bibinfo {author} {\bibfnamefont {G.}~\bibnamefont
  {Tsutsui}}, \bibinfo {author} {\bibfnamefont {S.}~\bibnamefont {Mochizuki}},
  \bibinfo {author} {\bibfnamefont {N.}~\bibnamefont {Loubet}}, \bibinfo
  {author} {\bibfnamefont {S.~W.}\ \bibnamefont {Bedell}},\ and\ \bibinfo
  {author} {\bibfnamefont {D.~K.}\ \bibnamefont {Sadana}},\ }\href
  {https://doi.org/10.1063/1.5075637} {\bibfield  {journal} {\bibinfo
  {journal} {AIP Adv.}\ }\textbf {\bibinfo {volume} {9}},\ \bibinfo {pages}
  {030701} (\bibinfo {year} {2019})}\BibitemShut {NoStop}%
\bibitem [{\citenamefont {Fang}\ \emph {et~al.}(2011)\citenamefont {Fang},
  \citenamefont {Madsen}, \citenamefont {Carraro}, \citenamefont {Takei},
  \citenamefont {Kim}, \citenamefont {Plis}, \citenamefont {Chen},
  \citenamefont {Krishna}, \citenamefont {Chueh}, \citenamefont {Maboudian},\
  and\ \citenamefont {Javey}}]{Fang2011}%
  \BibitemOpen
  \bibfield  {author} {\bibinfo {author} {\bibfnamefont {H.}~\bibnamefont
  {Fang}}, \bibinfo {author} {\bibfnamefont {M.}~\bibnamefont {Madsen}},
  \bibinfo {author} {\bibfnamefont {C.}~\bibnamefont {Carraro}}, \bibinfo
  {author} {\bibfnamefont {K.}~\bibnamefont {Takei}}, \bibinfo {author}
  {\bibfnamefont {H.~S.}\ \bibnamefont {Kim}}, \bibinfo {author} {\bibfnamefont
  {E.}~\bibnamefont {Plis}}, \bibinfo {author} {\bibfnamefont {S.~Y.}\
  \bibnamefont {Chen}}, \bibinfo {author} {\bibfnamefont {S.}~\bibnamefont
  {Krishna}}, \bibinfo {author} {\bibfnamefont {Y.~L.}\ \bibnamefont {Chueh}},
  \bibinfo {author} {\bibfnamefont {R.}~\bibnamefont {Maboudian}},\ and\
  \bibinfo {author} {\bibfnamefont {A.}~\bibnamefont {Javey}},\ }\href
  {https://doi.org/10.1063/1.3537963} {\bibfield  {journal} {\bibinfo
  {journal} {Appl. Phys. Lett.}\ }\textbf {\bibinfo {volume} {98}},\ \bibinfo
  {pages} {012111} (\bibinfo {year} {2011})}\BibitemShut {NoStop}%
\bibitem [{\citenamefont {Bannow}\ \emph {et~al.}(2017)\citenamefont {Bannow},
  \citenamefont {Rosenow}, \citenamefont {Springer}, \citenamefont {Fischer},
  \citenamefont {Hader}, \citenamefont {Moloney}, \citenamefont {Tonner},\ and\
  \citenamefont {Koch}}]{Bannow2017}%
  \BibitemOpen
  \bibfield  {author} {\bibinfo {author} {\bibfnamefont {L.~C.}\ \bibnamefont
  {Bannow}}, \bibinfo {author} {\bibfnamefont {P.}~\bibnamefont {Rosenow}},
  \bibinfo {author} {\bibfnamefont {P.}~\bibnamefont {Springer}}, \bibinfo
  {author} {\bibfnamefont {E.~W.}\ \bibnamefont {Fischer}}, \bibinfo {author}
  {\bibfnamefont {J.}~\bibnamefont {Hader}}, \bibinfo {author} {\bibfnamefont
  {J.~V.}\ \bibnamefont {Moloney}}, \bibinfo {author} {\bibfnamefont
  {R.}~\bibnamefont {Tonner}},\ and\ \bibinfo {author} {\bibfnamefont {S.~W.}\
  \bibnamefont {Koch}},\ }\href {https://doi.org/10.1088/1361-651X/aa7478}
  {\bibfield  {journal} {\bibinfo  {journal} {Model. Simul. Mater. Sci. Eng.}\
  }\textbf {\bibinfo {volume} {25}},\ \bibinfo {pages} {065001} (\bibinfo
  {year} {2017})}\BibitemShut {NoStop}%
\bibitem [{\citenamefont {Anderson}\ and\ \citenamefont
  {Jones}(1991)}]{Anderson1991Optimized}%
  \BibitemOpen
  \bibfield  {author} {\bibinfo {author} {\bibfnamefont {N.~G.}\ \bibnamefont
  {Anderson}}\ and\ \bibinfo {author} {\bibfnamefont {S.~D.}\ \bibnamefont
  {Jones}},\ }\href {https://doi.org/10.1063/1.349115} {\bibfield  {journal}
  {\bibinfo  {journal} {J. Appl. Phys.}\ }\textbf {\bibinfo {volume} {70}},\
  \bibinfo {pages} {4342} (\bibinfo {year} {1991})}\BibitemShut {NoStop}%
\bibitem [{\citenamefont {Tan}\ \emph {et~al.}(2016)\citenamefont {Tan},
  \citenamefont {Povolotskyi}, \citenamefont {Kubis}, \citenamefont {Boykin},\
  and\ \citenamefont {Klimeck}}]{Tan2016Transferable}%
  \BibitemOpen
  \bibfield  {author} {\bibinfo {author} {\bibfnamefont {Y.}~\bibnamefont
  {Tan}}, \bibinfo {author} {\bibfnamefont {M.}~\bibnamefont {Povolotskyi}},
  \bibinfo {author} {\bibfnamefont {T.}~\bibnamefont {Kubis}}, \bibinfo
  {author} {\bibfnamefont {T.~B.}\ \bibnamefont {Boykin}},\ and\ \bibinfo
  {author} {\bibfnamefont {G.}~\bibnamefont {Klimeck}},\ }\href
  {https://doi.org/10.1103/PhysRevB.94.045311} {\bibfield  {journal} {\bibinfo
  {journal} {Phys. Rev. B}\ }\textbf {\bibinfo {volume} {94}},\ \bibinfo
  {pages} {045311} (\bibinfo {year} {2016})}\BibitemShut {NoStop}%
\bibitem [{\citenamefont {Jancu}\ \emph {et~al.}(1998)\citenamefont {Jancu},
  \citenamefont {Scholz}, \citenamefont {Beltram},\ and\ \citenamefont
  {Bassani}}]{Jancu1998}%
  \BibitemOpen
  \bibfield  {author} {\bibinfo {author} {\bibfnamefont {J.~M.}\ \bibnamefont
  {Jancu}}, \bibinfo {author} {\bibfnamefont {R.}~\bibnamefont {Scholz}},
  \bibinfo {author} {\bibfnamefont {F.}~\bibnamefont {Beltram}},\ and\ \bibinfo
  {author} {\bibfnamefont {F.}~\bibnamefont {Bassani}},\ }\href
  {https://doi.org/10.1103/PhysRevB.57.6493} {\bibfield  {journal} {\bibinfo
  {journal} {Phys. Rev. B}\ }\textbf {\bibinfo {volume} {57}},\ \bibinfo
  {pages} {6493} (\bibinfo {year} {1998})}\BibitemShut {NoStop}%
\bibitem [{\citenamefont {Rosenow}\ \emph {et~al.}(2018)\citenamefont
  {Rosenow}, \citenamefont {Bannow}, \citenamefont {Fischer}, \citenamefont
  {Stolz}, \citenamefont {Volz}, \citenamefont {Koch},\ and\ \citenamefont
  {Tonner}}]{Rosenow2018Ab}%
  \BibitemOpen
  \bibfield  {author} {\bibinfo {author} {\bibfnamefont {P.}~\bibnamefont
  {Rosenow}}, \bibinfo {author} {\bibfnamefont {L.~C.}\ \bibnamefont {Bannow}},
  \bibinfo {author} {\bibfnamefont {E.~W.}\ \bibnamefont {Fischer}}, \bibinfo
  {author} {\bibfnamefont {W.}~\bibnamefont {Stolz}}, \bibinfo {author}
  {\bibfnamefont {K.}~\bibnamefont {Volz}}, \bibinfo {author} {\bibfnamefont
  {S.~W.}\ \bibnamefont {Koch}},\ and\ \bibinfo {author} {\bibfnamefont
  {R.}~\bibnamefont {Tonner}},\ }\href
  {https://doi.org/10.1103/PhysRevB.97.075201} {\bibfield  {journal} {\bibinfo
  {journal} {Phys. Rev. B}\ }\textbf {\bibinfo {volume} {97}},\ \bibinfo
  {pages} {075201} (\bibinfo {year} {2018})}\BibitemShut {NoStop}%
\bibitem [{\citenamefont {Mondal}\ and\ \citenamefont
  {Tonner-Zech}(2023)}]{Mondal2022}%
  \BibitemOpen
  \bibfield  {author} {\bibinfo {author} {\bibfnamefont {B.}~\bibnamefont
  {Mondal}}\ and\ \bibinfo {author} {\bibfnamefont {R.}~\bibnamefont
  {Tonner-Zech}},\ }\href {https://doi.org/10.1088/1402-4896/acd08b} {\bibfield
   {journal} {\bibinfo  {journal} {Phys. Scr.}\ }\textbf {\bibinfo {volume}
  {98}},\ \bibinfo {pages} {065924} (\bibinfo {year} {2023})}\BibitemShut
  {NoStop}%
\bibitem [{\citenamefont {Kim}\ \emph {et~al.}(2010)\citenamefont {Kim},
  \citenamefont {Marsman}, \citenamefont {Kresse}, \citenamefont {Tran},\ and\
  \citenamefont {Blaha}}]{Kim2010Towards}%
  \BibitemOpen
  \bibfield  {author} {\bibinfo {author} {\bibfnamefont {Y.-S.}\ \bibnamefont
  {Kim}}, \bibinfo {author} {\bibfnamefont {M.}~\bibnamefont {Marsman}},
  \bibinfo {author} {\bibfnamefont {G.}~\bibnamefont {Kresse}}, \bibinfo
  {author} {\bibfnamefont {F.}~\bibnamefont {Tran}},\ and\ \bibinfo {author}
  {\bibfnamefont {P.}~\bibnamefont {Blaha}},\ }\href
  {https://doi.org/10.1103/PhysRevB.82.205212} {\bibfield  {journal} {\bibinfo
  {journal} {Phys. Rev. B}\ }\textbf {\bibinfo {volume} {82}},\ \bibinfo
  {pages} {205212} (\bibinfo {year} {2010})}\BibitemShut {NoStop}%
\bibitem [{\citenamefont {Tran}\ and\ \citenamefont
  {Blaha}(2009)}]{Tran2009Accurate}%
  \BibitemOpen
  \bibfield  {author} {\bibinfo {author} {\bibfnamefont {F.}~\bibnamefont
  {Tran}}\ and\ \bibinfo {author} {\bibfnamefont {P.}~\bibnamefont {Blaha}},\
  }\href {https://doi.org/10.1103/PhysRevLett.102.226401} {\bibfield  {journal}
  {\bibinfo  {journal} {Phys. Rev. Lett.}\ }\textbf {\bibinfo {volume} {102}},\
  \bibinfo {pages} {226401} (\bibinfo {year} {2009})}\BibitemShut {NoStop}%
\bibitem [{\citenamefont {Jiang}(2013)}]{Jiang2013Band}%
  \BibitemOpen
  \bibfield  {author} {\bibinfo {author} {\bibfnamefont {H.}~\bibnamefont
  {Jiang}},\ }\href {https://doi.org/10.1063/1.4798706} {\bibfield  {journal}
  {\bibinfo  {journal} {J. Chem. Phys.}\ }\textbf {\bibinfo {volume} {138}},\
  \bibinfo {pages} {134115} (\bibinfo {year} {2013})}\BibitemShut {NoStop}%
\bibitem [{\citenamefont {Rehman}\ \emph {et~al.}(2016)\citenamefont {Rehman},
  \citenamefont {Shafiq}, \citenamefont {{Saifullah}}, \citenamefont {Ahmad},
  \citenamefont {Jalali-Asadabadi}, \citenamefont {Maqbool}, \citenamefont
  {Khan}, \citenamefont {Rahnamaye-Aliabad},\ and\ \citenamefont
  {Ahmad}}]{Rehman2016Electronic}%
  \BibitemOpen
  \bibfield  {author} {\bibinfo {author} {\bibfnamefont {G.}~\bibnamefont
  {Rehman}}, \bibinfo {author} {\bibfnamefont {M.}~\bibnamefont {Shafiq}},
  \bibinfo {author} {\bibnamefont {{Saifullah}}}, \bibinfo {author}
  {\bibfnamefont {R.}~\bibnamefont {Ahmad}}, \bibinfo {author} {\bibfnamefont
  {S.}~\bibnamefont {Jalali-Asadabadi}}, \bibinfo {author} {\bibfnamefont
  {M.}~\bibnamefont {Maqbool}}, \bibinfo {author} {\bibfnamefont
  {I.}~\bibnamefont {Khan}}, \bibinfo {author} {\bibfnamefont {H.}~\bibnamefont
  {Rahnamaye-Aliabad}},\ and\ \bibinfo {author} {\bibfnamefont
  {I.}~\bibnamefont {Ahmad}},\ }\href
  {https://doi.org/10.1007/s11664-016-4492-7} {\bibfield  {journal} {\bibinfo
  {journal} {J. Electron. Mater.}\ }\textbf {\bibinfo {volume} {45}},\ \bibinfo
  {pages} {3314} (\bibinfo {year} {2016})}\BibitemShut {NoStop}%
\bibitem [{\citenamefont {Koller}\ \emph {et~al.}(2012)\citenamefont {Koller},
  \citenamefont {Tran},\ and\ \citenamefont {Blaha}}]{Koller2012Improving}%
  \BibitemOpen
  \bibfield  {author} {\bibinfo {author} {\bibfnamefont {D.}~\bibnamefont
  {Koller}}, \bibinfo {author} {\bibfnamefont {F.}~\bibnamefont {Tran}},\ and\
  \bibinfo {author} {\bibfnamefont {P.}~\bibnamefont {Blaha}},\ }\href
  {https://doi.org/10.1103/PhysRevB.85.155109} {\bibfield  {journal} {\bibinfo
  {journal} {Phys. Rev. B}\ }\textbf {\bibinfo {volume} {85}},\ \bibinfo
  {pages} {155109} (\bibinfo {year} {2012})}\BibitemShut {NoStop}%
\bibitem [{\citenamefont {Wang}\ \emph {et~al.}(1998)\citenamefont {Wang},
  \citenamefont {Bellaiche}, \citenamefont {Wei},\ and\ \citenamefont
  {Zunger}}]{Wang1998}%
  \BibitemOpen
  \bibfield  {author} {\bibinfo {author} {\bibfnamefont {L.~W.}\ \bibnamefont
  {Wang}}, \bibinfo {author} {\bibfnamefont {L.}~\bibnamefont {Bellaiche}},
  \bibinfo {author} {\bibfnamefont {S.~H.}\ \bibnamefont {Wei}},\ and\ \bibinfo
  {author} {\bibfnamefont {A.}~\bibnamefont {Zunger}},\ }\href
  {https://doi.org/10.1103/PhysRevLett.80.4725} {\bibfield  {journal} {\bibinfo
   {journal} {Phys. Rev. Lett.}\ }\textbf {\bibinfo {volume} {80}},\ \bibinfo
  {pages} {4725} (\bibinfo {year} {1998})}\BibitemShut {NoStop}%
\bibitem [{\citenamefont {Medeiros}\ \emph {et~al.}(2015)\citenamefont
  {Medeiros}, \citenamefont {Tsirkin}, \citenamefont {Stafstr{\"{o}}m},\ and\
  \citenamefont {Bj{\"{o}}rk}}]{Medeiros2015}%
  \BibitemOpen
  \bibfield  {author} {\bibinfo {author} {\bibfnamefont {P.~V.~C.}\
  \bibnamefont {Medeiros}}, \bibinfo {author} {\bibfnamefont {S.~S.}\
  \bibnamefont {Tsirkin}}, \bibinfo {author} {\bibfnamefont {S.}~\bibnamefont
  {Stafstr{\"{o}}m}},\ and\ \bibinfo {author} {\bibfnamefont {J.}~\bibnamefont
  {Bj{\"{o}}rk}},\ }\href {https://doi.org/10.1103/PhysRevB.91.041116}
  {\bibfield  {journal} {\bibinfo  {journal} {Phys. Rev. B}\ }\textbf {\bibinfo
  {volume} {91}},\ \bibinfo {pages} {041116(R)} (\bibinfo {year}
  {2015})}\BibitemShut {NoStop}%
\bibitem [{\citenamefont {Popescu}\ and\ \citenamefont
  {Zunger}(2010)}]{Popescu2010}%
  \BibitemOpen
  \bibfield  {author} {\bibinfo {author} {\bibfnamefont {V.}~\bibnamefont
  {Popescu}}\ and\ \bibinfo {author} {\bibfnamefont {A.}~\bibnamefont
  {Zunger}},\ }\href {https://doi.org/10.1103/PhysRevLett.104.236403}
  {\bibfield  {journal} {\bibinfo  {journal} {Phys. Rev. Lett.}\ }\textbf
  {\bibinfo {volume} {104}},\ \bibinfo {pages} {236403} (\bibinfo {year}
  {2010})}\BibitemShut {NoStop}%
\bibitem [{\citenamefont {Rubel}\ \emph {et~al.}(2014)\citenamefont {Rubel},
  \citenamefont {Bokhanchuk}, \citenamefont {Ahmed},\ and\ \citenamefont
  {Assmann}}]{Rubel2014}%
  \BibitemOpen
  \bibfield  {author} {\bibinfo {author} {\bibfnamefont {O.}~\bibnamefont
  {Rubel}}, \bibinfo {author} {\bibfnamefont {A.}~\bibnamefont {Bokhanchuk}},
  \bibinfo {author} {\bibfnamefont {S.~J.}\ \bibnamefont {Ahmed}},\ and\
  \bibinfo {author} {\bibfnamefont {E.}~\bibnamefont {Assmann}},\ }\href
  {https://doi.org/10.1103/PhysRevB.90.115202} {\bibfield  {journal} {\bibinfo
  {journal} {Phys. Rev. B}\ }\textbf {\bibinfo {volume} {90}},\ \bibinfo
  {pages} {115202} (\bibinfo {year} {2014})}\BibitemShut {NoStop}%
\bibitem [{\citenamefont {Popescu}\ and\ \citenamefont
  {Zunger}(2012)}]{Popescu2012}%
  \BibitemOpen
  \bibfield  {author} {\bibinfo {author} {\bibfnamefont {V.}~\bibnamefont
  {Popescu}}\ and\ \bibinfo {author} {\bibfnamefont {A.}~\bibnamefont
  {Zunger}},\ }\href {https://doi.org/10.1103/PhysRevB.85.085201} {\bibfield
  {journal} {\bibinfo  {journal} {Phys. Rev. B}\ }\textbf {\bibinfo {volume}
  {85}},\ \bibinfo {pages} {085201} (\bibinfo {year} {2012})}\BibitemShut
  {NoStop}%
\bibitem [{\citenamefont {Hinuma}\ \emph {et~al.}(2014)\citenamefont {Hinuma},
  \citenamefont {Gr{\"{u}}neis}, \citenamefont {Kresse},\ and\ \citenamefont
  {Oba}}]{Hinuma2014Band}%
  \BibitemOpen
  \bibfield  {author} {\bibinfo {author} {\bibfnamefont {Y.}~\bibnamefont
  {Hinuma}}, \bibinfo {author} {\bibfnamefont {A.}~\bibnamefont
  {Gr{\"{u}}neis}}, \bibinfo {author} {\bibfnamefont {G.}~\bibnamefont
  {Kresse}},\ and\ \bibinfo {author} {\bibfnamefont {F.}~\bibnamefont {Oba}},\
  }\href {https://doi.org/10.1103/PhysRevB.90.155405} {\bibfield  {journal}
  {\bibinfo  {journal} {Phys. Rev. B}\ }\textbf {\bibinfo {volume} {90}},\
  \bibinfo {pages} {155405} (\bibinfo {year} {2014})}\BibitemShut {NoStop}%
\bibitem [{\citenamefont {Craford}\ \emph {et~al.}(1973)\citenamefont
  {Craford}, \citenamefont {Keune}, \citenamefont {Groves},\ and\ \citenamefont
  {Herzog}}]{Craford1973}%
  \BibitemOpen
  \bibfield  {author} {\bibinfo {author} {\bibfnamefont {M.~G.}\ \bibnamefont
  {Craford}}, \bibinfo {author} {\bibfnamefont {D.~L.}\ \bibnamefont {Keune}},
  \bibinfo {author} {\bibfnamefont {W.~O.}\ \bibnamefont {Groves}},\ and\
  \bibinfo {author} {\bibfnamefont {A.~H.}\ \bibnamefont {Herzog}},\ }\href
  {https://doi.org/10.1007/BF02658108} {\bibfield  {journal} {\bibinfo
  {journal} {J. Electron. Mater.}\ }\textbf {\bibinfo {volume} {2}},\ \bibinfo
  {pages} {137} (\bibinfo {year} {1973})}\BibitemShut {NoStop}%
\bibitem [{\citenamefont {Henning}\ and\ \citenamefont
  {Thomas}(1983)}]{Henning1983}%
  \BibitemOpen
  \bibfield  {author} {\bibinfo {author} {\bibfnamefont {I.~D.}\ \bibnamefont
  {Henning}}\ and\ \bibinfo {author} {\bibfnamefont {H.}~\bibnamefont
  {Thomas}},\ }\href {https://doi.org/10.1002/PSSA.2210790230} {\bibfield
  {journal} {\bibinfo  {journal} {Phys. status solidi (a)}\ }\textbf {\bibinfo
  {volume} {79}},\ \bibinfo {pages} {567} (\bibinfo {year} {1983})}\BibitemShut
  {NoStop}%
\bibitem [{\citenamefont {Tanaka}\ and\ \citenamefont
  {Toyama}(1994)}]{Tanaka1994}%
  \BibitemOpen
  \bibfield  {author} {\bibinfo {author} {\bibfnamefont {Y.}~\bibnamefont
  {Tanaka}}\ and\ \bibinfo {author} {\bibfnamefont {T.}~\bibnamefont
  {Toyama}},\ }\href {https://doi.org/10.1109/16.297748} {\bibfield  {journal}
  {\bibinfo  {journal} {IEEE Trans. Electron Devices}\ }\textbf {\bibinfo
  {volume} {41}},\ \bibinfo {pages} {1475} (\bibinfo {year}
  {1994})}\BibitemShut {NoStop}%
\bibitem [{\citenamefont {Sato}\ and\ \citenamefont {Imai}(2002)}]{Sato2002}%
  \BibitemOpen
  \bibfield  {author} {\bibinfo {author} {\bibfnamefont {T.}~\bibnamefont
  {Sato}}\ and\ \bibinfo {author} {\bibfnamefont {M.}~\bibnamefont {Imai}},\
  }\href {https://doi.org/10.1143/JJAP.41.5995} {\bibfield  {journal} {\bibinfo
   {journal} {Jpn. J. Appl. Phys.}\ }\textbf {\bibinfo {volume} {41}},\
  \bibinfo {pages} {5995} (\bibinfo {year} {2002})}\BibitemShut {NoStop}%
\bibitem [{\citenamefont {Geisz}\ and\ \citenamefont
  {Friedman}(2002)}]{Geisz2002}%
  \BibitemOpen
  \bibfield  {author} {\bibinfo {author} {\bibfnamefont {J.~F.}\ \bibnamefont
  {Geisz}}\ and\ \bibinfo {author} {\bibfnamefont {D.~J.}\ \bibnamefont
  {Friedman}},\ }\href {https://doi.org/10.1088/0268-1242/17/8/305} {\bibfield
  {journal} {\bibinfo  {journal} {Semicond. Sci. Technol.}\ }\textbf {\bibinfo
  {volume} {17}},\ \bibinfo {pages} {769} (\bibinfo {year} {2002})}\BibitemShut
  {NoStop}%
\bibitem [{\citenamefont {Lang}\ \emph {et~al.}(2013)\citenamefont {Lang},
  \citenamefont {Faucher}, \citenamefont {Tomasulo}, \citenamefont
  {Nay~Yaung},\ and\ \citenamefont {Larry~Lee}}]{Lang2013}%
  \BibitemOpen
  \bibfield  {author} {\bibinfo {author} {\bibfnamefont {J.~R.}\ \bibnamefont
  {Lang}}, \bibinfo {author} {\bibfnamefont {J.}~\bibnamefont {Faucher}},
  \bibinfo {author} {\bibfnamefont {S.}~\bibnamefont {Tomasulo}}, \bibinfo
  {author} {\bibfnamefont {K.}~\bibnamefont {Nay~Yaung}},\ and\ \bibinfo
  {author} {\bibfnamefont {M.}~\bibnamefont {Larry~Lee}},\ }\href
  {https://doi.org/10.1063/1.4819456} {\bibfield  {journal} {\bibinfo
  {journal} {Appl. Phys. Lett.}\ }\textbf {\bibinfo {volume} {103}},\ \bibinfo
  {pages} {092102} (\bibinfo {year} {2013})}\BibitemShut {NoStop}%
\bibitem [{\citenamefont {Hayashi}\ \emph {et~al.}(1994)\citenamefont
  {Hayashi}, \citenamefont {Soga}, \citenamefont {Nishikawa}, \citenamefont
  {Jimbo},\ and\ \citenamefont {Umeno}}]{Hayashi1994}%
  \BibitemOpen
  \bibfield  {author} {\bibinfo {author} {\bibfnamefont {K.}~\bibnamefont
  {Hayashi}}, \bibinfo {author} {\bibfnamefont {T.}~\bibnamefont {Soga}},
  \bibinfo {author} {\bibfnamefont {H.}~\bibnamefont {Nishikawa}}, \bibinfo
  {author} {\bibfnamefont {T.}~\bibnamefont {Jimbo}},\ and\ \bibinfo {author}
  {\bibfnamefont {M.}~\bibnamefont {Umeno}},\ }\href
  {https://doi.org/10.1109/WCPEC.1994.520736} {\bibfield  {journal} {\bibinfo
  {journal} {Conf. Rec. IEEE Photovolt. Spec. Conf.}\ }\textbf {\bibinfo
  {volume} {2}},\ \bibinfo {pages} {1890} (\bibinfo {year} {1994})}\BibitemShut
  {NoStop}%
\bibitem [{\citenamefont {Grassman}\ \emph {et~al.}(2016)\citenamefont
  {Grassman}, \citenamefont {Chmielewski}, \citenamefont {Carnevale},
  \citenamefont {Carlin},\ and\ \citenamefont {Ringel}}]{Grassman2016}%
  \BibitemOpen
  \bibfield  {author} {\bibinfo {author} {\bibfnamefont {T.~J.}\ \bibnamefont
  {Grassman}}, \bibinfo {author} {\bibfnamefont {D.~J.}\ \bibnamefont
  {Chmielewski}}, \bibinfo {author} {\bibfnamefont {S.~D.}\ \bibnamefont
  {Carnevale}}, \bibinfo {author} {\bibfnamefont {J.~A.}\ \bibnamefont
  {Carlin}},\ and\ \bibinfo {author} {\bibfnamefont {S.~A.}\ \bibnamefont
  {Ringel}},\ }\href {https://doi.org/10.1109/JPHOTOV.2015.2493365} {\bibfield
  {journal} {\bibinfo  {journal} {IEEE J. Photovolt.}\ }\textbf {\bibinfo
  {volume} {6}},\ \bibinfo {pages} {326} (\bibinfo {year} {2016})}\BibitemShut
  {NoStop}%
\bibitem [{\citenamefont {Weyers}\ \emph {et~al.}(1992)\citenamefont {Weyers},
  \citenamefont {Michio~Sato},\ and\ \citenamefont
  {Hiroaki~Ando}}]{Weyers1992}%
  \BibitemOpen
  \bibfield  {author} {\bibinfo {author} {\bibfnamefont {M.}~\bibnamefont
  {Weyers}}, \bibinfo {author} {\bibfnamefont {M.~S.}\ \bibnamefont
  {Michio~Sato}},\ and\ \bibinfo {author} {\bibfnamefont {H.~A.}\ \bibnamefont
  {Hiroaki~Ando}},\ }\href {https://doi.org/10.1143/JJAP.31.L853} {\bibfield
  {journal} {\bibinfo  {journal} {Jpn. J. Appl. Phys.}\ }\textbf {\bibinfo
  {volume} {31}},\ \bibinfo {pages} {L853} (\bibinfo {year}
  {1992})}\BibitemShut {NoStop}%
\bibitem [{\citenamefont {Kunert}\ \emph {et~al.}(2006)\citenamefont {Kunert},
  \citenamefont {Volz}, \citenamefont {Koch},\ and\ \citenamefont
  {Stolz}}]{Kunert2006}%
  \BibitemOpen
  \bibfield  {author} {\bibinfo {author} {\bibfnamefont {B.}~\bibnamefont
  {Kunert}}, \bibinfo {author} {\bibfnamefont {K.}~\bibnamefont {Volz}},
  \bibinfo {author} {\bibfnamefont {J.}~\bibnamefont {Koch}},\ and\ \bibinfo
  {author} {\bibfnamefont {W.}~\bibnamefont {Stolz}},\ }\href
  {https://doi.org/10.1063/1.2200758} {\bibfield  {journal} {\bibinfo
  {journal} {Appl. Phys. Lett.}\ }\textbf {\bibinfo {volume} {88}},\ \bibinfo
  {pages} {182108} (\bibinfo {year} {2006})}\BibitemShut {NoStop}%
\bibitem [{\citenamefont {Zhao}\ \emph {et~al.}(2004)\citenamefont {Zhao},
  \citenamefont {Chen}, \citenamefont {Wang},\ and\ \citenamefont
  {Yoon}}]{Zhao2004}%
  \BibitemOpen
  \bibfield  {author} {\bibinfo {author} {\bibfnamefont {Y.}~\bibnamefont
  {Zhao}}, \bibinfo {author} {\bibfnamefont {G.}~\bibnamefont {Chen}}, \bibinfo
  {author} {\bibfnamefont {S.}~\bibnamefont {Wang}},\ and\ \bibinfo {author}
  {\bibfnamefont {S.~F.}\ \bibnamefont {Yoon}},\ }\href
  {https://doi.org/10.1016/J.TSF.2003.11.289} {\bibfield  {journal} {\bibinfo
  {journal} {Thin Solid Films}\ }\textbf {\bibinfo {volume} {450}},\ \bibinfo
  {pages} {352} (\bibinfo {year} {2004})}\BibitemShut {NoStop}%
\bibitem [{\citenamefont {Loualiche}\ \emph {et~al.}(1998)\citenamefont
  {Loualiche}, \citenamefont {Le~Corre}, \citenamefont {Salaun}, \citenamefont
  {Caulet}, \citenamefont {Lambert}, \citenamefont {Gauneau}, \citenamefont
  {Lecrosnier},\ and\ \citenamefont {Deveaud}}]{Loualiche1998}%
  \BibitemOpen
  \bibfield  {author} {\bibinfo {author} {\bibfnamefont {S.}~\bibnamefont
  {Loualiche}}, \bibinfo {author} {\bibfnamefont {A.}~\bibnamefont {Le~Corre}},
  \bibinfo {author} {\bibfnamefont {S.}~\bibnamefont {Salaun}}, \bibinfo
  {author} {\bibfnamefont {J.}~\bibnamefont {Caulet}}, \bibinfo {author}
  {\bibfnamefont {B.}~\bibnamefont {Lambert}}, \bibinfo {author} {\bibfnamefont
  {M.}~\bibnamefont {Gauneau}}, \bibinfo {author} {\bibfnamefont
  {D.}~\bibnamefont {Lecrosnier}},\ and\ \bibinfo {author} {\bibfnamefont
  {B.}~\bibnamefont {Deveaud}},\ }\href {https://doi.org/10.1063/1.105450}
  {\bibfield  {journal} {\bibinfo  {journal} {Appl. Phys. Lett.}\ }\textbf
  {\bibinfo {volume} {59}},\ \bibinfo {pages} {423} (\bibinfo {year}
  {1998})}\BibitemShut {NoStop}%
\bibitem [{\citenamefont {Shimomura}\ \emph {et~al.}(1996)\citenamefont
  {Shimomura}, \citenamefont {Anan},\ and\ \citenamefont
  {Sugou}}]{Shimomura1996}%
  \BibitemOpen
  \bibfield  {author} {\bibinfo {author} {\bibfnamefont {H.}~\bibnamefont
  {Shimomura}}, \bibinfo {author} {\bibfnamefont {T.}~\bibnamefont {Anan}},\
  and\ \bibinfo {author} {\bibfnamefont {S.}~\bibnamefont {Sugou}},\ }\href
  {https://doi.org/10.1016/0022-0248(95)00950-7} {\bibfield  {journal}
  {\bibinfo  {journal} {J. Cryst. Growth}\ }\textbf {\bibinfo {volume} {162}},\
  \bibinfo {pages} {121} (\bibinfo {year} {1996})}\BibitemShut {NoStop}%
\bibitem [{\citenamefont {Nakajima}\ \emph {et~al.}(2000)\citenamefont
  {Nakajima}, \citenamefont {Ujihara}, \citenamefont {Miyashita},\ and\
  \citenamefont {Sazaki}}]{Nakajima2000}%
  \BibitemOpen
  \bibfield  {author} {\bibinfo {author} {\bibfnamefont {K.}~\bibnamefont
  {Nakajima}}, \bibinfo {author} {\bibfnamefont {T.}~\bibnamefont {Ujihara}},
  \bibinfo {author} {\bibfnamefont {S.}~\bibnamefont {Miyashita}},\ and\
  \bibinfo {author} {\bibfnamefont {G.}~\bibnamefont {Sazaki}},\ }\href
  {https://doi.org/10.1016/S0022-0248(99)00735-6} {\bibfield  {journal}
  {\bibinfo  {journal} {J. Cryst. Growth}\ }\textbf {\bibinfo {volume} {209}},\
  \bibinfo {pages} {637} (\bibinfo {year} {2000})}\BibitemShut {NoStop}%
\bibitem [{\citenamefont {Russell}\ \emph {et~al.}(2016)\citenamefont
  {Russell}, \citenamefont {Andriotis}, \citenamefont {Menon}, \citenamefont
  {Jasinski}, \citenamefont {Martinez-Garcia},\ and\ \citenamefont
  {Sunkara}}]{Russell2016}%
  \BibitemOpen
  \bibfield  {author} {\bibinfo {author} {\bibfnamefont {H.~B.}\ \bibnamefont
  {Russell}}, \bibinfo {author} {\bibfnamefont {A.~N.}\ \bibnamefont
  {Andriotis}}, \bibinfo {author} {\bibfnamefont {M.}~\bibnamefont {Menon}},
  \bibinfo {author} {\bibfnamefont {J.~B.}\ \bibnamefont {Jasinski}}, \bibinfo
  {author} {\bibfnamefont {A.}~\bibnamefont {Martinez-Garcia}},\ and\ \bibinfo
  {author} {\bibfnamefont {M.~K.}\ \bibnamefont {Sunkara}},\ }\href
  {https://doi.org/10.1038/SREP20822} {\bibfield  {journal} {\bibinfo
  {journal} {Sci. Rep.}\ }\textbf {\bibinfo {volume} {6}},\ \bibinfo {pages}
  {20822} (\bibinfo {year} {2016})}\BibitemShut {NoStop}%
\bibitem [{\citenamefont {Jou}\ \emph {et~al.}(1988)\citenamefont {Jou},
  \citenamefont {Cherng}, \citenamefont {Jen},\ and\ \citenamefont
  {Stringfellow}}]{Jou1988}%
  \BibitemOpen
  \bibfield  {author} {\bibinfo {author} {\bibfnamefont {M.~J.}\ \bibnamefont
  {Jou}}, \bibinfo {author} {\bibfnamefont {Y.~T.}\ \bibnamefont {Cherng}},
  \bibinfo {author} {\bibfnamefont {H.~R.}\ \bibnamefont {Jen}},\ and\ \bibinfo
  {author} {\bibfnamefont {G.~B.}\ \bibnamefont {Stringfellow}},\ }\href
  {https://doi.org/10.1063/1.99413} {\bibfield  {journal} {\bibinfo  {journal}
  {Appl. Phys. Lett.}\ }\textbf {\bibinfo {volume} {52}},\ \bibinfo {pages}
  {549} (\bibinfo {year} {1988})}\BibitemShut {NoStop}%
\bibitem [{\citenamefont {Cherng}\ \emph {et~al.}(1984)\citenamefont {Cherng},
  \citenamefont {Cohen},\ and\ \citenamefont {Stringfellow}}]{Cherng1984}%
  \BibitemOpen
  \bibfield  {author} {\bibinfo {author} {\bibfnamefont {M.~J.}\ \bibnamefont
  {Cherng}}, \bibinfo {author} {\bibfnamefont {R.~M.}\ \bibnamefont {Cohen}},\
  and\ \bibinfo {author} {\bibfnamefont {G.~B.}\ \bibnamefont {Stringfellow}},\
  }\href {https://doi.org/10.1007/BF02657927} {\bibfield  {journal} {\bibinfo
  {journal} {J. Electron. Mater.}\ }\textbf {\bibinfo {volume} {13}},\ \bibinfo
  {pages} {799} (\bibinfo {year} {1984})}\BibitemShut {NoStop}%
\bibitem [{\citenamefont {Jen}\ \emph {et~al.}(1998)\citenamefont {Jen},
  \citenamefont {Cherng},\ and\ \citenamefont {Stringfellow}}]{Jen1998}%
  \BibitemOpen
  \bibfield  {author} {\bibinfo {author} {\bibfnamefont {H.~R.}\ \bibnamefont
  {Jen}}, \bibinfo {author} {\bibfnamefont {M.~J.}\ \bibnamefont {Cherng}},\
  and\ \bibinfo {author} {\bibfnamefont {G.~B.}\ \bibnamefont {Stringfellow}},\
  }\href {https://doi.org/10.1063/1.96830} {\bibfield  {journal} {\bibinfo
  {journal} {Appl. Phys. Lett.}\ }\textbf {\bibinfo {volume} {48}},\ \bibinfo
  {pages} {1603} (\bibinfo {year} {1998})}\BibitemShut {NoStop}%
\bibitem [{\citenamefont {Christian}\ \emph {et~al.}(2016)\citenamefont
  {Christian}, \citenamefont {Beaton}, \citenamefont {Mascarenhas},\ and\
  \citenamefont {Alberi}}]{Christian2016}%
  \BibitemOpen
  \bibfield  {author} {\bibinfo {author} {\bibfnamefont {T.~M.}\ \bibnamefont
  {Christian}}, \bibinfo {author} {\bibfnamefont {D.~A.}\ \bibnamefont
  {Beaton}}, \bibinfo {author} {\bibfnamefont {A.}~\bibnamefont
  {Mascarenhas}},\ and\ \bibinfo {author} {\bibfnamefont {K.}~\bibnamefont
  {Alberi}},\ }\href {https://doi.org/10.1117/12.2245432} {\bibfield  {journal}
  {\bibinfo  {journal} {Proc. SPIE 10174, Int. Symp. Clust. Nanomater.}\
  }\textbf {\bibinfo {volume} {10174}},\ \bibinfo {pages} {228} (\bibinfo
  {year} {2016})}\BibitemShut {NoStop}%
\bibitem [{\citenamefont {Christian}\ \emph {et~al.}(2015)\citenamefont
  {Christian}, \citenamefont {Beaton}, \citenamefont {Alberi}, \citenamefont
  {Fluegel},\ and\ \citenamefont {Mascarenhas}}]{Christian2015}%
  \BibitemOpen
  \bibfield  {author} {\bibinfo {author} {\bibfnamefont {T.~M.}\ \bibnamefont
  {Christian}}, \bibinfo {author} {\bibfnamefont {D.~A.}\ \bibnamefont
  {Beaton}}, \bibinfo {author} {\bibfnamefont {K.}~\bibnamefont {Alberi}},
  \bibinfo {author} {\bibfnamefont {B.}~\bibnamefont {Fluegel}},\ and\ \bibinfo
  {author} {\bibfnamefont {A.}~\bibnamefont {Mascarenhas}},\ }\href
  {https://doi.org/10.7567/APEX.8.061202} {\bibfield  {journal} {\bibinfo
  {journal} {Appl. Phys. Express}\ }\textbf {\bibinfo {volume} {8}},\ \bibinfo
  {pages} {061202} (\bibinfo {year} {2015})}\BibitemShut {NoStop}%
\bibitem [{\citenamefont {Sweeney}\ \emph {et~al.}(2011)\citenamefont
  {Sweeney}, \citenamefont {Batool}, \citenamefont {Hild}, \citenamefont
  {Jin},\ and\ \citenamefont {Hosea}}]{Sweeney2011}%
  \BibitemOpen
  \bibfield  {author} {\bibinfo {author} {\bibfnamefont {S.~J.}\ \bibnamefont
  {Sweeney}}, \bibinfo {author} {\bibfnamefont {Z.}~\bibnamefont {Batool}},
  \bibinfo {author} {\bibfnamefont {K.}~\bibnamefont {Hild}}, \bibinfo {author}
  {\bibfnamefont {S.~R.}\ \bibnamefont {Jin}},\ and\ \bibinfo {author}
  {\bibfnamefont {T.~J.}\ \bibnamefont {Hosea}},\ }in\ \href
  {https://doi.org/10.1109/ICTON.2011.5970829} {\emph {\bibinfo {booktitle}
  {2011 13th Int. Conf. Transparent Opt. Networks}}}\ (\bibinfo {year} {2011})\
  pp.\ \bibinfo {pages} {1--4}\BibitemShut {NoStop}%
\bibitem [{\citenamefont {Wang}\ \emph {et~al.}(2013)\citenamefont {Wang},
  \citenamefont {Song}, \citenamefont {Wang}, \citenamefont {Gu}, \citenamefont
  {Zhao}, \citenamefont {Chen}, \citenamefont {Ye}, \citenamefont {Zhou},
  \citenamefont {Kang}, \citenamefont {Li}, \citenamefont {Cao}, \citenamefont
  {Zhang}, \citenamefont {Shao}, \citenamefont {Gong},\ and\ \citenamefont
  {Zhang}}]{Wang2013}%
  \BibitemOpen
  \bibfield  {author} {\bibinfo {author} {\bibfnamefont {S.}~\bibnamefont
  {Wang}}, \bibinfo {author} {\bibfnamefont {Y.}~\bibnamefont {Song}}, \bibinfo
  {author} {\bibfnamefont {K.}~\bibnamefont {Wang}}, \bibinfo {author}
  {\bibfnamefont {Y.}~\bibnamefont {Gu}}, \bibinfo {author} {\bibfnamefont
  {H.}~\bibnamefont {Zhao}}, \bibinfo {author} {\bibfnamefont {X.}~\bibnamefont
  {Chen}}, \bibinfo {author} {\bibfnamefont {H.}~\bibnamefont {Ye}}, \bibinfo
  {author} {\bibfnamefont {H.}~\bibnamefont {Zhou}}, \bibinfo {author}
  {\bibfnamefont {C.}~\bibnamefont {Kang}}, \bibinfo {author} {\bibfnamefont
  {Y.}~\bibnamefont {Li}}, \bibinfo {author} {\bibfnamefont {C.}~\bibnamefont
  {Cao}}, \bibinfo {author} {\bibfnamefont {L.}~\bibnamefont {Zhang}}, \bibinfo
  {author} {\bibfnamefont {J.}~\bibnamefont {Shao}}, \bibinfo {author}
  {\bibfnamefont {Q.}~\bibnamefont {Gong}},\ and\ \bibinfo {author}
  {\bibfnamefont {Y.}~\bibnamefont {Zhang}},\ }in\ \href
  {https://doi.org/10.1364/ACPC.2013.AF3B.5} {\emph {\bibinfo {booktitle} {Asia
  Commun. Photonics Conf. 2013}}}\ (\bibinfo  {publisher} {OSA},\ \bibinfo
  {address} {Washington, D.C.},\ \bibinfo {year} {2013})\ p.\ \bibinfo {pages}
  {AF3B.5}\BibitemShut {NoStop}%
\bibitem [{\citenamefont {Cooke}\ \emph {et~al.}(2006)\citenamefont {Cooke},
  \citenamefont {Hegmann}, \citenamefont {Young},\ and\ \citenamefont
  {Tiedje}}]{Cooke2006}%
  \BibitemOpen
  \bibfield  {author} {\bibinfo {author} {\bibfnamefont {D.~G.}\ \bibnamefont
  {Cooke}}, \bibinfo {author} {\bibfnamefont {F.~A.}\ \bibnamefont {Hegmann}},
  \bibinfo {author} {\bibfnamefont {E.~C.}\ \bibnamefont {Young}},\ and\
  \bibinfo {author} {\bibfnamefont {T.}~\bibnamefont {Tiedje}},\ }\href
  {https://doi.org/10.1063/1.2349314} {\bibfield  {journal} {\bibinfo
  {journal} {Appl. Phys. Lett.}\ }\textbf {\bibinfo {volume} {89}},\ \bibinfo
  {pages} {122103} (\bibinfo {year} {2006})}\BibitemShut {NoStop}%
\bibitem [{\citenamefont {Kresse}\ and\ \citenamefont
  {Hafner}(1993)}]{Kresse1993Ab}%
  \BibitemOpen
  \bibfield  {author} {\bibinfo {author} {\bibfnamefont {G.}~\bibnamefont
  {Kresse}}\ and\ \bibinfo {author} {\bibfnamefont {J.}~\bibnamefont
  {Hafner}},\ }\href {https://doi.org/10.1103/PhysRevB.47.558} {\bibfield
  {journal} {\bibinfo  {journal} {Phys. Rev. B}\ }\textbf {\bibinfo {volume}
  {47}},\ \bibinfo {pages} {558} (\bibinfo {year} {1993})}\BibitemShut
  {NoStop}%
\bibitem [{\citenamefont {Kresse}\ and\ \citenamefont
  {Hafner}(1994)}]{Kresse1994Ab}%
  \BibitemOpen
  \bibfield  {author} {\bibinfo {author} {\bibfnamefont {G.}~\bibnamefont
  {Kresse}}\ and\ \bibinfo {author} {\bibfnamefont {J.}~\bibnamefont
  {Hafner}},\ }\href {https://doi.org/10.1103/PhysRevB.49.14251} {\bibfield
  {journal} {\bibinfo  {journal} {Phys. Rev. B}\ }\textbf {\bibinfo {volume}
  {49}},\ \bibinfo {pages} {14251} (\bibinfo {year} {1994})}\BibitemShut
  {NoStop}%
\bibitem [{\citenamefont {Kresse}\ and\ \citenamefont
  {Furthm{\"{u}}ller}(1996{\natexlab{a}})}]{Kresse1996Efficient}%
  \BibitemOpen
  \bibfield  {author} {\bibinfo {author} {\bibfnamefont {G.}~\bibnamefont
  {Kresse}}\ and\ \bibinfo {author} {\bibfnamefont {J.}~\bibnamefont
  {Furthm{\"{u}}ller}},\ }\href {https://doi.org/10.1103/PhysRevB.54.11169}
  {\bibfield  {journal} {\bibinfo  {journal} {Phys. Rev. B}\ }\textbf {\bibinfo
  {volume} {54}},\ \bibinfo {pages} {11169} (\bibinfo {year}
  {1996}{\natexlab{a}})}\BibitemShut {NoStop}%
\bibitem [{\citenamefont {Kresse}\ and\ \citenamefont
  {Furthm{\"{u}}ller}(1996{\natexlab{b}})}]{Kresse1996Efficiency}%
  \BibitemOpen
  \bibfield  {author} {\bibinfo {author} {\bibfnamefont {G.}~\bibnamefont
  {Kresse}}\ and\ \bibinfo {author} {\bibfnamefont {J.}~\bibnamefont
  {Furthm{\"{u}}ller}},\ }\href {https://doi.org/10.1016/0927-0256(96)00008-0}
  {\bibfield  {journal} {\bibinfo  {journal} {Comput. Mater. Sci.}\ }\textbf
  {\bibinfo {volume} {6}},\ \bibinfo {pages} {15} (\bibinfo {year}
  {1996}{\natexlab{b}})}\BibitemShut {NoStop}%
\bibitem [{\citenamefont {Kresse}\ and\ \citenamefont
  {Joubert}(1999)}]{Kresse1999From}%
  \BibitemOpen
  \bibfield  {author} {\bibinfo {author} {\bibfnamefont {G.}~\bibnamefont
  {Kresse}}\ and\ \bibinfo {author} {\bibfnamefont {D.}~\bibnamefont
  {Joubert}},\ }\href {https://doi.org/10.1103/PhysRevB.59.1758} {\bibfield
  {journal} {\bibinfo  {journal} {Phys. Rev. B}\ }\textbf {\bibinfo {volume}
  {59}},\ \bibinfo {pages} {1758} (\bibinfo {year} {1999})}\BibitemShut
  {NoStop}%
\bibitem [{\citenamefont {Bl{\"{o}}chl}(1994)}]{Blochl1994Projector}%
  \BibitemOpen
  \bibfield  {author} {\bibinfo {author} {\bibfnamefont {P.~E.}\ \bibnamefont
  {Bl{\"{o}}chl}},\ }\href {https://doi.org/10.1103/PhysRevB.50.17953}
  {\bibfield  {journal} {\bibinfo  {journal} {Phys. Rev. B}\ }\textbf {\bibinfo
  {volume} {50}},\ \bibinfo {pages} {17953} (\bibinfo {year}
  {1994})}\BibitemShut {NoStop}%
\bibitem [{\citenamefont {Zunger}\ \emph {et~al.}(1990)\citenamefont {Zunger},
  \citenamefont {Wei}, \citenamefont {Ferreira},\ and\ \citenamefont
  {Bernard}}]{Zunger1990Special}%
  \BibitemOpen
  \bibfield  {author} {\bibinfo {author} {\bibfnamefont {A.}~\bibnamefont
  {Zunger}}, \bibinfo {author} {\bibfnamefont {S.~H.}\ \bibnamefont {Wei}},
  \bibinfo {author} {\bibfnamefont {L.~G.}\ \bibnamefont {Ferreira}},\ and\
  \bibinfo {author} {\bibfnamefont {J.~E.}\ \bibnamefont {Bernard}},\ }\href
  {https://doi.org/10.1103/PhysRevLett.65.353} {\bibfield  {journal} {\bibinfo
  {journal} {Phys. Rev. Lett.}\ }\textbf {\bibinfo {volume} {65}},\ \bibinfo
  {pages} {353} (\bibinfo {year} {1990})}\BibitemShut {NoStop}%
\bibitem [{\citenamefont {van~de Walle}\ \emph {et~al.}(2002)\citenamefont
  {van~de Walle}, \citenamefont {Asta},\ and\ \citenamefont
  {Ceder}}]{VandeWalle2002}%
  \BibitemOpen
  \bibfield  {author} {\bibinfo {author} {\bibfnamefont {A.}~\bibnamefont
  {van~de Walle}}, \bibinfo {author} {\bibfnamefont {M.}~\bibnamefont {Asta}},\
  and\ \bibinfo {author} {\bibfnamefont {G.}~\bibnamefont {Ceder}},\ }\href
  {https://doi.org/10.1016/S0364-5916(02)80006-2} {\bibfield  {journal}
  {\bibinfo  {journal} {Calphad}\ }\textbf {\bibinfo {volume} {26}},\ \bibinfo
  {pages} {539} (\bibinfo {year} {2002})}\BibitemShut {NoStop}%
\bibitem [{\citenamefont {van~de Walle}(2009)}]{VandeWalle2009}%
  \BibitemOpen
  \bibfield  {author} {\bibinfo {author} {\bibfnamefont {A.}~\bibnamefont
  {van~de Walle}},\ }\href {https://doi.org/10.1016/j.calphad.2008.12.005}
  {\bibfield  {journal} {\bibinfo  {journal} {Calphad}\ }\textbf {\bibinfo
  {volume} {33}},\ \bibinfo {pages} {266} (\bibinfo {year} {2009})}\BibitemShut
  {NoStop}%
\bibitem [{\citenamefont {van~de Walle}\ \emph {et~al.}(2013)\citenamefont
  {van~de Walle}, \citenamefont {Tiwary}, \citenamefont {de~Jong},
  \citenamefont {Olmsted}, \citenamefont {Asta}, \citenamefont {Dick},
  \citenamefont {Shin}, \citenamefont {Wang}, \citenamefont {Chen},\ and\
  \citenamefont {Liu}}]{VanDeWalle2013}%
  \BibitemOpen
  \bibfield  {author} {\bibinfo {author} {\bibfnamefont {A.}~\bibnamefont
  {van~de Walle}}, \bibinfo {author} {\bibfnamefont {P.}~\bibnamefont
  {Tiwary}}, \bibinfo {author} {\bibfnamefont {M.}~\bibnamefont {de~Jong}},
  \bibinfo {author} {\bibfnamefont {D.}~\bibnamefont {Olmsted}}, \bibinfo
  {author} {\bibfnamefont {M.}~\bibnamefont {Asta}}, \bibinfo {author}
  {\bibfnamefont {A.}~\bibnamefont {Dick}}, \bibinfo {author} {\bibfnamefont
  {D.}~\bibnamefont {Shin}}, \bibinfo {author} {\bibfnamefont {Y.}~\bibnamefont
  {Wang}}, \bibinfo {author} {\bibfnamefont {L.-Q.}\ \bibnamefont {Chen}},\
  and\ \bibinfo {author} {\bibfnamefont {Z.-K.}\ \bibnamefont {Liu}},\ }\href
  {https://doi.org/10.1016/j.calphad.2013.06.006} {\bibfield  {journal}
  {\bibinfo  {journal} {Calphad}\ }\textbf {\bibinfo {volume} {42}},\ \bibinfo
  {pages} {13} (\bibinfo {year} {2013})}\BibitemShut {NoStop}%
\bibitem [{\citenamefont {Perdew}\ \emph {et~al.}(1996)\citenamefont {Perdew},
  \citenamefont {Burke},\ and\ \citenamefont
  {Ernzerhof}}]{Perdew1996Generalized}%
  \BibitemOpen
  \bibfield  {author} {\bibinfo {author} {\bibfnamefont {J.~P.}\ \bibnamefont
  {Perdew}}, \bibinfo {author} {\bibfnamefont {K.}~\bibnamefont {Burke}},\ and\
  \bibinfo {author} {\bibfnamefont {M.}~\bibnamefont {Ernzerhof}},\ }\href
  {https://doi.org/10.1103/PhysRevLett.77.3865} {\bibfield  {journal} {\bibinfo
   {journal} {Phys. Rev. Lett.}\ }\textbf {\bibinfo {volume} {77}},\ \bibinfo
  {pages} {3865} (\bibinfo {year} {1996})}\BibitemShut {NoStop}%
\bibitem [{\citenamefont {Grimme}\ \emph {et~al.}(2010)\citenamefont {Grimme},
  \citenamefont {Antony}, \citenamefont {Ehrlich},\ and\ \citenamefont
  {Krieg}}]{Grimme2010Consistent}%
  \BibitemOpen
  \bibfield  {author} {\bibinfo {author} {\bibfnamefont {S.}~\bibnamefont
  {Grimme}}, \bibinfo {author} {\bibfnamefont {J.}~\bibnamefont {Antony}},
  \bibinfo {author} {\bibfnamefont {S.}~\bibnamefont {Ehrlich}},\ and\ \bibinfo
  {author} {\bibfnamefont {H.}~\bibnamefont {Krieg}},\ }\href
  {https://doi.org/10.1063/1.3382344} {\bibfield  {journal} {\bibinfo
  {journal} {J. Chem. Phys.}\ }\textbf {\bibinfo {volume} {132}},\ \bibinfo
  {pages} {154104} (\bibinfo {year} {2010})}\BibitemShut {NoStop}%
\bibitem [{\citenamefont {Grimme}\ \emph {et~al.}(2011)\citenamefont {Grimme},
  \citenamefont {Ehrlich},\ and\ \citenamefont {Goerigk}}]{Grimme2011Effect}%
  \BibitemOpen
  \bibfield  {author} {\bibinfo {author} {\bibfnamefont {S.}~\bibnamefont
  {Grimme}}, \bibinfo {author} {\bibfnamefont {S.}~\bibnamefont {Ehrlich}},\
  and\ \bibinfo {author} {\bibfnamefont {L.}~\bibnamefont {Goerigk}},\ }\href
  {https://doi.org/10.1002/jcc.21759} {\bibfield  {journal} {\bibinfo
  {journal} {J. Comput. Chem.}\ }\textbf {\bibinfo {volume} {32}},\ \bibinfo
  {pages} {1456} (\bibinfo {year} {2011})}\BibitemShut {NoStop}%
\bibitem [{\citenamefont {Monkhorst}\ and\ \citenamefont
  {Pack}(1976)}]{Monkhorst1976}%
  \BibitemOpen
  \bibfield  {author} {\bibinfo {author} {\bibfnamefont {H.~J.}\ \bibnamefont
  {Monkhorst}}\ and\ \bibinfo {author} {\bibfnamefont {J.~D.}\ \bibnamefont
  {Pack}},\ }\href {https://doi.org/10.1103/PhysRevB.13.5188} {\bibfield
  {journal} {\bibinfo  {journal} {Phys. Rev. B}\ }\textbf {\bibinfo {volume}
  {13}},\ \bibinfo {pages} {5188} (\bibinfo {year} {1976})}\BibitemShut
  {NoStop}%
\bibitem [{Sup()}]{Supp_info}%
  \BibitemOpen
  \href@noop {} {\bibinfo {title} {{See Supplemental Material for the effective
  bandstructures of GaAsP, GaAsN, and GaAsN under selected isotropic strain
  values; a visual representation of the assessment of uncertainty in the
  bandgap nature near the direct-indirect transition region; the DFT calculated
  bandgap values for all the systems described in the manuscript; the details
  of the interpolation procedure; and the bandgap phase diagrams. The
  Supplemental Material also contains Refs.~[45, 52, 95–103].}}}\BibitemShut
  {Stop}%
\bibitem [{\citenamefont {Medeiros}\ \emph {et~al.}(2014)\citenamefont
  {Medeiros}, \citenamefont {Stafstr{\"{o}}m},\ and\ \citenamefont
  {Bj{\"{o}}rk}}]{Medeiros2014}%
  \BibitemOpen
  \bibfield  {author} {\bibinfo {author} {\bibfnamefont {P.~V.~C.}\
  \bibnamefont {Medeiros}}, \bibinfo {author} {\bibfnamefont {S.}~\bibnamefont
  {Stafstr{\"{o}}m}},\ and\ \bibinfo {author} {\bibfnamefont {J.}~\bibnamefont
  {Bj{\"{o}}rk}},\ }\href {https://doi.org/10.1103/PhysRevB.89.041407}
  {\bibfield  {journal} {\bibinfo  {journal} {Phys. Rev. B}\ }\textbf {\bibinfo
  {volume} {89}},\ \bibinfo {pages} {041407(R)} (\bibinfo {year}
  {2014})}\BibitemShut {NoStop}%
\bibitem [{\citenamefont {Virtanen}\ \emph {et~al.}(2020)\citenamefont
  {Virtanen}, \citenamefont {Gommers}, \citenamefont {Oliphant}, \citenamefont
  {Haberland}, \citenamefont {Reddy}, \citenamefont {Cournapeau}, \citenamefont
  {Burovski}, \citenamefont {Peterson}, \citenamefont {Weckesser},
  \citenamefont {Bright}, \citenamefont {van~der Walt}, \citenamefont {Brett},
  \citenamefont {Wilson}, \citenamefont {Millman}, \citenamefont {Mayorov},
  \citenamefont {Nelson}, \citenamefont {Jones}, \citenamefont {Kern},
  \citenamefont {Larson}, \citenamefont {Carey}, \citenamefont {Polat},
  \citenamefont {Feng}, \citenamefont {Moore}, \citenamefont {VanderPlas},
  \citenamefont {Laxalde}, \citenamefont {Perktold}, \citenamefont {Cimrman},
  \citenamefont {Henriksen}, \citenamefont {Quintero}, \citenamefont {Harris},
  \citenamefont {Archibald}, \citenamefont {Ribeiro}, \citenamefont
  {Pedregosa}, \citenamefont {van Mulbregt}, \citenamefont {Vijaykumar},
  \citenamefont {Bardelli}, \citenamefont {Rothberg}, \citenamefont {Hilboll},
  \citenamefont {Kloeckner}, \citenamefont {Scopatz}, \citenamefont {Lee},
  \citenamefont {Rokem}, \citenamefont {Woods}, \citenamefont {Fulton},
  \citenamefont {Masson}, \citenamefont {H{\"{a}}ggstr{\"{o}}m}, \citenamefont
  {Fitzgerald}, \citenamefont {Nicholson}, \citenamefont {Hagen}, \citenamefont
  {Pasechnik}, \citenamefont {Olivetti}, \citenamefont {Martin}, \citenamefont
  {Wieser}, \citenamefont {Silva}, \citenamefont {Lenders}, \citenamefont
  {Wilhelm}, \citenamefont {Young}, \citenamefont {Price}, \citenamefont
  {Ingold}, \citenamefont {Allen}, \citenamefont {Lee}, \citenamefont {Audren},
  \citenamefont {Probst}, \citenamefont {Dietrich}, \citenamefont {Silterra},
  \citenamefont {Webber}, \citenamefont {Slavi{\v{c}}}, \citenamefont
  {Nothman}, \citenamefont {Buchner}, \citenamefont {Kulick}, \citenamefont
  {Sch{\"{o}}nberger}, \citenamefont {de~Miranda~Cardoso}, \citenamefont
  {Reimer}, \citenamefont {Harrington}, \citenamefont {Rodr{\'{i}}guez},
  \citenamefont {Nunez-Iglesias}, \citenamefont {Kuczynski}, \citenamefont
  {Tritz}, \citenamefont {Thoma}, \citenamefont {Newville}, \citenamefont
  {K{\"{u}}mmerer}, \citenamefont {Bolingbroke}, \citenamefont {Tartre},
  \citenamefont {Pak}, \citenamefont {Smith}, \citenamefont {Nowaczyk},
  \citenamefont {Shebanov}, \citenamefont {Pavlyk}, \citenamefont {Brodtkorb},
  \citenamefont {Lee}, \citenamefont {McGibbon}, \citenamefont {Feldbauer},
  \citenamefont {Lewis}, \citenamefont {Tygier}, \citenamefont {Sievert},
  \citenamefont {Vigna}, \citenamefont {Peterson}, \citenamefont {More},
  \citenamefont {Pudlik}, \citenamefont {Oshima}, \citenamefont {Pingel},
  \citenamefont {Robitaille}, \citenamefont {Spura}, \citenamefont {Jones},
  \citenamefont {Cera}, \citenamefont {Leslie}, \citenamefont {Zito},
  \citenamefont {Krauss}, \citenamefont {Upadhyay}, \citenamefont {Halchenko},\
  and\ \citenamefont {V{\'{a}}zquez-Baeza}}]{SciPyNMeth2020}%
  \BibitemOpen
  \bibfield  {author} {\bibinfo {author} {\bibfnamefont {P.}~\bibnamefont
  {Virtanen}}, \bibinfo {author} {\bibfnamefont {R.}~\bibnamefont {Gommers}},
  \bibinfo {author} {\bibfnamefont {T.~E.}\ \bibnamefont {Oliphant}}, \bibinfo
  {author} {\bibfnamefont {M.}~\bibnamefont {Haberland}}, \bibinfo {author}
  {\bibfnamefont {T.}~\bibnamefont {Reddy}}, \bibinfo {author} {\bibfnamefont
  {D.}~\bibnamefont {Cournapeau}}, \bibinfo {author} {\bibfnamefont
  {E.}~\bibnamefont {Burovski}}, \bibinfo {author} {\bibfnamefont
  {P.}~\bibnamefont {Peterson}}, \bibinfo {author} {\bibfnamefont
  {W.}~\bibnamefont {Weckesser}}, \bibinfo {author} {\bibfnamefont
  {J.}~\bibnamefont {Bright}}, \bibinfo {author} {\bibfnamefont {S.~J.}\
  \bibnamefont {van~der Walt}}, \bibinfo {author} {\bibfnamefont
  {M.}~\bibnamefont {Brett}}, \bibinfo {author} {\bibfnamefont
  {J.}~\bibnamefont {Wilson}}, \bibinfo {author} {\bibfnamefont {K.~J.}\
  \bibnamefont {Millman}}, \bibinfo {author} {\bibfnamefont {N.}~\bibnamefont
  {Mayorov}}, \bibinfo {author} {\bibfnamefont {A.~R.~J.}\ \bibnamefont
  {Nelson}}, \bibinfo {author} {\bibfnamefont {E.}~\bibnamefont {Jones}},
  \bibinfo {author} {\bibfnamefont {R.}~\bibnamefont {Kern}}, \bibinfo {author}
  {\bibfnamefont {E.}~\bibnamefont {Larson}}, \bibinfo {author} {\bibfnamefont
  {C.~J.}\ \bibnamefont {Carey}}, \bibinfo {author} {\bibfnamefont
  {I.}~\bibnamefont {Polat}}, \bibinfo {author} {\bibfnamefont
  {Y.}~\bibnamefont {Feng}}, \bibinfo {author} {\bibfnamefont {E.~W.}\
  \bibnamefont {Moore}}, \bibinfo {author} {\bibfnamefont {J.}~\bibnamefont
  {VanderPlas}}, \bibinfo {author} {\bibfnamefont {D.}~\bibnamefont {Laxalde}},
  \bibinfo {author} {\bibfnamefont {J.}~\bibnamefont {Perktold}}, \bibinfo
  {author} {\bibfnamefont {R.}~\bibnamefont {Cimrman}}, \bibinfo {author}
  {\bibfnamefont {I.}~\bibnamefont {Henriksen}}, \bibinfo {author}
  {\bibfnamefont {E.~A.}\ \bibnamefont {Quintero}}, \bibinfo {author}
  {\bibfnamefont {C.~R.}\ \bibnamefont {Harris}}, \bibinfo {author}
  {\bibfnamefont {A.~M.}\ \bibnamefont {Archibald}}, \bibinfo {author}
  {\bibfnamefont {A.~H.}\ \bibnamefont {Ribeiro}}, \bibinfo {author}
  {\bibfnamefont {F.}~\bibnamefont {Pedregosa}}, \bibinfo {author}
  {\bibfnamefont {P.}~\bibnamefont {van Mulbregt}}, \bibinfo {author}
  {\bibfnamefont {A.}~\bibnamefont {Vijaykumar}}, \bibinfo {author}
  {\bibfnamefont {A.~P.}\ \bibnamefont {Bardelli}}, \bibinfo {author}
  {\bibfnamefont {A.}~\bibnamefont {Rothberg}}, \bibinfo {author}
  {\bibfnamefont {A.}~\bibnamefont {Hilboll}}, \bibinfo {author} {\bibfnamefont
  {A.}~\bibnamefont {Kloeckner}}, \bibinfo {author} {\bibfnamefont
  {A.}~\bibnamefont {Scopatz}}, \bibinfo {author} {\bibfnamefont
  {A.}~\bibnamefont {Lee}}, \bibinfo {author} {\bibfnamefont {A.}~\bibnamefont
  {Rokem}}, \bibinfo {author} {\bibfnamefont {C.~N.}\ \bibnamefont {Woods}},
  \bibinfo {author} {\bibfnamefont {C.}~\bibnamefont {Fulton}}, \bibinfo
  {author} {\bibfnamefont {C.}~\bibnamefont {Masson}}, \bibinfo {author}
  {\bibfnamefont {C.}~\bibnamefont {H{\"{a}}ggstr{\"{o}}m}}, \bibinfo {author}
  {\bibfnamefont {C.}~\bibnamefont {Fitzgerald}}, \bibinfo {author}
  {\bibfnamefont {D.~A.}\ \bibnamefont {Nicholson}}, \bibinfo {author}
  {\bibfnamefont {D.~R.}\ \bibnamefont {Hagen}}, \bibinfo {author}
  {\bibfnamefont {D.~V.}\ \bibnamefont {Pasechnik}}, \bibinfo {author}
  {\bibfnamefont {E.}~\bibnamefont {Olivetti}}, \bibinfo {author}
  {\bibfnamefont {E.}~\bibnamefont {Martin}}, \bibinfo {author} {\bibfnamefont
  {E.}~\bibnamefont {Wieser}}, \bibinfo {author} {\bibfnamefont
  {F.}~\bibnamefont {Silva}}, \bibinfo {author} {\bibfnamefont
  {F.}~\bibnamefont {Lenders}}, \bibinfo {author} {\bibfnamefont
  {F.}~\bibnamefont {Wilhelm}}, \bibinfo {author} {\bibfnamefont
  {G.}~\bibnamefont {Young}}, \bibinfo {author} {\bibfnamefont {G.~A.}\
  \bibnamefont {Price}}, \bibinfo {author} {\bibfnamefont {G.-L.}\ \bibnamefont
  {Ingold}}, \bibinfo {author} {\bibfnamefont {G.~E.}\ \bibnamefont {Allen}},
  \bibinfo {author} {\bibfnamefont {G.~R.}\ \bibnamefont {Lee}}, \bibinfo
  {author} {\bibfnamefont {H.}~\bibnamefont {Audren}}, \bibinfo {author}
  {\bibfnamefont {I.}~\bibnamefont {Probst}}, \bibinfo {author} {\bibfnamefont
  {J.~P.}\ \bibnamefont {Dietrich}}, \bibinfo {author} {\bibfnamefont
  {J.}~\bibnamefont {Silterra}}, \bibinfo {author} {\bibfnamefont {J.~T.}\
  \bibnamefont {Webber}}, \bibinfo {author} {\bibfnamefont {J.}~\bibnamefont
  {Slavi{\v{c}}}}, \bibinfo {author} {\bibfnamefont {J.}~\bibnamefont
  {Nothman}}, \bibinfo {author} {\bibfnamefont {J.}~\bibnamefont {Buchner}},
  \bibinfo {author} {\bibfnamefont {J.}~\bibnamefont {Kulick}}, \bibinfo
  {author} {\bibfnamefont {J.~L.}\ \bibnamefont {Sch{\"{o}}nberger}}, \bibinfo
  {author} {\bibfnamefont {J.~V.}\ \bibnamefont {de~Miranda~Cardoso}}, \bibinfo
  {author} {\bibfnamefont {J.}~\bibnamefont {Reimer}}, \bibinfo {author}
  {\bibfnamefont {J.}~\bibnamefont {Harrington}}, \bibinfo {author}
  {\bibfnamefont {J.~L.~C.}\ \bibnamefont {Rodr{\'{i}}guez}}, \bibinfo {author}
  {\bibfnamefont {J.}~\bibnamefont {Nunez-Iglesias}}, \bibinfo {author}
  {\bibfnamefont {J.}~\bibnamefont {Kuczynski}}, \bibinfo {author}
  {\bibfnamefont {K.}~\bibnamefont {Tritz}}, \bibinfo {author} {\bibfnamefont
  {M.}~\bibnamefont {Thoma}}, \bibinfo {author} {\bibfnamefont
  {M.}~\bibnamefont {Newville}}, \bibinfo {author} {\bibfnamefont
  {M.}~\bibnamefont {K{\"{u}}mmerer}}, \bibinfo {author} {\bibfnamefont
  {M.}~\bibnamefont {Bolingbroke}}, \bibinfo {author} {\bibfnamefont
  {M.}~\bibnamefont {Tartre}}, \bibinfo {author} {\bibfnamefont
  {M.}~\bibnamefont {Pak}}, \bibinfo {author} {\bibfnamefont {N.~J.}\
  \bibnamefont {Smith}}, \bibinfo {author} {\bibfnamefont {N.}~\bibnamefont
  {Nowaczyk}}, \bibinfo {author} {\bibfnamefont {N.}~\bibnamefont {Shebanov}},
  \bibinfo {author} {\bibfnamefont {O.}~\bibnamefont {Pavlyk}}, \bibinfo
  {author} {\bibfnamefont {P.~A.}\ \bibnamefont {Brodtkorb}}, \bibinfo {author}
  {\bibfnamefont {P.}~\bibnamefont {Lee}}, \bibinfo {author} {\bibfnamefont
  {R.~T.}\ \bibnamefont {McGibbon}}, \bibinfo {author} {\bibfnamefont
  {R.}~\bibnamefont {Feldbauer}}, \bibinfo {author} {\bibfnamefont
  {S.}~\bibnamefont {Lewis}}, \bibinfo {author} {\bibfnamefont
  {S.}~\bibnamefont {Tygier}}, \bibinfo {author} {\bibfnamefont
  {S.}~\bibnamefont {Sievert}}, \bibinfo {author} {\bibfnamefont
  {S.}~\bibnamefont {Vigna}}, \bibinfo {author} {\bibfnamefont
  {S.}~\bibnamefont {Peterson}}, \bibinfo {author} {\bibfnamefont
  {S.}~\bibnamefont {More}}, \bibinfo {author} {\bibfnamefont {T.}~\bibnamefont
  {Pudlik}}, \bibinfo {author} {\bibfnamefont {T.}~\bibnamefont {Oshima}},
  \bibinfo {author} {\bibfnamefont {T.~J.}\ \bibnamefont {Pingel}}, \bibinfo
  {author} {\bibfnamefont {T.~P.}\ \bibnamefont {Robitaille}}, \bibinfo
  {author} {\bibfnamefont {T.}~\bibnamefont {Spura}}, \bibinfo {author}
  {\bibfnamefont {T.~R.}\ \bibnamefont {Jones}}, \bibinfo {author}
  {\bibfnamefont {T.}~\bibnamefont {Cera}}, \bibinfo {author} {\bibfnamefont
  {T.}~\bibnamefont {Leslie}}, \bibinfo {author} {\bibfnamefont
  {T.}~\bibnamefont {Zito}}, \bibinfo {author} {\bibfnamefont {T.}~\bibnamefont
  {Krauss}}, \bibinfo {author} {\bibfnamefont {U.}~\bibnamefont {Upadhyay}},
  \bibinfo {author} {\bibfnamefont {Y.~O.}\ \bibnamefont {Halchenko}},\ and\
  \bibinfo {author} {\bibfnamefont {Y.}~\bibnamefont {V{\'{a}}zquez-Baeza}},\
  }\href {https://doi.org/10.1038/s41592-019-0686-2} {\bibfield  {journal}
  {\bibinfo  {journal} {Nat. Methods}\ }\textbf {\bibinfo {volume} {17}},\
  \bibinfo {pages} {261} (\bibinfo {year} {2020})}\BibitemShut {NoStop}%
\bibitem [{\citenamefont {Nielson}(1983)}]{Nielson1983}%
  \BibitemOpen
  \bibfield  {author} {\bibinfo {author} {\bibfnamefont {G.~M.}\ \bibnamefont
  {Nielson}},\ }\href {https://doi.org/10.1090/S0025-5718-1983-0679444-7}
  {\bibfield  {journal} {\bibinfo  {journal} {Math. Comput.}\ }\textbf
  {\bibinfo {volume} {40}},\ \bibinfo {pages} {253} (\bibinfo {year}
  {1983})}\BibitemShut {NoStop}%
\bibitem [{\citenamefont {Renka}\ and\ \citenamefont
  {Cline}(1984)}]{Renka1984}%
  \BibitemOpen
  \bibfield  {author} {\bibinfo {author} {\bibfnamefont {R.}~\bibnamefont
  {Renka}}\ and\ \bibinfo {author} {\bibfnamefont {A.}~\bibnamefont {Cline}},\
  }\href {https://doi.org/10.1216/RMJ-1984-14-1-223} {\bibfield  {journal}
  {\bibinfo  {journal} {Rocky Mt. J. Math.}\ }\textbf {\bibinfo {volume}
  {14}},\ \bibinfo {pages} {223} (\bibinfo {year} {1984})}\BibitemShut
  {NoStop}%
\bibitem [{\citenamefont {{Qhull code for convex hull, delaunay triangulation,
  Voronoi diagram and halfspace intersection about a
  point}}()}]{Www.qhull.org}%
  \BibitemOpen
  \bibfield  {author} {\bibinfo {author} {\bibnamefont {{Qhull code for convex
  hull, delaunay triangulation, Voronoi diagram and halfspace intersection
  about a point}}},\ }\href {http://www.qhull.org/} {\bibinfo {title}
  {{www.qhull.org}}}\BibitemShut {NoStop}%
\bibitem [{\citenamefont {Alfeld}(1984)}]{Alfeld1984}%
  \BibitemOpen
  \bibfield  {author} {\bibinfo {author} {\bibfnamefont {P.}~\bibnamefont
  {Alfeld}},\ }\href {https://doi.org/10.1016/0167-8396(84)90029-3} {\bibfield
  {journal} {\bibinfo  {journal} {Comput. Aided Geom. Des.}\ }\textbf {\bibinfo
  {volume} {1}},\ \bibinfo {pages} {169} (\bibinfo {year} {1984})}\BibitemShut
  {NoStop}%
\bibitem [{\citenamefont {Farin}(1986)}]{Farin1986}%
  \BibitemOpen
  \bibfield  {author} {\bibinfo {author} {\bibfnamefont {G.}~\bibnamefont
  {Farin}},\ }\href {https://doi.org/10.1016/0167-8396(86)90016-6} {\bibfield
  {journal} {\bibinfo  {journal} {Comput. Aided Geom. Des.}\ }\textbf {\bibinfo
  {volume} {3}},\ \bibinfo {pages} {83} (\bibinfo {year} {1986})}\BibitemShut
  {NoStop}%
\bibitem [{\citenamefont {Hunter}(2007)}]{Hunter:2007}%
  \BibitemOpen
  \bibfield  {author} {\bibinfo {author} {\bibfnamefont {J.~D.}\ \bibnamefont
  {Hunter}},\ }\href {https://doi.org/10.1109/MCSE.2007.55} {\bibfield
  {journal} {\bibinfo  {journal} {Comput. Sci. Eng.}\ }\textbf {\bibinfo
  {volume} {9}},\ \bibinfo {pages} {90} (\bibinfo {year} {2007})}\BibitemShut
  {NoStop}%
\bibitem [{\citenamefont {Caswell}\ \emph {et~al.}(2022)\citenamefont
  {Caswell}, \citenamefont {Lee}, \citenamefont {Droettboom}, \citenamefont
  {de~Andrade}, \citenamefont {Hoffmann}, \citenamefont {Klymak}, \citenamefont
  {Hunter}, \citenamefont {Firing}, \citenamefont {Stansby}, \citenamefont
  {Varoquaux}, \citenamefont {Nielsen}, \citenamefont {Root}, \citenamefont
  {May}, \citenamefont {Elson}, \citenamefont {Sepp{\"{a}}nen}, \citenamefont
  {Dale}, \citenamefont {Lee}, \citenamefont {McDougall}, \citenamefont
  {Straw}, \citenamefont {Hobson}, \citenamefont {{hannah}}, \citenamefont
  {Gustafsson}, \citenamefont {Lucas}, \citenamefont {Gohlke}, \citenamefont
  {Vincent}, \citenamefont {Yu}, \citenamefont {Ma}, \citenamefont {Silvester},
  \citenamefont {Moad},\ and\ \citenamefont
  {Kniazev}}]{thomas_a_caswell_2022_7084615}%
  \BibitemOpen
  \bibfield  {author} {\bibinfo {author} {\bibfnamefont {T.~A.}\ \bibnamefont
  {Caswell}}, \bibinfo {author} {\bibfnamefont {A.}~\bibnamefont {Lee}},
  \bibinfo {author} {\bibfnamefont {M.}~\bibnamefont {Droettboom}}, \bibinfo
  {author} {\bibfnamefont {E.~S.}\ \bibnamefont {de~Andrade}}, \bibinfo
  {author} {\bibfnamefont {T.}~\bibnamefont {Hoffmann}}, \bibinfo {author}
  {\bibfnamefont {J.}~\bibnamefont {Klymak}}, \bibinfo {author} {\bibfnamefont
  {J.}~\bibnamefont {Hunter}}, \bibinfo {author} {\bibfnamefont
  {E.}~\bibnamefont {Firing}}, \bibinfo {author} {\bibfnamefont
  {D.}~\bibnamefont {Stansby}}, \bibinfo {author} {\bibfnamefont
  {N.}~\bibnamefont {Varoquaux}}, \bibinfo {author} {\bibfnamefont {J.~H.}\
  \bibnamefont {Nielsen}}, \bibinfo {author} {\bibfnamefont {B.}~\bibnamefont
  {Root}}, \bibinfo {author} {\bibfnamefont {R.}~\bibnamefont {May}}, \bibinfo
  {author} {\bibfnamefont {P.}~\bibnamefont {Elson}}, \bibinfo {author}
  {\bibfnamefont {J.~K.}\ \bibnamefont {Sepp{\"{a}}nen}}, \bibinfo {author}
  {\bibfnamefont {D.}~\bibnamefont {Dale}}, \bibinfo {author} {\bibfnamefont
  {J.-J.}\ \bibnamefont {Lee}}, \bibinfo {author} {\bibfnamefont
  {D.}~\bibnamefont {McDougall}}, \bibinfo {author} {\bibfnamefont
  {A.}~\bibnamefont {Straw}}, \bibinfo {author} {\bibfnamefont
  {P.}~\bibnamefont {Hobson}}, \bibinfo {author} {\bibnamefont {{hannah}}},
  \bibinfo {author} {\bibfnamefont {O.}~\bibnamefont {Gustafsson}}, \bibinfo
  {author} {\bibfnamefont {G.}~\bibnamefont {Lucas}}, \bibinfo {author}
  {\bibfnamefont {C.}~\bibnamefont {Gohlke}}, \bibinfo {author} {\bibfnamefont
  {A.~F.}\ \bibnamefont {Vincent}}, \bibinfo {author} {\bibfnamefont {T.~S.}\
  \bibnamefont {Yu}}, \bibinfo {author} {\bibfnamefont {E.}~\bibnamefont {Ma}},
  \bibinfo {author} {\bibfnamefont {S.}~\bibnamefont {Silvester}}, \bibinfo
  {author} {\bibfnamefont {C.}~\bibnamefont {Moad}},\ and\ \bibinfo {author}
  {\bibfnamefont {N.}~\bibnamefont {Kniazev}},\ }\href
  {https://doi.org/10.5281/zenodo.7084615} {\bibinfo {title}
  {{matplotlib/matplotlib: REL: v3.6.0}}} (\bibinfo {year} {2022})\BibitemShut
  {NoStop}%
\bibitem [{\citenamefont {Ku}\ \emph {et~al.}(2010)\citenamefont {Ku},
  \citenamefont {Berlijn},\ and\ \citenamefont {Lee}}]{Ku2010}%
  \BibitemOpen
  \bibfield  {author} {\bibinfo {author} {\bibfnamefont {W.}~\bibnamefont
  {Ku}}, \bibinfo {author} {\bibfnamefont {T.}~\bibnamefont {Berlijn}},\ and\
  \bibinfo {author} {\bibfnamefont {C.-C.}\ \bibnamefont {Lee}},\ }\href
  {https://doi.org/10.1103/PhysRevLett.104.216401} {\bibfield  {journal}
  {\bibinfo  {journal} {Phys. Rev. Lett.}\ }\textbf {\bibinfo {volume} {104}},\
  \bibinfo {pages} {216401} (\bibinfo {year} {2010})}\BibitemShut {NoStop}%
\bibitem [{\citenamefont {Yang}\ \emph {et~al.}(2018)\citenamefont {Yang},
  \citenamefont {Yang}, \citenamefont {Derunova}, \citenamefont {Parkin},
  \citenamefont {Yan},\ and\ \citenamefont {Ali}}]{Yang2018}%
  \BibitemOpen
  \bibfield  {author} {\bibinfo {author} {\bibfnamefont {S.~Y.}\ \bibnamefont
  {Yang}}, \bibinfo {author} {\bibfnamefont {H.}~\bibnamefont {Yang}}, \bibinfo
  {author} {\bibfnamefont {E.}~\bibnamefont {Derunova}}, \bibinfo {author}
  {\bibfnamefont {S.~S.}\ \bibnamefont {Parkin}}, \bibinfo {author}
  {\bibfnamefont {B.}~\bibnamefont {Yan}},\ and\ \bibinfo {author}
  {\bibfnamefont {M.~N.}\ \bibnamefont {Ali}},\ }\href
  {https://doi.org/10.1080/23746149.2017.1414631} {\bibfield  {journal}
  {\bibinfo  {journal} {Adv. Phys. X}\ }\textbf {\bibinfo {volume} {3}},\
  \bibinfo {pages} {265} (\bibinfo {year} {2018})}\BibitemShut {NoStop}%
\bibitem [{\citenamefont {Goodrich}\ \emph {et~al.}(2019)\citenamefont
  {Goodrich}, \citenamefont {Borovac}, \citenamefont {Tan},\ and\ \citenamefont
  {Tansu}}]{Goodrich2019}%
  \BibitemOpen
  \bibfield  {author} {\bibinfo {author} {\bibfnamefont {J.~C.}\ \bibnamefont
  {Goodrich}}, \bibinfo {author} {\bibfnamefont {D.}~\bibnamefont {Borovac}},
  \bibinfo {author} {\bibfnamefont {C.~K.}\ \bibnamefont {Tan}},\ and\ \bibinfo
  {author} {\bibfnamefont {N.}~\bibnamefont {Tansu}},\ }\href
  {https://doi.org/10.1038/s41598-019-41286-y} {\bibfield  {journal} {\bibinfo
  {journal} {Sci. Rep.}\ }\textbf {\bibinfo {volume} {9}},\ \bibinfo {pages}
  {5128} (\bibinfo {year} {2019})}\BibitemShut {NoStop}%
\bibitem [{\citenamefont {Wei}\ and\ \citenamefont {Zunger}(1996)}]{Wei1996}%
  \BibitemOpen
  \bibfield  {author} {\bibinfo {author} {\bibfnamefont {S.-H.}\ \bibnamefont
  {Wei}}\ and\ \bibinfo {author} {\bibfnamefont {A.}~\bibnamefont {Zunger}},\
  }\href {https://doi.org/10.1103/PhysRevLett.76.664} {\bibfield  {journal}
  {\bibinfo  {journal} {Phys. Rev. Lett.}\ }\textbf {\bibinfo {volume} {76}},\
  \bibinfo {pages} {664} (\bibinfo {year} {1996})}\BibitemShut {NoStop}%
\bibitem [{\citenamefont {Wu}\ \emph {et~al.}(2004)\citenamefont {Wu},
  \citenamefont {Walukiewicz}, \citenamefont {Yu}, \citenamefont {Denlinger},
  \citenamefont {Shan}, \citenamefont {Ager}, \citenamefont {Kimura},
  \citenamefont {Tang},\ and\ \citenamefont {Kuech}}]{Wu2004}%
  \BibitemOpen
  \bibfield  {author} {\bibinfo {author} {\bibfnamefont {J.}~\bibnamefont
  {Wu}}, \bibinfo {author} {\bibfnamefont {W.}~\bibnamefont {Walukiewicz}},
  \bibinfo {author} {\bibfnamefont {K.~M.}\ \bibnamefont {Yu}}, \bibinfo
  {author} {\bibfnamefont {J.~D.}\ \bibnamefont {Denlinger}}, \bibinfo {author}
  {\bibfnamefont {W.}~\bibnamefont {Shan}}, \bibinfo {author} {\bibfnamefont
  {J.~W.}\ \bibnamefont {Ager}}, \bibinfo {author} {\bibfnamefont
  {A.}~\bibnamefont {Kimura}}, \bibinfo {author} {\bibfnamefont {H.~F.}\
  \bibnamefont {Tang}},\ and\ \bibinfo {author} {\bibfnamefont {T.~F.}\
  \bibnamefont {Kuech}},\ }\href {https://doi.org/10.1103/PhysRevB.70.115214}
  {\bibfield  {journal} {\bibinfo  {journal} {Phys. Rev. B}\ }\textbf {\bibinfo
  {volume} {70}},\ \bibinfo {pages} {115214} (\bibinfo {year}
  {2004})}\BibitemShut {NoStop}%
\bibitem [{\citenamefont {Misiewicz}\ \emph {et~al.}(2005)\citenamefont
  {Misiewicz}, \citenamefont {Kudrawiec},\ and\ \citenamefont
  {Sek}}]{Misiewicz2005}%
  \BibitemOpen
  \bibfield  {author} {\bibinfo {author} {\bibfnamefont {J.}~\bibnamefont
  {Misiewicz}}, \bibinfo {author} {\bibfnamefont {R.}~\bibnamefont
  {Kudrawiec}},\ and\ \bibinfo {author} {\bibfnamefont {G.}~\bibnamefont
  {Sek}},\ }in\ \href {https://doi.org/10.1016/b978-008044502-1/50009-3} {\emph
  {\bibinfo {booktitle} {Dilute Nitride Semiconductors}}}\ (\bibinfo
  {publisher} {Elsevier},\ \bibinfo {year} {2005})\ pp.\ \bibinfo {pages}
  {279--324}\BibitemShut {NoStop}%
\bibitem [{\citenamefont {Capizzi}\ \emph {et~al.}(1981)\citenamefont
  {Capizzi}, \citenamefont {Modesti}, \citenamefont {Martelli},\ and\
  \citenamefont {Frova}}]{Capizzi1981}%
  \BibitemOpen
  \bibfield  {author} {\bibinfo {author} {\bibfnamefont {M.}~\bibnamefont
  {Capizzi}}, \bibinfo {author} {\bibfnamefont {S.}~\bibnamefont {Modesti}},
  \bibinfo {author} {\bibfnamefont {F.}~\bibnamefont {Martelli}},\ and\
  \bibinfo {author} {\bibfnamefont {A.}~\bibnamefont {Frova}},\ }\href
  {https://doi.org/10.1016/0038-1098(81)90684-0} {\bibfield  {journal}
  {\bibinfo  {journal} {Solid State Commun.}\ }\textbf {\bibinfo {volume}
  {39}},\ \bibinfo {pages} {333} (\bibinfo {year} {1981})}\BibitemShut
  {NoStop}%
\bibitem [{\citenamefont {Str{\"{o}}mberg}\ \emph {et~al.}(2020)\citenamefont
  {Str{\"{o}}mberg}, \citenamefont {Omanakuttan}, \citenamefont {Liu},
  \citenamefont {Mu}, \citenamefont {Xu}, \citenamefont {Lourdudoss},\ and\
  \citenamefont {Sun}}]{Stromberg2020}%
  \BibitemOpen
  \bibfield  {author} {\bibinfo {author} {\bibfnamefont {A.}~\bibnamefont
  {Str{\"{o}}mberg}}, \bibinfo {author} {\bibfnamefont {G.}~\bibnamefont
  {Omanakuttan}}, \bibinfo {author} {\bibfnamefont {Y.}~\bibnamefont {Liu}},
  \bibinfo {author} {\bibfnamefont {T.}~\bibnamefont {Mu}}, \bibinfo {author}
  {\bibfnamefont {Z.}~\bibnamefont {Xu}}, \bibinfo {author} {\bibfnamefont
  {S.}~\bibnamefont {Lourdudoss}},\ and\ \bibinfo {author} {\bibfnamefont
  {Y.-T.}\ \bibnamefont {Sun}},\ }\href
  {https://doi.org/10.1016/j.jcrysgro.2020.125623} {\bibfield  {journal}
  {\bibinfo  {journal} {J. Cryst. Growth}\ }\textbf {\bibinfo {volume} {540}},\
  \bibinfo {pages} {125623} (\bibinfo {year} {2020})}\BibitemShut {NoStop}%
\bibitem [{\citenamefont {Nattermann}\ \emph {et~al.}(2017)\citenamefont
  {Nattermann}, \citenamefont {Beyer}, \citenamefont {Ludewig}, \citenamefont
  {Hepp}, \citenamefont {Sterzer},\ and\ \citenamefont
  {Volz}}]{Nattermann2017}%
  \BibitemOpen
  \bibfield  {author} {\bibinfo {author} {\bibfnamefont {L.}~\bibnamefont
  {Nattermann}}, \bibinfo {author} {\bibfnamefont {A.}~\bibnamefont {Beyer}},
  \bibinfo {author} {\bibfnamefont {P.}~\bibnamefont {Ludewig}}, \bibinfo
  {author} {\bibfnamefont {T.}~\bibnamefont {Hepp}}, \bibinfo {author}
  {\bibfnamefont {E.}~\bibnamefont {Sterzer}},\ and\ \bibinfo {author}
  {\bibfnamefont {K.}~\bibnamefont {Volz}},\ }\href
  {https://doi.org/10.1016/j.jcrysgro.2017.02.021} {\bibfield  {journal}
  {\bibinfo  {journal} {J. Cryst. Growth}\ }\textbf {\bibinfo {volume} {463}},\
  \bibinfo {pages} {151} (\bibinfo {year} {2017})}\BibitemShut {NoStop}%
\bibitem [{\citenamefont {Ludewig}\ \emph {et~al.}(2013)\citenamefont
  {Ludewig}, \citenamefont {Knaub}, \citenamefont {Stolz},\ and\ \citenamefont
  {Volz}}]{Ludewig2013}%
  \BibitemOpen
  \bibfield  {author} {\bibinfo {author} {\bibfnamefont {P.}~\bibnamefont
  {Ludewig}}, \bibinfo {author} {\bibfnamefont {N.}~\bibnamefont {Knaub}},
  \bibinfo {author} {\bibfnamefont {W.}~\bibnamefont {Stolz}},\ and\ \bibinfo
  {author} {\bibfnamefont {K.}~\bibnamefont {Volz}},\ }\href
  {https://doi.org/10.1016/J.JCRYSGRO.2012.07.002} {\bibfield  {journal}
  {\bibinfo  {journal} {J. Cryst. Growth}\ }\textbf {\bibinfo {volume} {370}},\
  \bibinfo {pages} {186} (\bibinfo {year} {2013})}\BibitemShut {NoStop}%
\bibitem [{\citenamefont {Ludewig}\ \emph {et~al.}(2017)\citenamefont
  {Ludewig}, \citenamefont {Diederich}, \citenamefont {Jandieri},\ and\
  \citenamefont {Stolz}}]{Ludewig2017}%
  \BibitemOpen
  \bibfield  {author} {\bibinfo {author} {\bibfnamefont {P.}~\bibnamefont
  {Ludewig}}, \bibinfo {author} {\bibfnamefont {M.}~\bibnamefont {Diederich}},
  \bibinfo {author} {\bibfnamefont {K.}~\bibnamefont {Jandieri}},\ and\
  \bibinfo {author} {\bibfnamefont {W.}~\bibnamefont {Stolz}},\ }\href
  {https://doi.org/10.1016/j.jcrysgro.2017.03.003} {\bibfield  {journal}
  {\bibinfo  {journal} {J. Cryst. Growth}\ }\textbf {\bibinfo {volume} {467}},\
  \bibinfo {pages} {61} (\bibinfo {year} {2017})}\BibitemShut {NoStop}%
\bibitem [{\citenamefont {Lewis}\ \emph {et~al.}(2012)\citenamefont {Lewis},
  \citenamefont {Masnadi-Shirazi},\ and\ \citenamefont {Tiedje}}]{Lewis2012}%
  \BibitemOpen
  \bibfield  {author} {\bibinfo {author} {\bibfnamefont {R.~B.}\ \bibnamefont
  {Lewis}}, \bibinfo {author} {\bibfnamefont {M.}~\bibnamefont
  {Masnadi-Shirazi}},\ and\ \bibinfo {author} {\bibfnamefont {T.}~\bibnamefont
  {Tiedje}},\ }\href {https://doi.org/10.1063/1.4748172} {\bibfield  {journal}
  {\bibinfo  {journal} {Appl. Phys. Lett.}\ }\textbf {\bibinfo {volume}
  {101}},\ \bibinfo {pages} {082112} (\bibinfo {year} {2012})}\BibitemShut
  {NoStop}%
\bibitem [{\citenamefont {Masnadi-Shirazi}\ \emph {et~al.}(2014)\citenamefont
  {Masnadi-Shirazi}, \citenamefont {Lewis}, \citenamefont {Bahrami-Yekta},
  \citenamefont {Tiedje}, \citenamefont {Chicoine},\ and\ \citenamefont
  {Servati}}]{Masnadi-Shirazi2014}%
  \BibitemOpen
  \bibfield  {author} {\bibinfo {author} {\bibfnamefont {M.}~\bibnamefont
  {Masnadi-Shirazi}}, \bibinfo {author} {\bibfnamefont {R.~B.}\ \bibnamefont
  {Lewis}}, \bibinfo {author} {\bibfnamefont {V.}~\bibnamefont
  {Bahrami-Yekta}}, \bibinfo {author} {\bibfnamefont {T.}~\bibnamefont
  {Tiedje}}, \bibinfo {author} {\bibfnamefont {M.}~\bibnamefont {Chicoine}},\
  and\ \bibinfo {author} {\bibfnamefont {P.}~\bibnamefont {Servati}},\ }\href
  {https://doi.org/10.1063/1.4904081} {\bibfield  {journal} {\bibinfo
  {journal} {J. Appl. Phys.}\ }\textbf {\bibinfo {volume} {116}},\ \bibinfo
  {pages} {223506} (\bibinfo {year} {2014})}\BibitemShut {NoStop}%
\bibitem [{\citenamefont {Ludewig}\ \emph {et~al.}(2014)\citenamefont
  {Ludewig}, \citenamefont {Bushell}, \citenamefont {Nattermann}, \citenamefont
  {Knaub}, \citenamefont {Stolz},\ and\ \citenamefont
  {Volz}}]{Ludewig2014GrowthMOVPE}%
  \BibitemOpen
  \bibfield  {author} {\bibinfo {author} {\bibfnamefont {P.}~\bibnamefont
  {Ludewig}}, \bibinfo {author} {\bibfnamefont {Z.~L.}\ \bibnamefont
  {Bushell}}, \bibinfo {author} {\bibfnamefont {L.}~\bibnamefont {Nattermann}},
  \bibinfo {author} {\bibfnamefont {N.}~\bibnamefont {Knaub}}, \bibinfo
  {author} {\bibfnamefont {W.}~\bibnamefont {Stolz}},\ and\ \bibinfo {author}
  {\bibfnamefont {K.}~\bibnamefont {Volz}},\ }\href
  {https://doi.org/10.1016/J.JCRYSGRO.2014.03.041} {\bibfield  {journal}
  {\bibinfo  {journal} {J. Cryst. Growth}\ }\textbf {\bibinfo {volume} {396}},\
  \bibinfo {pages} {95} (\bibinfo {year} {2014})}\BibitemShut {NoStop}%
\bibitem [{\citenamefont {Mohmad}\ \emph {et~al.}(2011)\citenamefont {Mohmad},
  \citenamefont {Bastiman}, \citenamefont {Hunter}, \citenamefont {Ng},
  \citenamefont {Sweeney},\ and\ \citenamefont
  {David}}]{Mohmad2011TheGaAs1xBix}%
  \BibitemOpen
  \bibfield  {author} {\bibinfo {author} {\bibfnamefont {A.~R.}\ \bibnamefont
  {Mohmad}}, \bibinfo {author} {\bibfnamefont {F.}~\bibnamefont {Bastiman}},
  \bibinfo {author} {\bibfnamefont {C.~J.}\ \bibnamefont {Hunter}}, \bibinfo
  {author} {\bibfnamefont {J.~S.}\ \bibnamefont {Ng}}, \bibinfo {author}
  {\bibfnamefont {S.~J.}\ \bibnamefont {Sweeney}},\ and\ \bibinfo {author}
  {\bibfnamefont {J.~P.}\ \bibnamefont {David}},\ }\href
  {https://doi.org/10.1063/1.3617461} {\bibfield  {journal} {\bibinfo
  {journal} {Appl. Phys. Lett.}\ }\textbf {\bibinfo {volume} {99}},\ \bibinfo
  {pages} {042107} (\bibinfo {year} {2011})}\BibitemShut {NoStop}%
\bibitem [{\citenamefont {Cipriano}\ \emph {et~al.}(2020)\citenamefont
  {Cipriano}, \citenamefont {Di~Liberto}, \citenamefont {Tosoni},\ and\
  \citenamefont {Pacchioni}}]{Cipriano2020}%
  \BibitemOpen
  \bibfield  {author} {\bibinfo {author} {\bibfnamefont {L.~A.}\ \bibnamefont
  {Cipriano}}, \bibinfo {author} {\bibfnamefont {G.}~\bibnamefont
  {Di~Liberto}}, \bibinfo {author} {\bibfnamefont {S.}~\bibnamefont {Tosoni}},\
  and\ \bibinfo {author} {\bibfnamefont {G.}~\bibnamefont {Pacchioni}},\ }\href
  {https://doi.org/10.1039/D0NR03577G} {\bibfield  {journal} {\bibinfo
  {journal} {Nanoscale}\ }\textbf {\bibinfo {volume} {12}},\ \bibinfo {pages}
  {17494} (\bibinfo {year} {2020})}\BibitemShut {NoStop}%
\end{thebibliography}%


\providecommand{\noopsort}[1]{}\providecommand{\singleletter}[1]{#1}
\begin{thebibliography}{11}%
\makeatletter
\providecommand \@ifxundefined [1]{%
 \@ifx{#1\undefined}
}%
\providecommand \@ifnum [1]{%
 \ifnum #1\expandafter \@firstoftwo
 \else \expandafter \@secondoftwo
 \fi
}%
\providecommand \@ifx [1]{%
 \ifx #1\expandafter \@firstoftwo
 \else \expandafter \@secondoftwo
 \fi
}%
\providecommand \natexlab [1]{#1}%
\providecommand \enquote  [1]{``#1''}%
\providecommand \bibnamefont  [1]{#1}%
\providecommand \bibfnamefont [1]{#1}%
\providecommand \citenamefont [1]{#1}%
\providecommand \href@noop [0]{\@secondoftwo}%
\providecommand \href [0]{\begingroup \@sanitize@url \@href}%
\providecommand \@href[1]{\@@startlink{#1}\@@href}%
\providecommand \@@href[1]{\endgroup#1\@@endlink}%
\providecommand \@sanitize@url [0]{\catcode `\\12\catcode `\$12\catcode
  `\&12\catcode `\#12\catcode `\^12\catcode `\_12\catcode `\%12\relax}%
\providecommand \@@startlink[1]{}%
\providecommand \@@endlink[0]{}%
\providecommand \url  [0]{\begingroup\@sanitize@url \@url }%
\providecommand \@url [1]{\endgroup\@href {#1}{\urlprefix }}%
\providecommand \urlprefix  [0]{URL }%
\providecommand \Eprint [0]{\href }%
\providecommand \doibase [0]{https://doi.org/}%
\providecommand \selectlanguage [0]{\@gobble}%
\providecommand \bibinfo  [0]{\@secondoftwo}%
\providecommand \bibfield  [0]{\@secondoftwo}%
\providecommand \translation [1]{[#1]}%
\providecommand \BibitemOpen [0]{}%
\providecommand \bibitemStop [0]{}%
\providecommand \bibitemNoStop [0]{.\EOS\space}%
\providecommand \EOS [0]{\spacefactor3000\relax}%
\providecommand \BibitemShut  [1]{\csname bibitem#1\endcsname}%
\let\auto@bib@innerbib\@empty
\bibitem [{\citenamefont {Medeiros}\ \emph {et~al.}(2015)\citenamefont
  {Medeiros}, \citenamefont {Tsirkin}, \citenamefont {Stafstr{\"{o}}m},\ and\
  \citenamefont {Bj{\"{o}}rk}}]{Medeiros2015Sup}%
  \BibitemOpen
  \bibfield  {author} {\bibinfo {author} {\bibfnamefont {P.~V.~C.}\
  \bibnamefont {Medeiros}}, \bibinfo {author} {\bibfnamefont {S.~S.}\
  \bibnamefont {Tsirkin}}, \bibinfo {author} {\bibfnamefont {S.}~\bibnamefont
  {Stafstr{\"{o}}m}},\ and\ \bibinfo {author} {\bibfnamefont {J.}~\bibnamefont
  {Bj{\"{o}}rk}},\ }\bibfield  {title} {\bibinfo {title} {{Unfolding spinor
  wave functions and expectation values of general operators: Introducing the
  unfolding-density operator}},\ }\href
  {https://doi.org/10.1103/PhysRevB.91.041116} {\bibfield  {journal} {\bibinfo
  {journal} {Physical Review B}\ }\textbf {\bibinfo {volume} {91}},\ \bibinfo
  {pages} {041116(R)} (\bibinfo {year} {2015})}\BibitemShut {NoStop}%
\bibitem [{\citenamefont {Medeiros}\ \emph {et~al.}(2014)\citenamefont
  {Medeiros}, \citenamefont {Stafstr{\"{o}}m},\ and\ \citenamefont
  {Bj{\"{o}}rk}}]{Medeiros2014Sup}%
  \BibitemOpen
  \bibfield  {author} {\bibinfo {author} {\bibfnamefont {P.~V.~C.}\
  \bibnamefont {Medeiros}}, \bibinfo {author} {\bibfnamefont {S.}~\bibnamefont
  {Stafstr{\"{o}}m}},\ and\ \bibinfo {author} {\bibfnamefont {J.}~\bibnamefont
  {Bj{\"{o}}rk}},\ }\bibfield  {title} {\bibinfo {title} {{Effects of extrinsic
  and intrinsic perturbations on the electronic structure of graphene:
  Retaining an effective primitive cell band structure by band unfolding}},\
  }\href {https://doi.org/10.1103/PhysRevB.89.041407} {\bibfield  {journal}
  {\bibinfo  {journal} {Physical Review B}\ }\textbf {\bibinfo {volume} {89}},\
  \bibinfo {pages} {041407(R)} (\bibinfo {year} {2014})}\BibitemShut {NoStop}%
\bibitem [{\citenamefont {Mondal}\ and\ \citenamefont
  {Tonner-Zech}(2023)}]{Mondal2022Sup}%
  \BibitemOpen
  \bibfield  {author} {\bibinfo {author} {\bibfnamefont {B.}~\bibnamefont
  {Mondal}}\ and\ \bibinfo {author} {\bibfnamefont {R.}~\bibnamefont
  {Tonner-Zech}},\ }\bibfield  {title} {\bibinfo {title} {{Systematic
  strain-induced bandgap tuning in binary III-V semiconductors from density
  functional theory}},\ }\href {https://doi.org/10.1088/1402-4896/acd08b}
  {\bibfield  {journal} {\bibinfo  {journal} {Phys. Scr.}\ }\textbf {\bibinfo
  {volume} {98}},\ \bibinfo {pages} {065924} (\bibinfo {year}
  {2023})}\BibitemShut {NoStop}%
\bibitem [{\citenamefont {Virtanen}\ \emph {et~al.}(2020)\citenamefont
  {Virtanen}, \citenamefont {Gommers}, \citenamefont {Oliphant}, \citenamefont
  {Haberland}, \citenamefont {Reddy}, \citenamefont {Cournapeau}, \citenamefont
  {Burovski}, \citenamefont {Peterson}, \citenamefont {Weckesser},
  \citenamefont {Bright}, \citenamefont {van~der Walt}, \citenamefont {Brett},
  \citenamefont {Wilson}, \citenamefont {Millman}, \citenamefont {Mayorov},
  \citenamefont {Nelson}, \citenamefont {Jones}, \citenamefont {Kern},
  \citenamefont {Larson}, \citenamefont {Carey}, \citenamefont {Polat},
  \citenamefont {Feng}, \citenamefont {Moore}, \citenamefont {VanderPlas},
  \citenamefont {Laxalde}, \citenamefont {Perktold}, \citenamefont {Cimrman},
  \citenamefont {Henriksen}, \citenamefont {Quintero}, \citenamefont {Harris},
  \citenamefont {Archibald}, \citenamefont {Ribeiro}, \citenamefont
  {Pedregosa}, \citenamefont {van Mulbregt}, \citenamefont {Vijaykumar},
  \citenamefont {Bardelli}, \citenamefont {Rothberg}, \citenamefont {Hilboll},
  \citenamefont {Kloeckner}, \citenamefont {Scopatz}, \citenamefont {Lee},
  \citenamefont {Rokem}, \citenamefont {Woods}, \citenamefont {Fulton},
  \citenamefont {Masson}, \citenamefont {H{\"{a}}ggstr{\"{o}}m}, \citenamefont
  {Fitzgerald}, \citenamefont {Nicholson}, \citenamefont {Hagen}, \citenamefont
  {Pasechnik}, \citenamefont {Olivetti}, \citenamefont {Martin}, \citenamefont
  {Wieser}, \citenamefont {Silva}, \citenamefont {Lenders}, \citenamefont
  {Wilhelm}, \citenamefont {Young}, \citenamefont {Price}, \citenamefont
  {Ingold}, \citenamefont {Allen}, \citenamefont {Lee}, \citenamefont {Audren},
  \citenamefont {Probst}, \citenamefont {Dietrich}, \citenamefont {Silterra},
  \citenamefont {Webber}, \citenamefont {Slavi{\v{c}}}, \citenamefont
  {Nothman}, \citenamefont {Buchner}, \citenamefont {Kulick}, \citenamefont
  {Sch{\"{o}}nberger}, \citenamefont {de~Miranda~Cardoso}, \citenamefont
  {Reimer}, \citenamefont {Harrington}, \citenamefont {Rodr{\'{i}}guez},
  \citenamefont {Nunez-Iglesias}, \citenamefont {Kuczynski}, \citenamefont
  {Tritz}, \citenamefont {Thoma}, \citenamefont {Newville}, \citenamefont
  {K{\"{u}}mmerer}, \citenamefont {Bolingbroke}, \citenamefont {Tartre},
  \citenamefont {Pak}, \citenamefont {Smith}, \citenamefont {Nowaczyk},
  \citenamefont {Shebanov}, \citenamefont {Pavlyk}, \citenamefont {Brodtkorb},
  \citenamefont {Lee}, \citenamefont {McGibbon}, \citenamefont {Feldbauer},
  \citenamefont {Lewis}, \citenamefont {Tygier}, \citenamefont {Sievert},
  \citenamefont {Vigna}, \citenamefont {Peterson}, \citenamefont {More},
  \citenamefont {Pudlik}, \citenamefont {Oshima}, \citenamefont {Pingel},
  \citenamefont {Robitaille}, \citenamefont {Spura}, \citenamefont {Jones},
  \citenamefont {Cera}, \citenamefont {Leslie}, \citenamefont {Zito},
  \citenamefont {Krauss}, \citenamefont {Upadhyay}, \citenamefont {Halchenko},\
  and\ \citenamefont {V{\'{a}}zquez-Baeza}}]{SciPyNMeth2020}%
  \BibitemOpen
  \bibfield  {author} {\bibinfo {author} {\bibfnamefont {P.}~\bibnamefont
  {Virtanen}}, \bibinfo {author} {\bibfnamefont {R.}~\bibnamefont {Gommers}},
  \bibinfo {author} {\bibfnamefont {T.~E.}\ \bibnamefont {Oliphant}}, \bibinfo
  {author} {\bibfnamefont {M.}~\bibnamefont {Haberland}}, \bibinfo {author}
  {\bibfnamefont {T.}~\bibnamefont {Reddy}}, \bibinfo {author} {\bibfnamefont
  {D.}~\bibnamefont {Cournapeau}}, \bibinfo {author} {\bibfnamefont
  {E.}~\bibnamefont {Burovski}}, \bibinfo {author} {\bibfnamefont
  {P.}~\bibnamefont {Peterson}}, \bibinfo {author} {\bibfnamefont
  {W.}~\bibnamefont {Weckesser}}, \bibinfo {author} {\bibfnamefont
  {J.}~\bibnamefont {Bright}}, \bibinfo {author} {\bibfnamefont {S.~J.}\
  \bibnamefont {van~der Walt}}, \bibinfo {author} {\bibfnamefont
  {M.}~\bibnamefont {Brett}}, \bibinfo {author} {\bibfnamefont
  {J.}~\bibnamefont {Wilson}}, \bibinfo {author} {\bibfnamefont {K.~J.}\
  \bibnamefont {Millman}}, \bibinfo {author} {\bibfnamefont {N.}~\bibnamefont
  {Mayorov}}, \bibinfo {author} {\bibfnamefont {A.~R.~J.}\ \bibnamefont
  {Nelson}}, \bibinfo {author} {\bibfnamefont {E.}~\bibnamefont {Jones}},
  \bibinfo {author} {\bibfnamefont {R.}~\bibnamefont {Kern}}, \bibinfo {author}
  {\bibfnamefont {E.}~\bibnamefont {Larson}}, \bibinfo {author} {\bibfnamefont
  {C.~J.}\ \bibnamefont {Carey}}, \bibinfo {author} {\bibfnamefont
  {I.}~\bibnamefont {Polat}}, \bibinfo {author} {\bibfnamefont
  {Y.}~\bibnamefont {Feng}}, \bibinfo {author} {\bibfnamefont {E.~W.}\
  \bibnamefont {Moore}}, \bibinfo {author} {\bibfnamefont {J.}~\bibnamefont
  {VanderPlas}}, \bibinfo {author} {\bibfnamefont {D.}~\bibnamefont {Laxalde}},
  \bibinfo {author} {\bibfnamefont {J.}~\bibnamefont {Perktold}}, \bibinfo
  {author} {\bibfnamefont {R.}~\bibnamefont {Cimrman}}, \bibinfo {author}
  {\bibfnamefont {I.}~\bibnamefont {Henriksen}}, \bibinfo {author}
  {\bibfnamefont {E.~A.}\ \bibnamefont {Quintero}}, \bibinfo {author}
  {\bibfnamefont {C.~R.}\ \bibnamefont {Harris}}, \bibinfo {author}
  {\bibfnamefont {A.~M.}\ \bibnamefont {Archibald}}, \bibinfo {author}
  {\bibfnamefont {A.~H.}\ \bibnamefont {Ribeiro}}, \bibinfo {author}
  {\bibfnamefont {F.}~\bibnamefont {Pedregosa}}, \bibinfo {author}
  {\bibfnamefont {P.}~\bibnamefont {van Mulbregt}}, \bibinfo {author}
  {\bibfnamefont {A.}~\bibnamefont {Vijaykumar}}, \bibinfo {author}
  {\bibfnamefont {A.~P.}\ \bibnamefont {Bardelli}}, \bibinfo {author}
  {\bibfnamefont {A.}~\bibnamefont {Rothberg}}, \bibinfo {author}
  {\bibfnamefont {A.}~\bibnamefont {Hilboll}}, \bibinfo {author} {\bibfnamefont
  {A.}~\bibnamefont {Kloeckner}}, \bibinfo {author} {\bibfnamefont
  {A.}~\bibnamefont {Scopatz}}, \bibinfo {author} {\bibfnamefont
  {A.}~\bibnamefont {Lee}}, \bibinfo {author} {\bibfnamefont {A.}~\bibnamefont
  {Rokem}}, \bibinfo {author} {\bibfnamefont {C.~N.}\ \bibnamefont {Woods}},
  \bibinfo {author} {\bibfnamefont {C.}~\bibnamefont {Fulton}}, \bibinfo
  {author} {\bibfnamefont {C.}~\bibnamefont {Masson}}, \bibinfo {author}
  {\bibfnamefont {C.}~\bibnamefont {H{\"{a}}ggstr{\"{o}}m}}, \bibinfo {author}
  {\bibfnamefont {C.}~\bibnamefont {Fitzgerald}}, \bibinfo {author}
  {\bibfnamefont {D.~A.}\ \bibnamefont {Nicholson}}, \bibinfo {author}
  {\bibfnamefont {D.~R.}\ \bibnamefont {Hagen}}, \bibinfo {author}
  {\bibfnamefont {D.~V.}\ \bibnamefont {Pasechnik}}, \bibinfo {author}
  {\bibfnamefont {E.}~\bibnamefont {Olivetti}}, \bibinfo {author}
  {\bibfnamefont {E.}~\bibnamefont {Martin}}, \bibinfo {author} {\bibfnamefont
  {E.}~\bibnamefont {Wieser}}, \bibinfo {author} {\bibfnamefont
  {F.}~\bibnamefont {Silva}}, \bibinfo {author} {\bibfnamefont
  {F.}~\bibnamefont {Lenders}}, \bibinfo {author} {\bibfnamefont
  {F.}~\bibnamefont {Wilhelm}}, \bibinfo {author} {\bibfnamefont
  {G.}~\bibnamefont {Young}}, \bibinfo {author} {\bibfnamefont {G.~A.}\
  \bibnamefont {Price}}, \bibinfo {author} {\bibfnamefont {G.-L.}\ \bibnamefont
  {Ingold}}, \bibinfo {author} {\bibfnamefont {G.~E.}\ \bibnamefont {Allen}},
  \bibinfo {author} {\bibfnamefont {G.~R.}\ \bibnamefont {Lee}}, \bibinfo
  {author} {\bibfnamefont {H.}~\bibnamefont {Audren}}, \bibinfo {author}
  {\bibfnamefont {I.}~\bibnamefont {Probst}}, \bibinfo {author} {\bibfnamefont
  {J.~P.}\ \bibnamefont {Dietrich}}, \bibinfo {author} {\bibfnamefont
  {J.}~\bibnamefont {Silterra}}, \bibinfo {author} {\bibfnamefont {J.~T.}\
  \bibnamefont {Webber}}, \bibinfo {author} {\bibfnamefont {J.}~\bibnamefont
  {Slavi{\v{c}}}}, \bibinfo {author} {\bibfnamefont {J.}~\bibnamefont
  {Nothman}}, \bibinfo {author} {\bibfnamefont {J.}~\bibnamefont {Buchner}},
  \bibinfo {author} {\bibfnamefont {J.}~\bibnamefont {Kulick}}, \bibinfo
  {author} {\bibfnamefont {J.~L.}\ \bibnamefont {Sch{\"{o}}nberger}}, \bibinfo
  {author} {\bibfnamefont {J.~V.}\ \bibnamefont {de~Miranda~Cardoso}}, \bibinfo
  {author} {\bibfnamefont {J.}~\bibnamefont {Reimer}}, \bibinfo {author}
  {\bibfnamefont {J.}~\bibnamefont {Harrington}}, \bibinfo {author}
  {\bibfnamefont {J.~L.~C.}\ \bibnamefont {Rodr{\'{i}}guez}}, \bibinfo {author}
  {\bibfnamefont {J.}~\bibnamefont {Nunez-Iglesias}}, \bibinfo {author}
  {\bibfnamefont {J.}~\bibnamefont {Kuczynski}}, \bibinfo {author}
  {\bibfnamefont {K.}~\bibnamefont {Tritz}}, \bibinfo {author} {\bibfnamefont
  {M.}~\bibnamefont {Thoma}}, \bibinfo {author} {\bibfnamefont
  {M.}~\bibnamefont {Newville}}, \bibinfo {author} {\bibfnamefont
  {M.}~\bibnamefont {K{\"{u}}mmerer}}, \bibinfo {author} {\bibfnamefont
  {M.}~\bibnamefont {Bolingbroke}}, \bibinfo {author} {\bibfnamefont
  {M.}~\bibnamefont {Tartre}}, \bibinfo {author} {\bibfnamefont
  {M.}~\bibnamefont {Pak}}, \bibinfo {author} {\bibfnamefont {N.~J.}\
  \bibnamefont {Smith}}, \bibinfo {author} {\bibfnamefont {N.}~\bibnamefont
  {Nowaczyk}}, \bibinfo {author} {\bibfnamefont {N.}~\bibnamefont {Shebanov}},
  \bibinfo {author} {\bibfnamefont {O.}~\bibnamefont {Pavlyk}}, \bibinfo
  {author} {\bibfnamefont {P.~A.}\ \bibnamefont {Brodtkorb}}, \bibinfo {author}
  {\bibfnamefont {P.}~\bibnamefont {Lee}}, \bibinfo {author} {\bibfnamefont
  {R.~T.}\ \bibnamefont {McGibbon}}, \bibinfo {author} {\bibfnamefont
  {R.}~\bibnamefont {Feldbauer}}, \bibinfo {author} {\bibfnamefont
  {S.}~\bibnamefont {Lewis}}, \bibinfo {author} {\bibfnamefont
  {S.}~\bibnamefont {Tygier}}, \bibinfo {author} {\bibfnamefont
  {S.}~\bibnamefont {Sievert}}, \bibinfo {author} {\bibfnamefont
  {S.}~\bibnamefont {Vigna}}, \bibinfo {author} {\bibfnamefont
  {S.}~\bibnamefont {Peterson}}, \bibinfo {author} {\bibfnamefont
  {S.}~\bibnamefont {More}}, \bibinfo {author} {\bibfnamefont {T.}~\bibnamefont
  {Pudlik}}, \bibinfo {author} {\bibfnamefont {T.}~\bibnamefont {Oshima}},
  \bibinfo {author} {\bibfnamefont {T.~J.}\ \bibnamefont {Pingel}}, \bibinfo
  {author} {\bibfnamefont {T.~P.}\ \bibnamefont {Robitaille}}, \bibinfo
  {author} {\bibfnamefont {T.}~\bibnamefont {Spura}}, \bibinfo {author}
  {\bibfnamefont {T.~R.}\ \bibnamefont {Jones}}, \bibinfo {author}
  {\bibfnamefont {T.}~\bibnamefont {Cera}}, \bibinfo {author} {\bibfnamefont
  {T.}~\bibnamefont {Leslie}}, \bibinfo {author} {\bibfnamefont
  {T.}~\bibnamefont {Zito}}, \bibinfo {author} {\bibfnamefont {T.}~\bibnamefont
  {Krauss}}, \bibinfo {author} {\bibfnamefont {U.}~\bibnamefont {Upadhyay}},
  \bibinfo {author} {\bibfnamefont {Y.~O.}\ \bibnamefont {Halchenko}},\ and\
  \bibinfo {author} {\bibfnamefont {Y.}~\bibnamefont {V{\'{a}}zquez-Baeza}},\
  }\bibfield  {title} {\bibinfo {title} {{SciPy 1.0: fundamental algorithms for
  scientific computing in Python}},\ }\href
  {https://doi.org/10.1038/s41592-019-0686-2} {\bibfield  {journal} {\bibinfo
  {journal} {Nature Methods}\ }\textbf {\bibinfo {volume} {17}},\ \bibinfo
  {pages} {261} (\bibinfo {year} {2020})}\BibitemShut {NoStop}%
\bibitem [{\citenamefont {{Qhull code for convex hull, delaunay triangulation,
  Voronoi diagram and halfspace intersection about a
  point}}()}]{Www.qhull.org}%
  \BibitemOpen
  \bibfield  {author} {\bibinfo {author} {\bibnamefont {{Qhull code for convex
  hull, delaunay triangulation, Voronoi diagram and halfspace intersection
  about a point}}},\ }\href {http://www.qhull.org/} {\bibinfo {title}
  {http://www.qhull.org/}}\BibitemShut {NoStop}%
\bibitem [{\citenamefont {Alfeld}(1984)}]{Alfeld1984}%
  \BibitemOpen
  \bibfield  {author} {\bibinfo {author} {\bibfnamefont {P.}~\bibnamefont
  {Alfeld}},\ }\bibfield  {title} {\bibinfo {title} {{A trivariate
  clough—tocher scheme for tetrahedral data}},\ }\href
  {https://doi.org/10.1016/0167-8396(84)90029-3} {\bibfield  {journal}
  {\bibinfo  {journal} {Computer Aided Geometric Design}\ }\textbf {\bibinfo
  {volume} {1}},\ \bibinfo {pages} {169} (\bibinfo {year} {1984})}\BibitemShut
  {NoStop}%
\bibitem [{\citenamefont {Farin}(1986)}]{Farin1986}%
  \BibitemOpen
  \bibfield  {author} {\bibinfo {author} {\bibfnamefont {G.}~\bibnamefont
  {Farin}},\ }\bibfield  {title} {\bibinfo {title} {{Triangular
  Bernstein-B{\'{e}}zier patches}},\ }\href
  {https://doi.org/10.1016/0167-8396(86)90016-6} {\bibfield  {journal}
  {\bibinfo  {journal} {Computer Aided Geometric Design}\ }\textbf {\bibinfo
  {volume} {3}},\ \bibinfo {pages} {83} (\bibinfo {year} {1986})}\BibitemShut
  {NoStop}%
\bibitem [{\citenamefont {Nielson}(1983)}]{Nielson1983}%
  \BibitemOpen
  \bibfield  {author} {\bibinfo {author} {\bibfnamefont {G.~M.}\ \bibnamefont
  {Nielson}},\ }\bibfield  {title} {\bibinfo {title} {{A method for
  interpolating scattered data based upon a minimum norm network}},\ }\href
  {https://doi.org/10.1090/S0025-5718-1983-0679444-7} {\bibfield  {journal}
  {\bibinfo  {journal} {Mathematics of Computation}\ }\textbf {\bibinfo
  {volume} {40}},\ \bibinfo {pages} {253} (\bibinfo {year} {1983})}\BibitemShut
  {NoStop}%
\bibitem [{\citenamefont {Renka}\ and\ \citenamefont
  {Cline}(1984)}]{Renka1984}%
  \BibitemOpen
  \bibfield  {author} {\bibinfo {author} {\bibfnamefont {R.}~\bibnamefont
  {Renka}}\ and\ \bibinfo {author} {\bibfnamefont {A.}~\bibnamefont {Cline}},\
  }\bibfield  {title} {\bibinfo {title} {{A triangle-based $C^1$ interpolation
  method}},\ }\href {https://doi.org/10.1216/RMJ-1984-14-1-223} {\bibfield
  {journal} {\bibinfo  {journal} {Rocky Mountain Journal of Mathematics}\
  }\textbf {\bibinfo {volume} {14}},\ \bibinfo {pages} {223} (\bibinfo {year}
  {1984})}\BibitemShut {NoStop}%
\bibitem [{\citenamefont {Hunter}(2007)}]{Hunter:2007}%
  \BibitemOpen
  \bibfield  {author} {\bibinfo {author} {\bibfnamefont {J.~D.}\ \bibnamefont
  {Hunter}},\ }\bibfield  {title} {\bibinfo {title} {{Matplotlib: A 2D Graphics
  Environment}},\ }\href {https://doi.org/10.1109/MCSE.2007.55} {\bibfield
  {journal} {\bibinfo  {journal} {Computing in Science {\&} Engineering}\
  }\textbf {\bibinfo {volume} {9}},\ \bibinfo {pages} {90} (\bibinfo {year}
  {2007})}\BibitemShut {NoStop}%
\bibitem [{\citenamefont {Caswell}\ \emph {et~al.}(2022)\citenamefont
  {Caswell}, \citenamefont {Lee}, \citenamefont {Droettboom}, \citenamefont
  {de~Andrade}, \citenamefont {Hoffmann}, \citenamefont {Klymak}, \citenamefont
  {Hunter}, \citenamefont {Firing}, \citenamefont {Stansby}, \citenamefont
  {Varoquaux}, \citenamefont {Nielsen}, \citenamefont {Root}, \citenamefont
  {May}, \citenamefont {Elson}, \citenamefont {Sepp{\"{a}}nen}, \citenamefont
  {Dale}, \citenamefont {Lee}, \citenamefont {McDougall}, \citenamefont
  {Straw}, \citenamefont {Hobson}, \citenamefont {{hannah}}, \citenamefont
  {Gustafsson}, \citenamefont {Lucas}, \citenamefont {Gohlke}, \citenamefont
  {Vincent}, \citenamefont {Yu}, \citenamefont {Ma}, \citenamefont {Silvester},
  \citenamefont {Moad},\ and\ \citenamefont
  {Kniazev}}]{thomas_a_caswell_2022_7084615}%
  \BibitemOpen
  \bibfield  {author} {\bibinfo {author} {\bibfnamefont {T.~A.}\ \bibnamefont
  {Caswell}}, \bibinfo {author} {\bibfnamefont {A.}~\bibnamefont {Lee}},
  \bibinfo {author} {\bibfnamefont {M.}~\bibnamefont {Droettboom}}, \bibinfo
  {author} {\bibfnamefont {E.~S.}\ \bibnamefont {de~Andrade}}, \bibinfo
  {author} {\bibfnamefont {T.}~\bibnamefont {Hoffmann}}, \bibinfo {author}
  {\bibfnamefont {J.}~\bibnamefont {Klymak}}, \bibinfo {author} {\bibfnamefont
  {J.}~\bibnamefont {Hunter}}, \bibinfo {author} {\bibfnamefont
  {E.}~\bibnamefont {Firing}}, \bibinfo {author} {\bibfnamefont
  {D.}~\bibnamefont {Stansby}}, \bibinfo {author} {\bibfnamefont
  {N.}~\bibnamefont {Varoquaux}}, \bibinfo {author} {\bibfnamefont {J.~H.}\
  \bibnamefont {Nielsen}}, \bibinfo {author} {\bibfnamefont {B.}~\bibnamefont
  {Root}}, \bibinfo {author} {\bibfnamefont {R.}~\bibnamefont {May}}, \bibinfo
  {author} {\bibfnamefont {P.}~\bibnamefont {Elson}}, \bibinfo {author}
  {\bibfnamefont {J.~K.}\ \bibnamefont {Sepp{\"{a}}nen}}, \bibinfo {author}
  {\bibfnamefont {D.}~\bibnamefont {Dale}}, \bibinfo {author} {\bibfnamefont
  {J.-J.}\ \bibnamefont {Lee}}, \bibinfo {author} {\bibfnamefont
  {D.}~\bibnamefont {McDougall}}, \bibinfo {author} {\bibfnamefont
  {A.}~\bibnamefont {Straw}}, \bibinfo {author} {\bibfnamefont
  {P.}~\bibnamefont {Hobson}}, \bibinfo {author} {\bibnamefont {{hannah}}},
  \bibinfo {author} {\bibfnamefont {O.}~\bibnamefont {Gustafsson}}, \bibinfo
  {author} {\bibfnamefont {G.}~\bibnamefont {Lucas}}, \bibinfo {author}
  {\bibfnamefont {C.}~\bibnamefont {Gohlke}}, \bibinfo {author} {\bibfnamefont
  {A.~F.}\ \bibnamefont {Vincent}}, \bibinfo {author} {\bibfnamefont {T.~S.}\
  \bibnamefont {Yu}}, \bibinfo {author} {\bibfnamefont {E.}~\bibnamefont {Ma}},
  \bibinfo {author} {\bibfnamefont {S.}~\bibnamefont {Silvester}}, \bibinfo
  {author} {\bibfnamefont {C.}~\bibnamefont {Moad}},\ and\ \bibinfo {author}
  {\bibfnamefont {N.}~\bibnamefont {Kniazev}},\ }\href
  {https://doi.org/10.5281/zenodo.7084615} {\bibinfo {title}
  {{matplotlib/matplotlib: REL: v3.6.0}}} (\bibinfo {year} {2022})\BibitemShut
  {NoStop}%
\end{thebibliography}%

\end{document}


\preprint{APS/123-QED}

\title{Supplemental Material \\ \vspace{2cm} Accurate first-principles bandgap predictions \\in strain-engineered ternary III-V semiconductors}

\author{Badal Mondal}
\affiliation{Wilhelm-Ostwald-Institut f\"ur Physikalische und Theoretische Chemie, Universit\"at Leipzig, 04103 Leipzig, Germany}%
\affiliation{Fachbereich Physik, Philipps-Universit\"at Marburg, 35032 Marburg, Germany}

\author{Marcel Kr\"oner}%
\affiliation{Material Science Center and Department of Physics, Philipps-Universit\"at Marburg, D-35043 Marburg, Germany}%

\author{Thilo Hepp}%
\affiliation{Material Science Center and Department of Physics, Philipps-Universit\"at Marburg, D-35043 Marburg, Germany}%

\author{Kerstin Volz}%
\affiliation{Material Science Center and Department of Physics, Philipps-Universit\"at Marburg, D-35043 Marburg, Germany}%

\author{Ralf Tonner-Zech}
\email{ralf.tonner@uni-leipzig.de}
\affiliation{Wilhelm-Ostwald-Institut f\"ur Physikalische und Theoretische Chemie, Universit\"at Leipzig, 04103 Leipzig, Germany}%
\date{\today}
\maketitle
\pagebreak
\vspace*{\fill}
\section{\label{secSI:ebsgaasp}Effective bandstructures of G\lowercase{a}A\lowercase{s}$_{0.963}$P$_{0.037}$ under isotropic strain}
\begin{figure}[htb]
\centering
  \subfloat[Strain = 0.0\%]{\label{fig:figS1a}\includegraphics[width=3.4in]{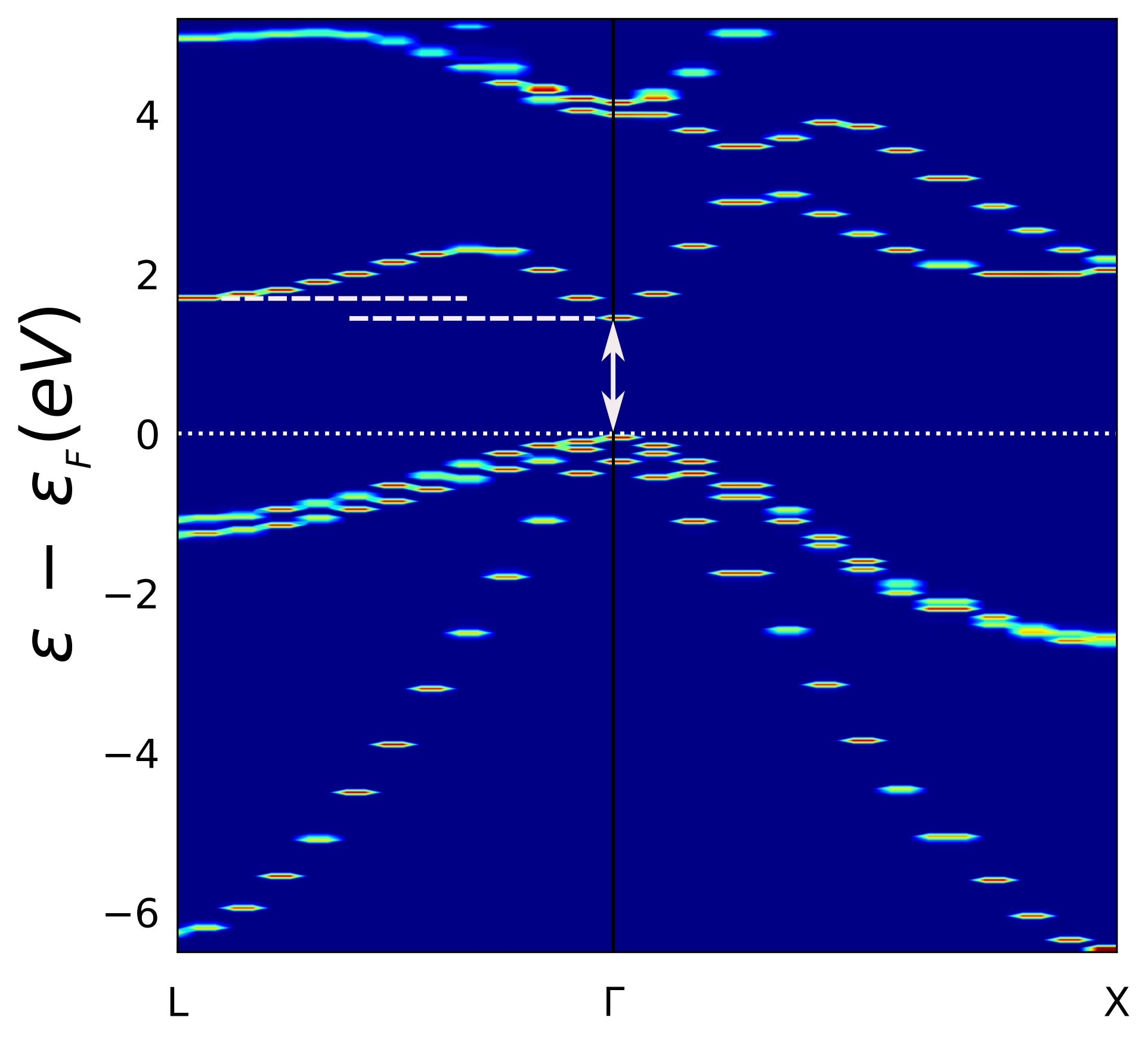}}
  \hspace{0.2in}
  \subfloat[Strain = --1.3\%]{\label{fig:figS1b}\includegraphics[width=3.4in]{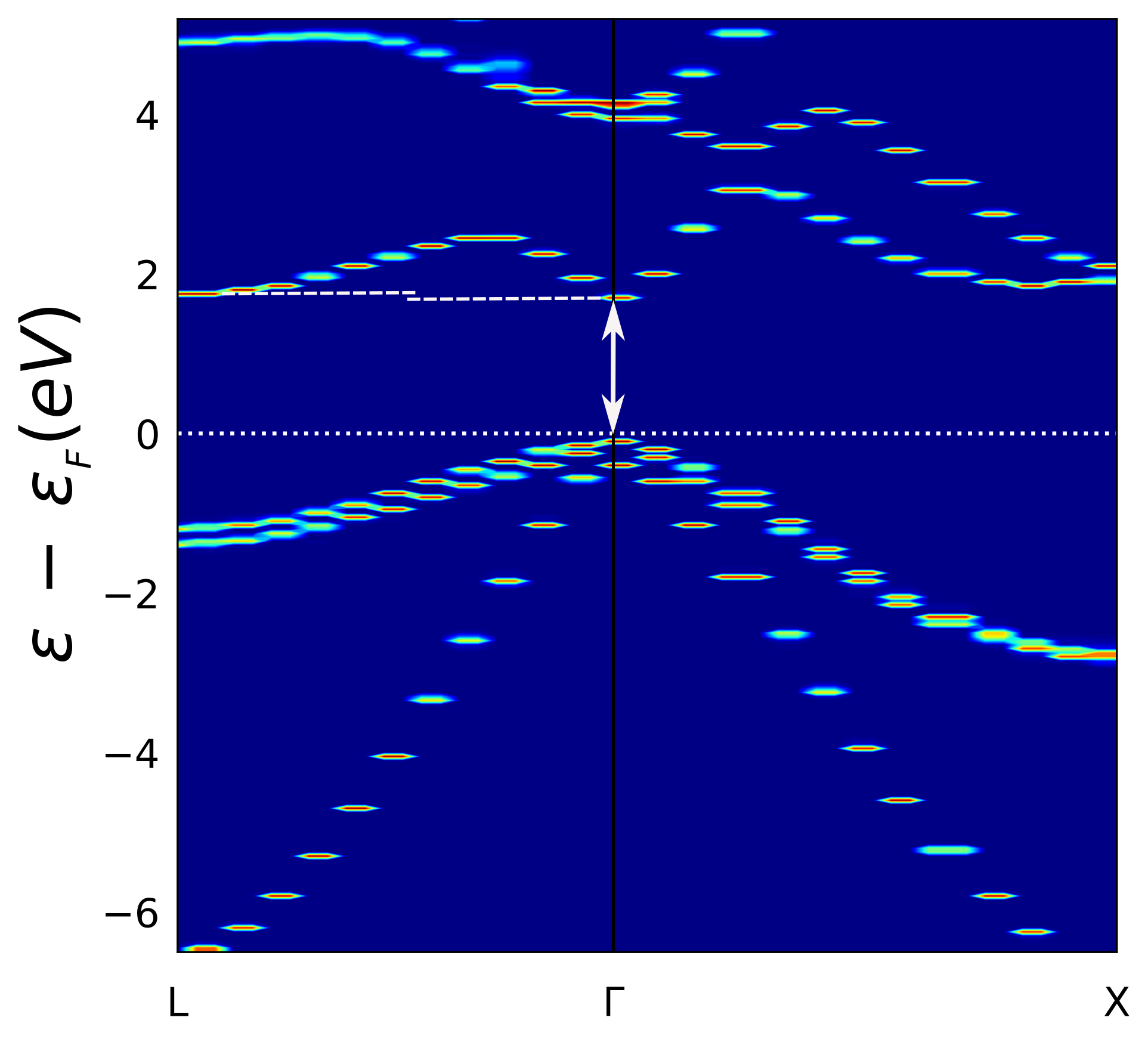}} \\
  \subfloat[Strain = --1.5\%]{\label{fig:figS1c}\includegraphics[width=3.4in]{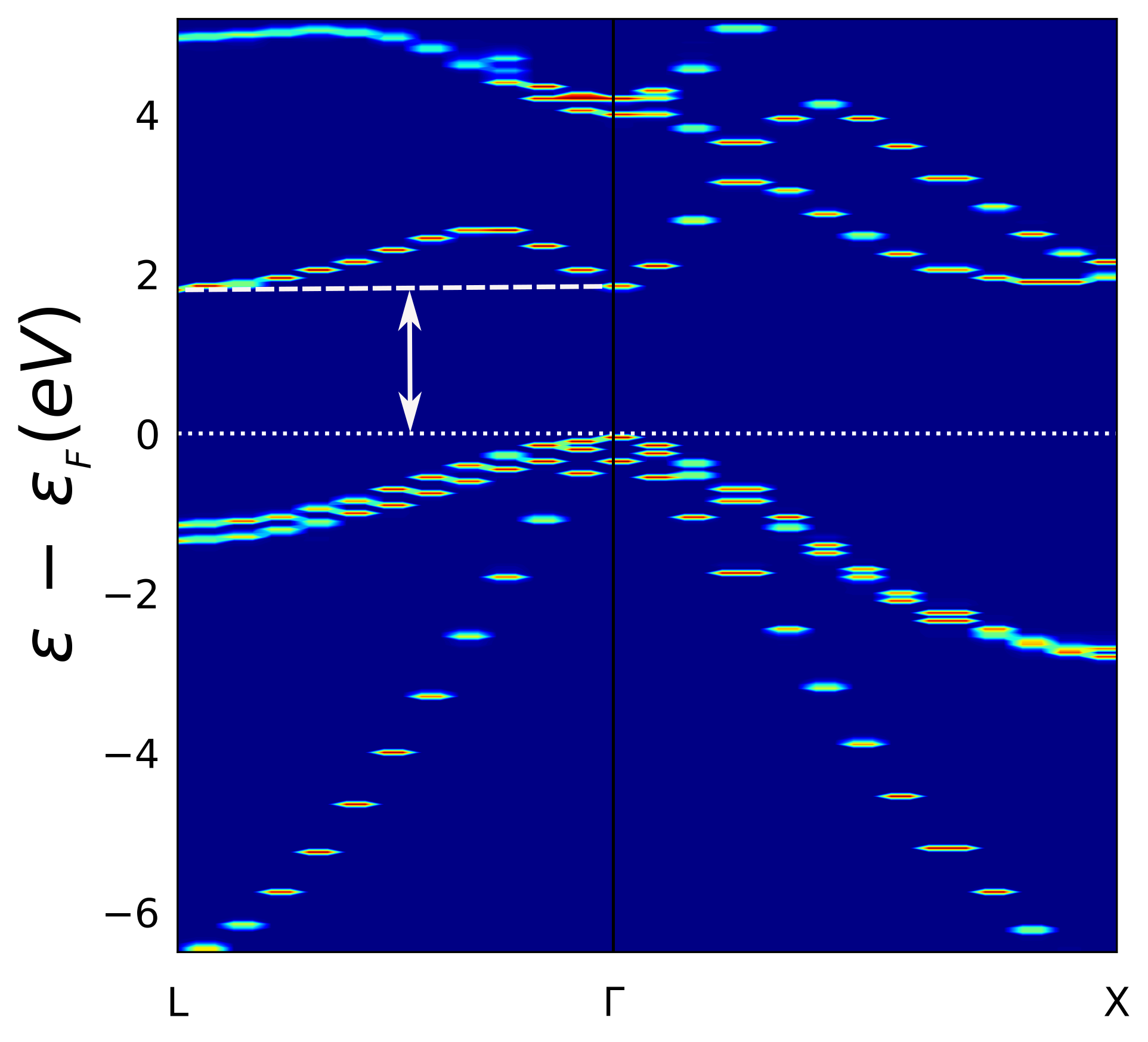}}
  \hspace{0.2in}
  \subfloat[Strain = --2.0\%]{\label{fig:figS1d}\includegraphics[width=3.4in]{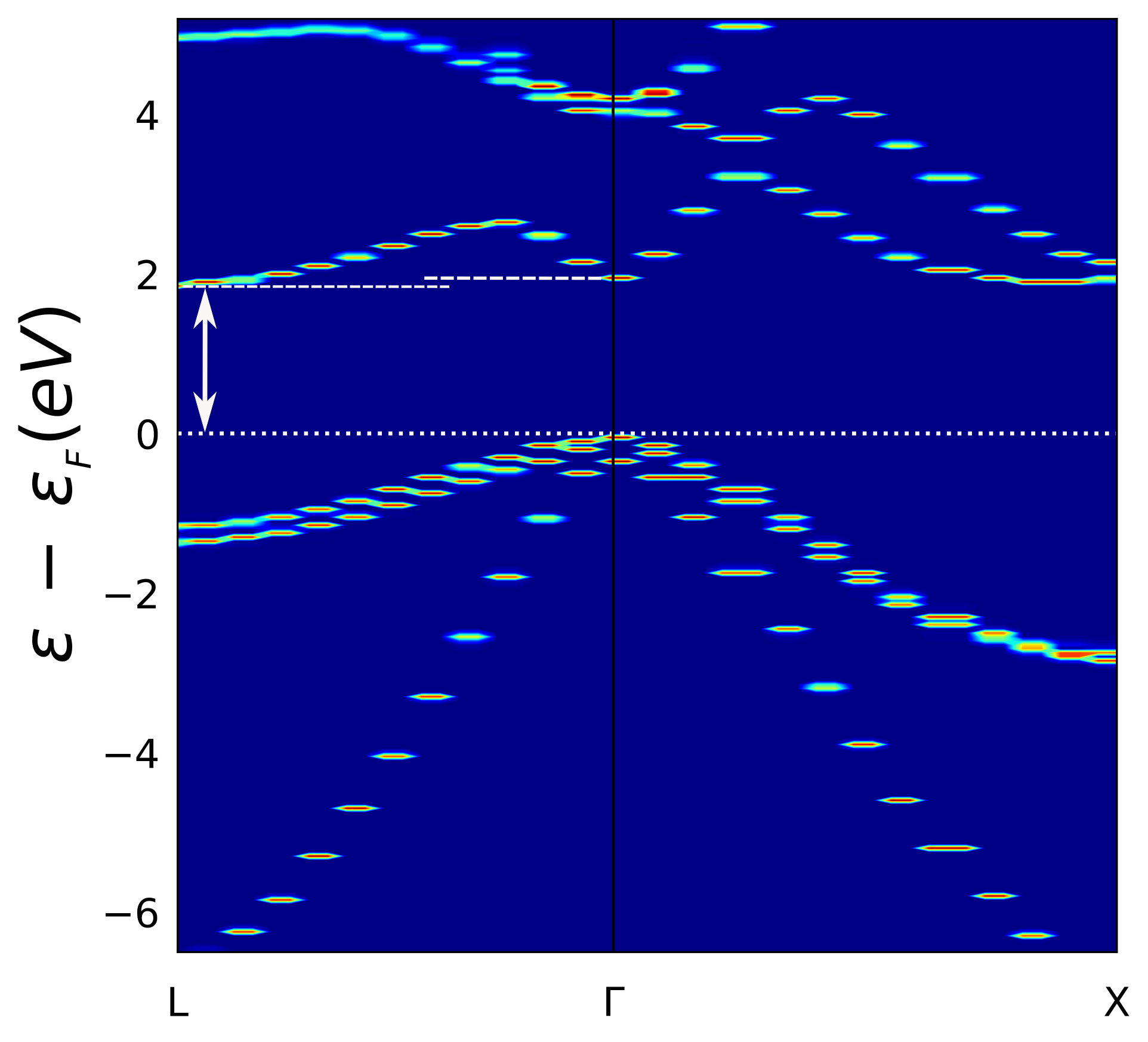}}
  \caption{Effective bandstructures of GaAs$_{0.963}$P$_{0.037}$ under isotropic strain. The bandstructures were plotted using `BandUP' \cite{Medeiros2015Sup,Medeiros2014Sup}. In (a) and (b), the bandgaps are direct. (c) The conduction band minima are degenerate at the $\Gamma$- and L-point. This is the direct-indirect transition point. (d) The bandgap becomes an indirect bandgap. The positive and negative strains correspond to the tensile and compressive strains, respectively. For simplicity, only the L$\rightarrow \Gamma \rightarrow $ X high symmetry path is shown.}
  \label{fig:figS1}
\end{figure}
\vspace*{\fill}
\pagebreak
\vspace*{\fill}
\section{\label{secSII:bandgapnature}Determining the nature of bandgap using Bloch spectral weights}
\begin{figure}[h]
    \centering
    \includegraphics[width=7in]{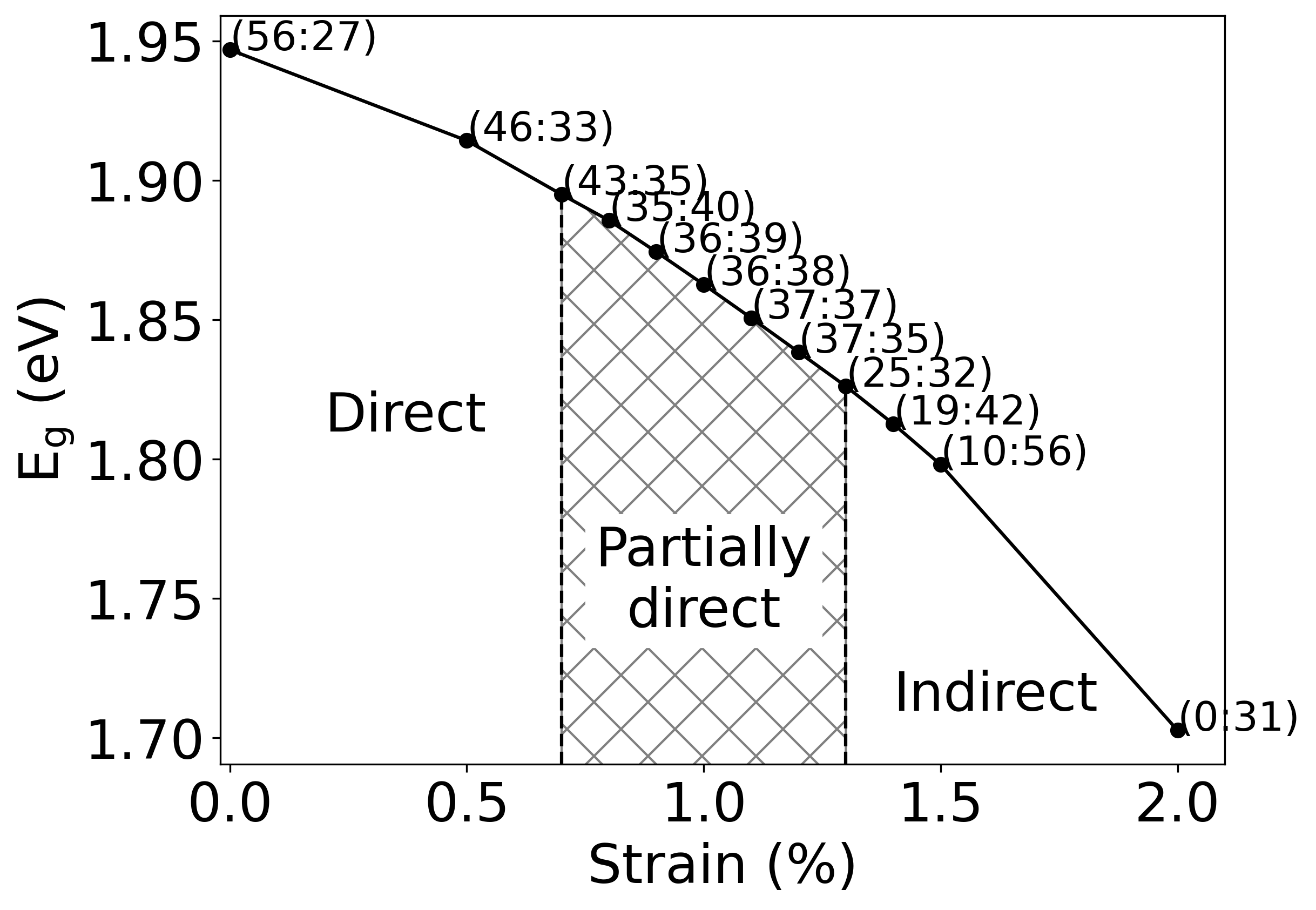}
    \caption{Variation of the bandgap (E$_{\text{g}}$) under biaxial tensile strain for GaAs$_{0.63}$P$_{0.37}$. The Bloch spectral weights (BSW) are given for the conduction band at the $\Gamma$ and L or X point [$\Gamma$ : (L or X)]. The vertical lines separate regions where the CBM changes character. As we are only interested in the direct and indirect character of the bandgap, here, we do not explicitly show the L and X BSWs.}
    \label{fig:figS2}
\end{figure}
\vspace*{\fill}
\pagebreak
\vspace*{\fill}
\section{\label{secSIII:ebsgaasn}Effective bandstructures of G\lowercase{a}A\lowercase{s}$_{0.995}$N$_{0.005}$ under isotropic strain}
\begin{figure}[h]
\centering
  \subfloat[Strain = 0.0\%]{\label{fig:figS3a}\includegraphics[width=3.4in]{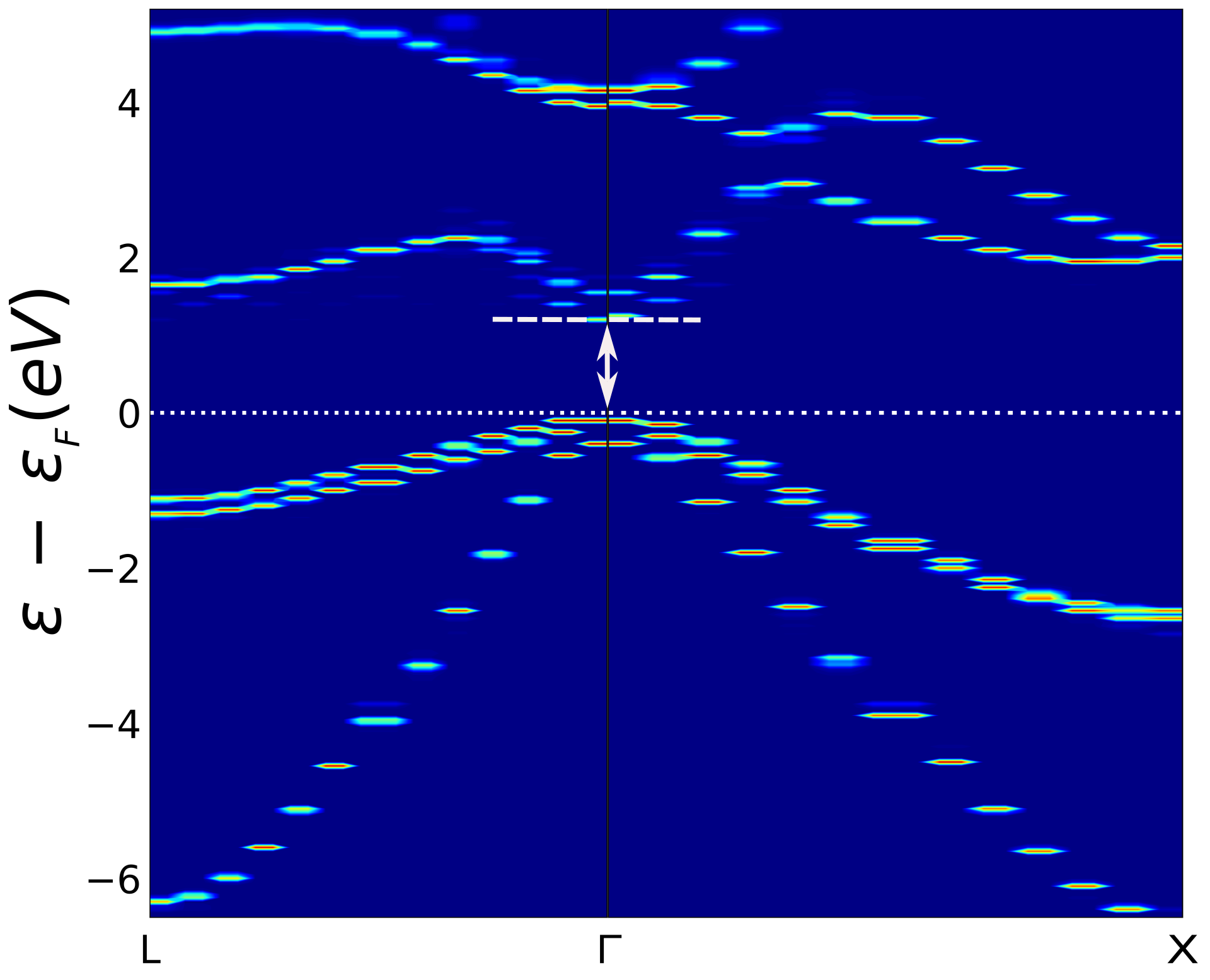}}\\
  \subfloat[Strain = --4.5\%]{\label{fig:figS3b}\includegraphics[width=3.4in]{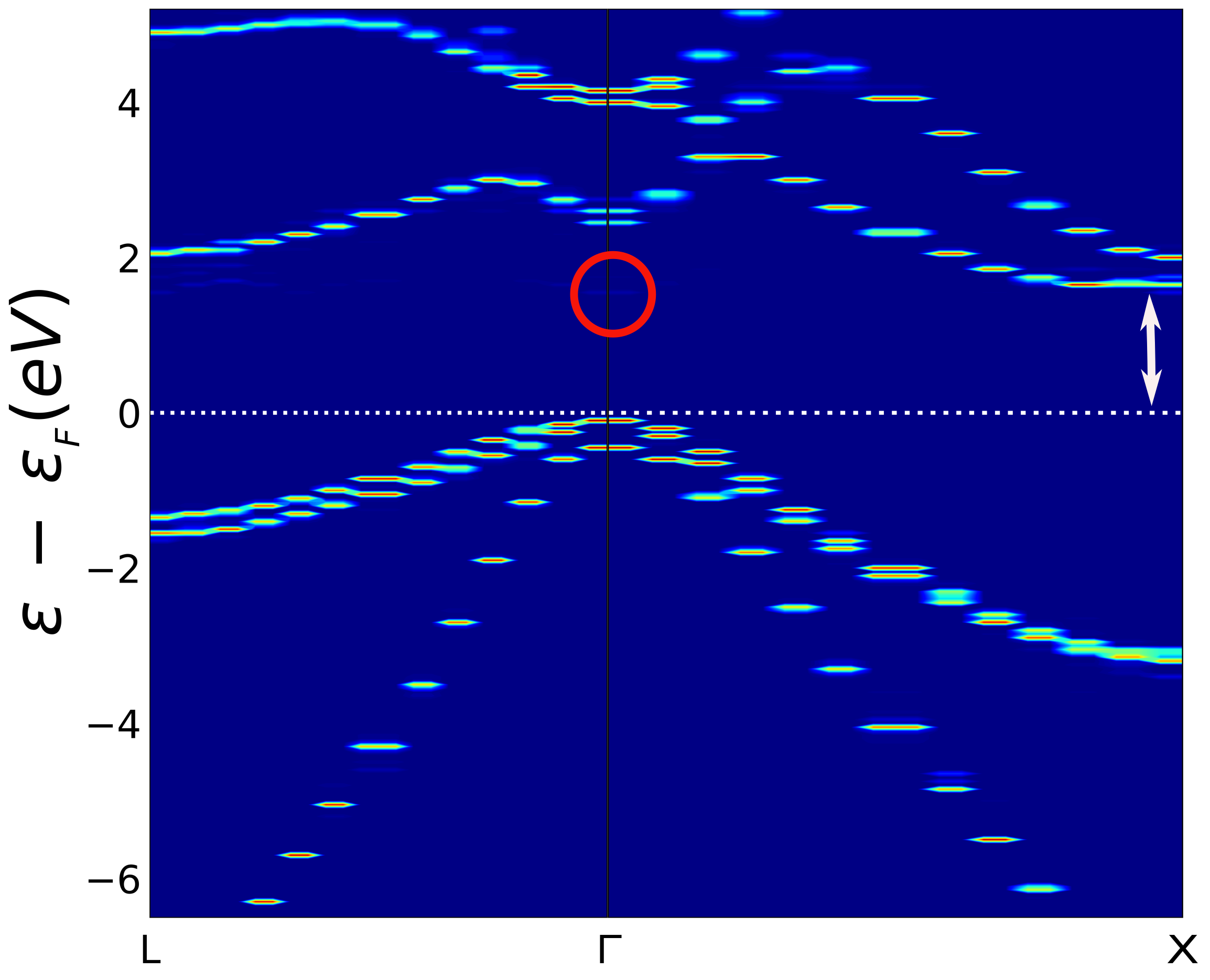}} 
  \hspace{0.2in}
  \subfloat[Strain = +4.5\%]{\label{fig:figS3c}\includegraphics[width=3.4in]{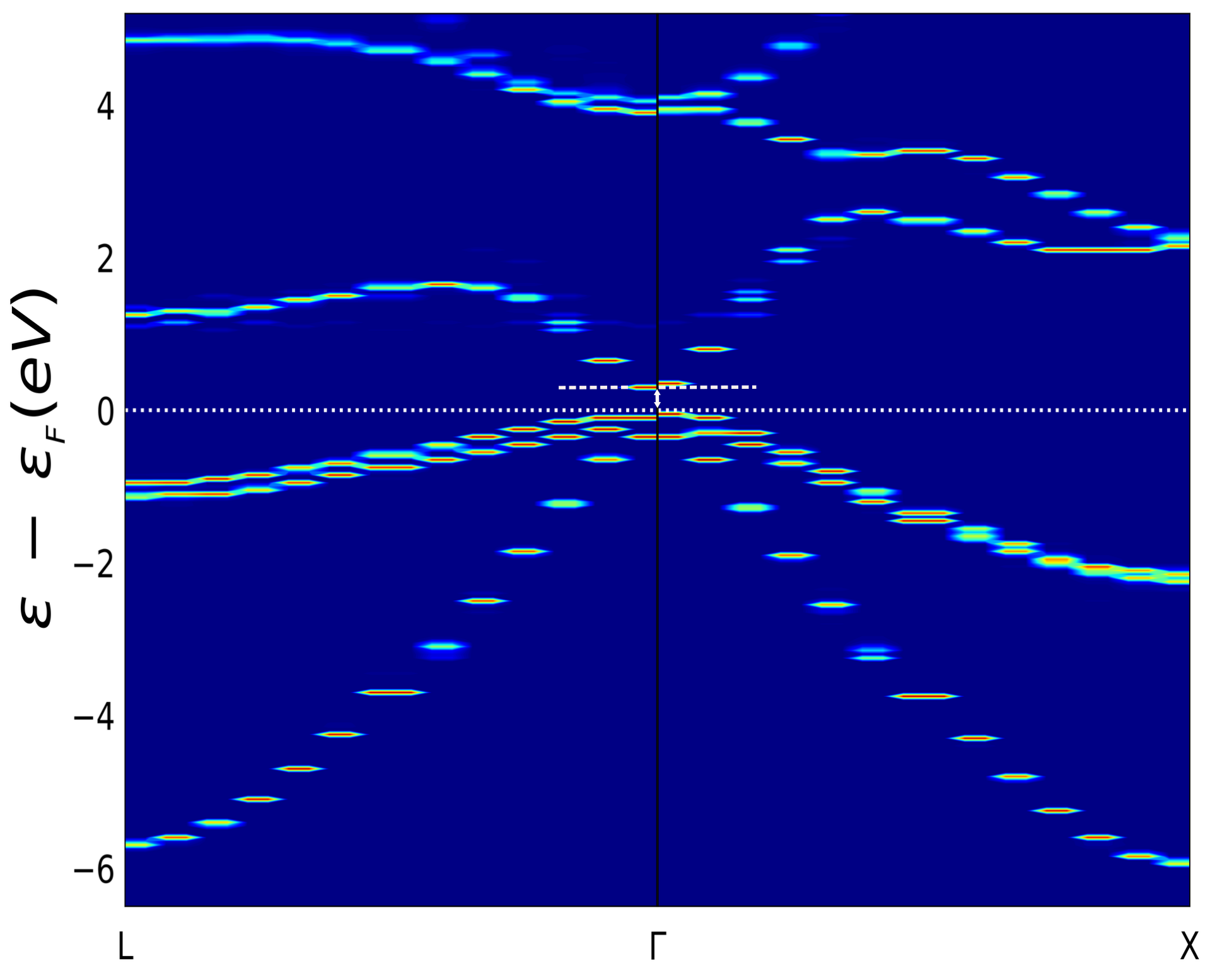}}

  \caption{Effective bandstructures of GaAs$_{0.995}$N$_{0.005}$ under isotropic strain. The bandstructures were plotted using `BandUP' \cite{Medeiros2015Sup,Medeiros2014Sup}. In (a) and (c), the bandgaps are direct. In (b), the Bloch spectral weight of the `N-defect state' at the conduction band $\Gamma$-point is very small, indicated by red circle. Therefore, the bandgap is categorized as an indirect bandgap. The positive and negative strains correspond to the tensile and compressive strains, respectively. For simplicity, only the L$\rightarrow \Gamma \rightarrow $ X high symmetry path is shown.}
  \label{fig:figS3}
\end{figure}
\vspace*{\fill}
\clearpage
\vspace*{\fill}
\begin{figure}
\centering
  \subfloat[Strain = 0.0\%]{\label{fig:figS4a}\includegraphics[width=3.4in]{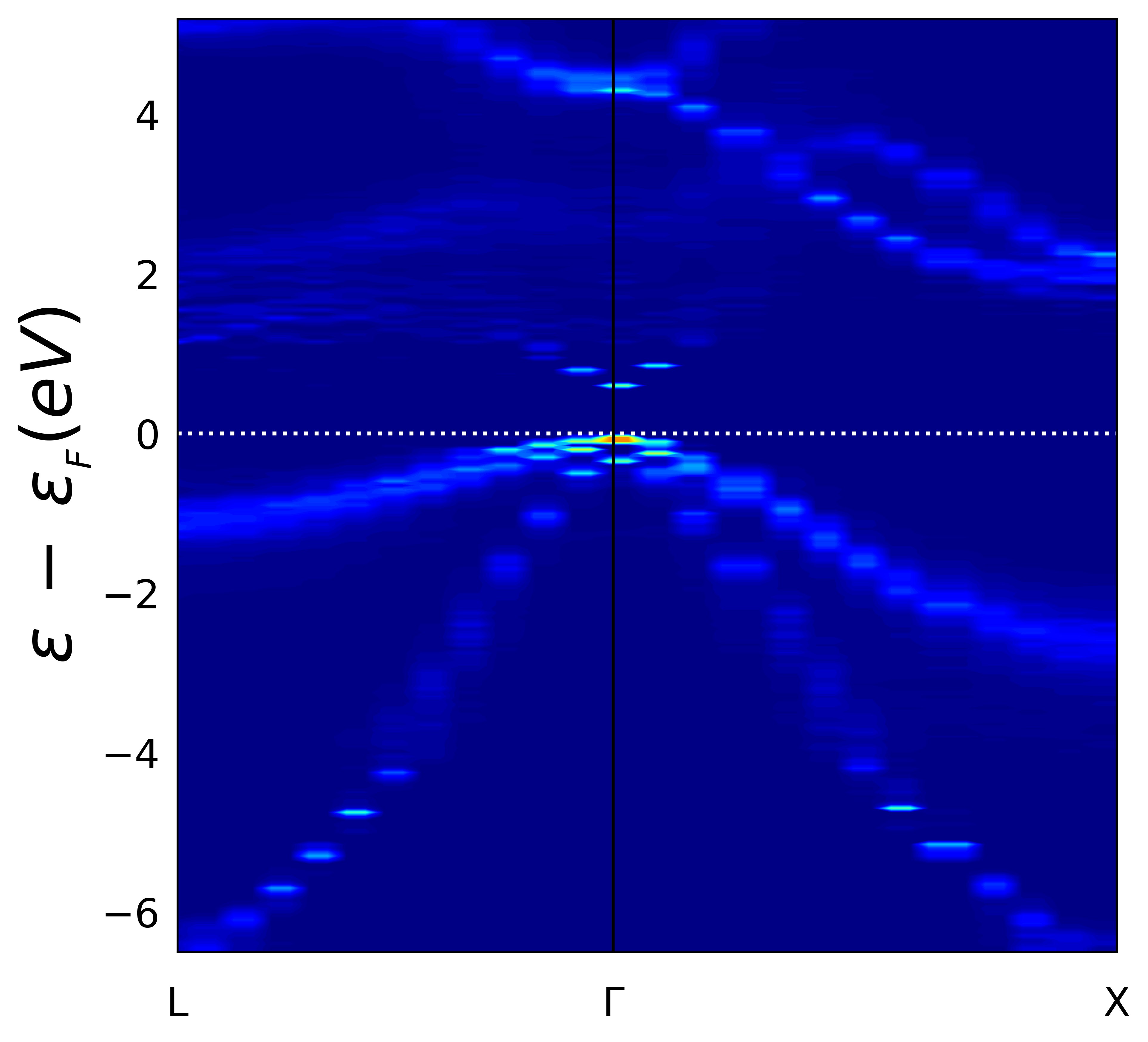}}
  \hspace{0.2in}
  \subfloat[Strain = --5.0\%]{\label{fig:figS4b}\includegraphics[width=3.4in]{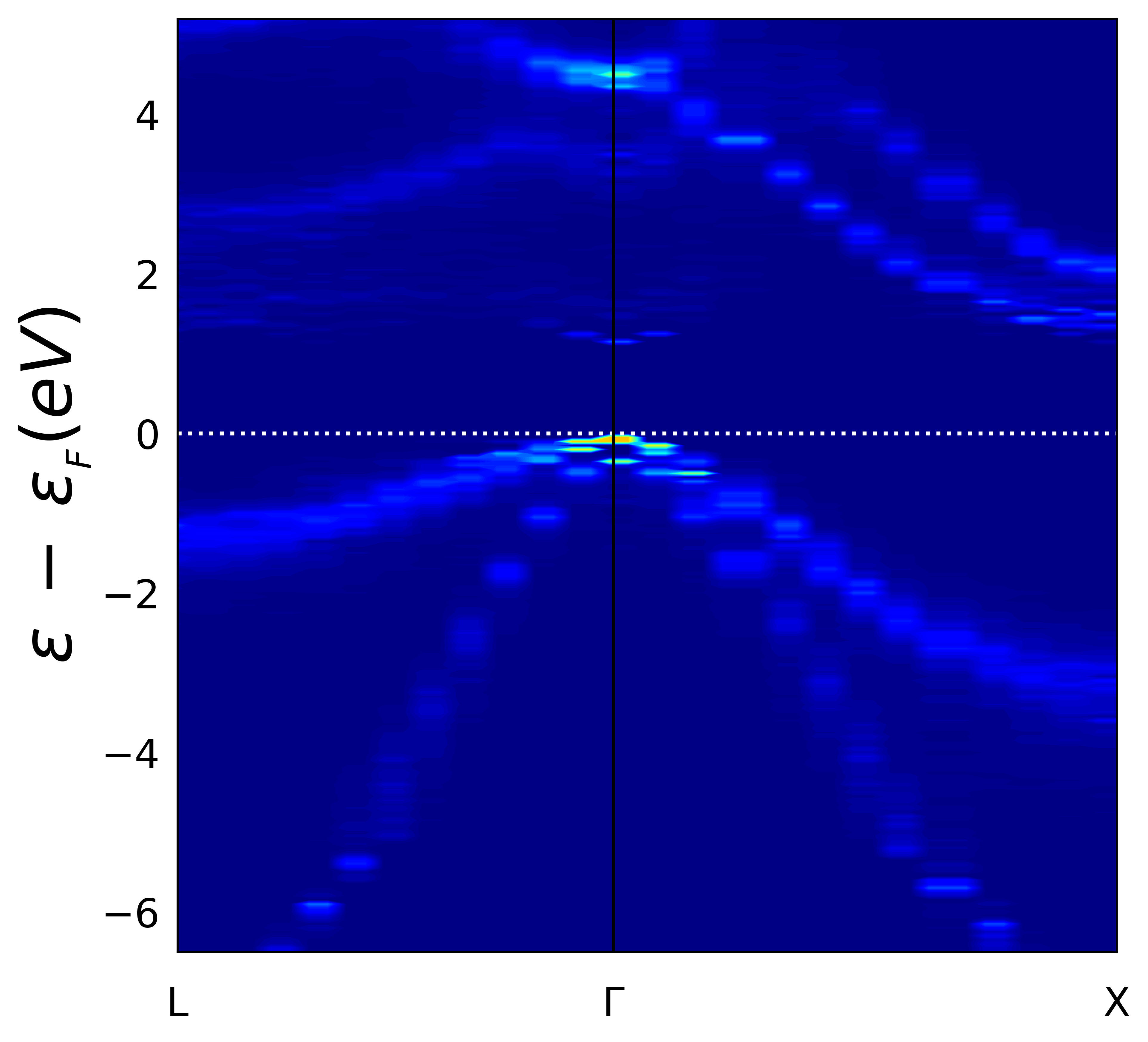}} 
  \caption{Effective bandstructures of GaAs$_{0.889}$N$_{0.111}$ under isotropic strain. The bandstructures were plotted using `BandUP' \cite{Medeiros2015Sup,Medeiros2014Sup}. The bands are strongly dispersed for high N concentrations. The positive and negative strains correspond to the tensile and compressive strains, respectively. For simplicity, only the L$\rightarrow \Gamma \rightarrow $ X high symmetry path is shown.}
  \label{fig:figS4}
\end{figure}

\vspace*{\fill}
\pagebreak
\vspace*{\fill}
\section{\label{secSIV:bandredefinition}Conduction band redefinition based on minimum cut-off Bloch spectral weights criterion}
\begin{figure}[h]
    \centering
    \includegraphics[width=7in]{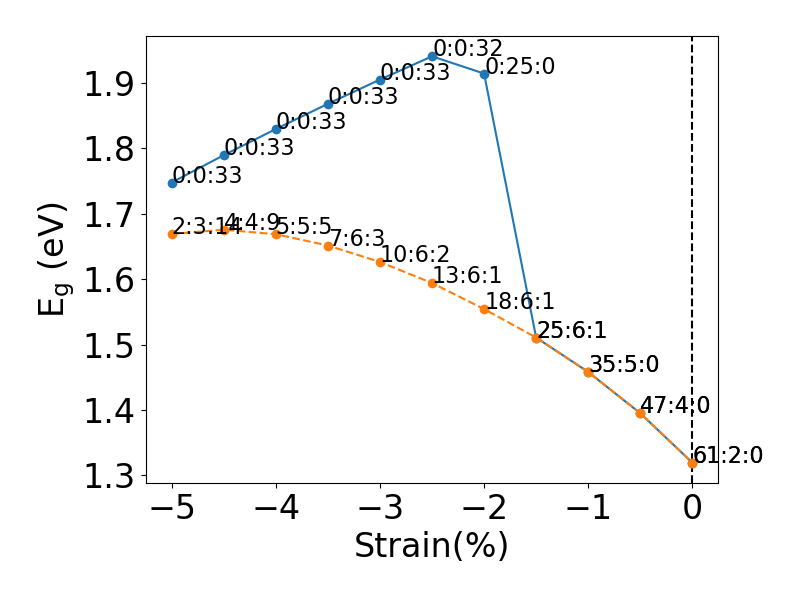}
    \caption{Variation of the bandgap (E$_{\text{g}}$) under biaxial compressive strain for GaAs$_{0.995}$N$_{0.005}$. The folded supercell conduction bands' $\Gamma$-, L-, and X-BSWs are given at each point. The blue and orange lines are the E$_{\text{g}}$ with and without redefinition of the CBM. 20\% BSW is the cut-off criterion for redefining an eigenstate.}
    \label{fig:figS5}
\end{figure}
\vspace*{\fill}
\pagebreak
\vspace*{\fill}
\section{\label{secSV:dftvalues}The density functional theory bandgap values (without interpolation)}

\begin{figure}[h!]
\centering
  \subfloat{\label{fig:figS6_}\includegraphics[scale=0.6]{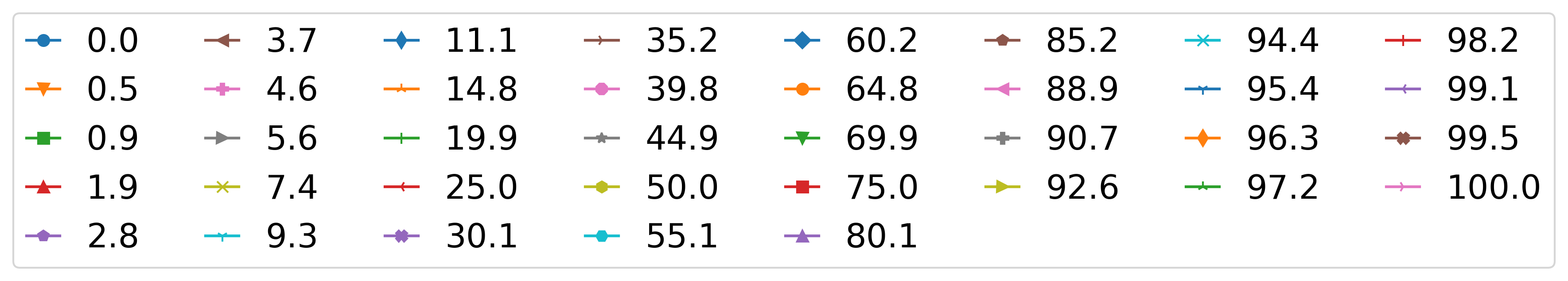}}\\  
  \setcounter{subfigure}{0}
  \subfloat[Isotropic strain for GaAsP]{%
    \includegraphics[width=3.4in]{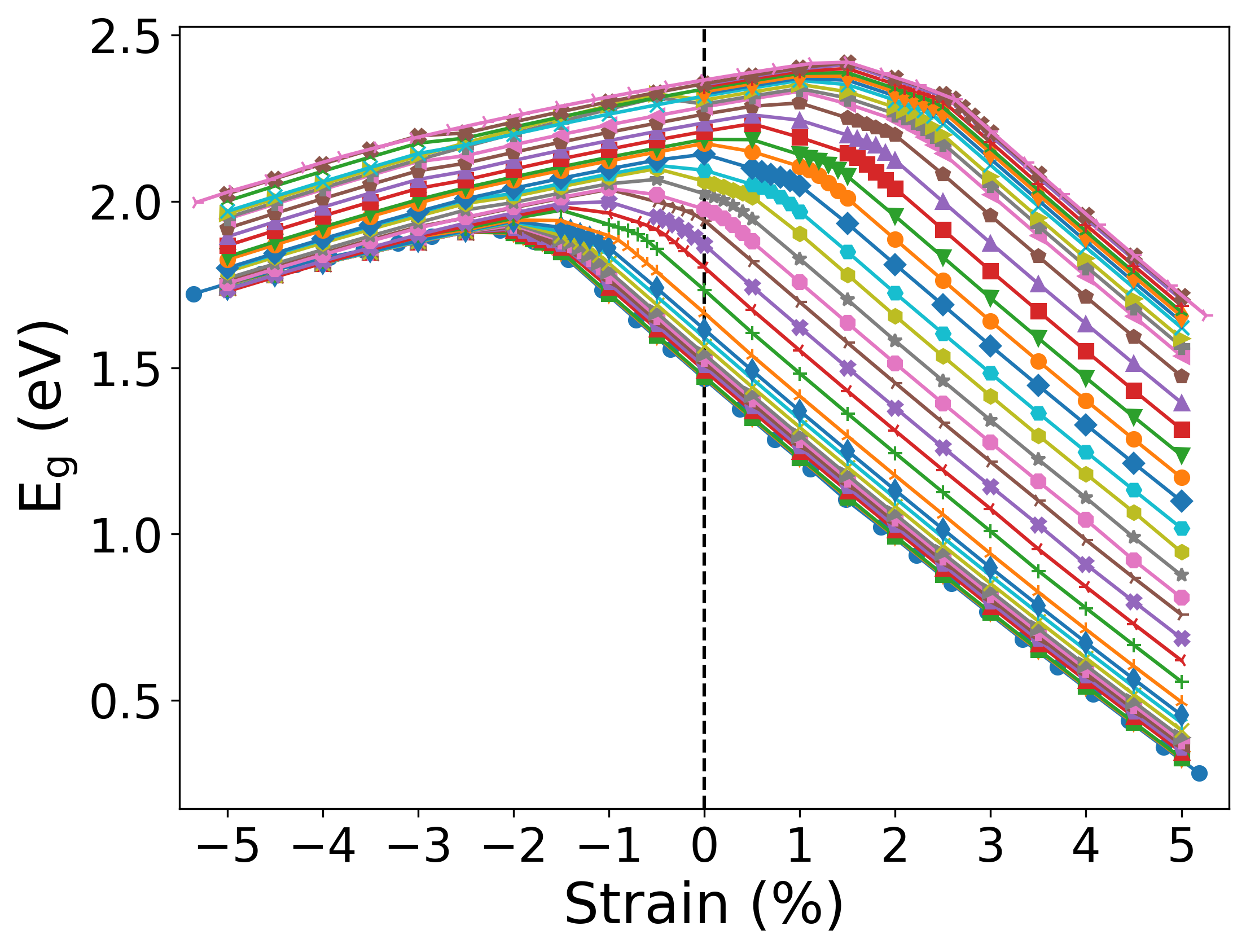}
    \hspace{0.2in}
    \includegraphics[width=3.4in]{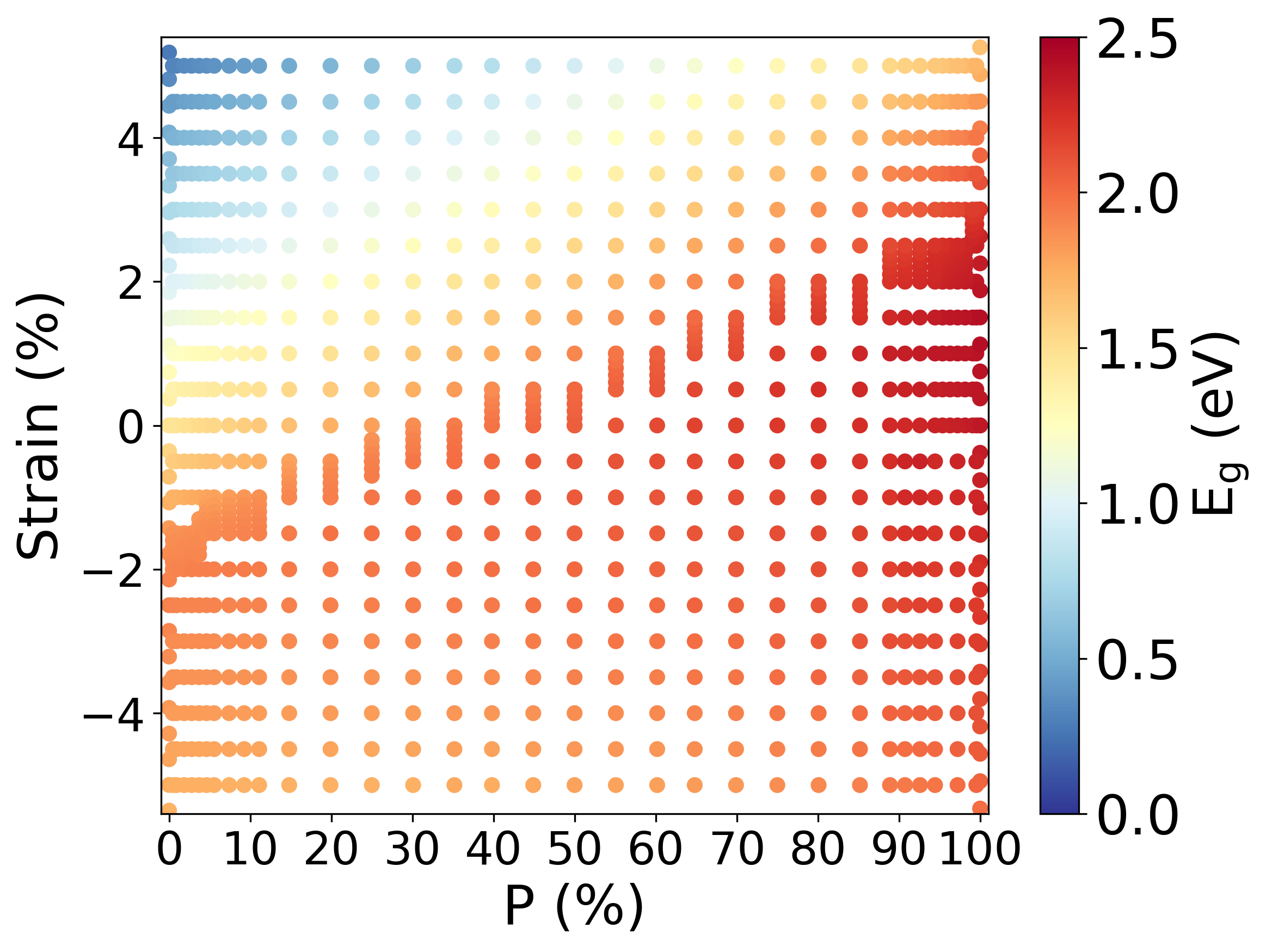}
    \label{fig:figS6a}
   } \\ %
   \subfloat[Biaxial strain for GaAsP]{%
    \includegraphics[width=3.4in]{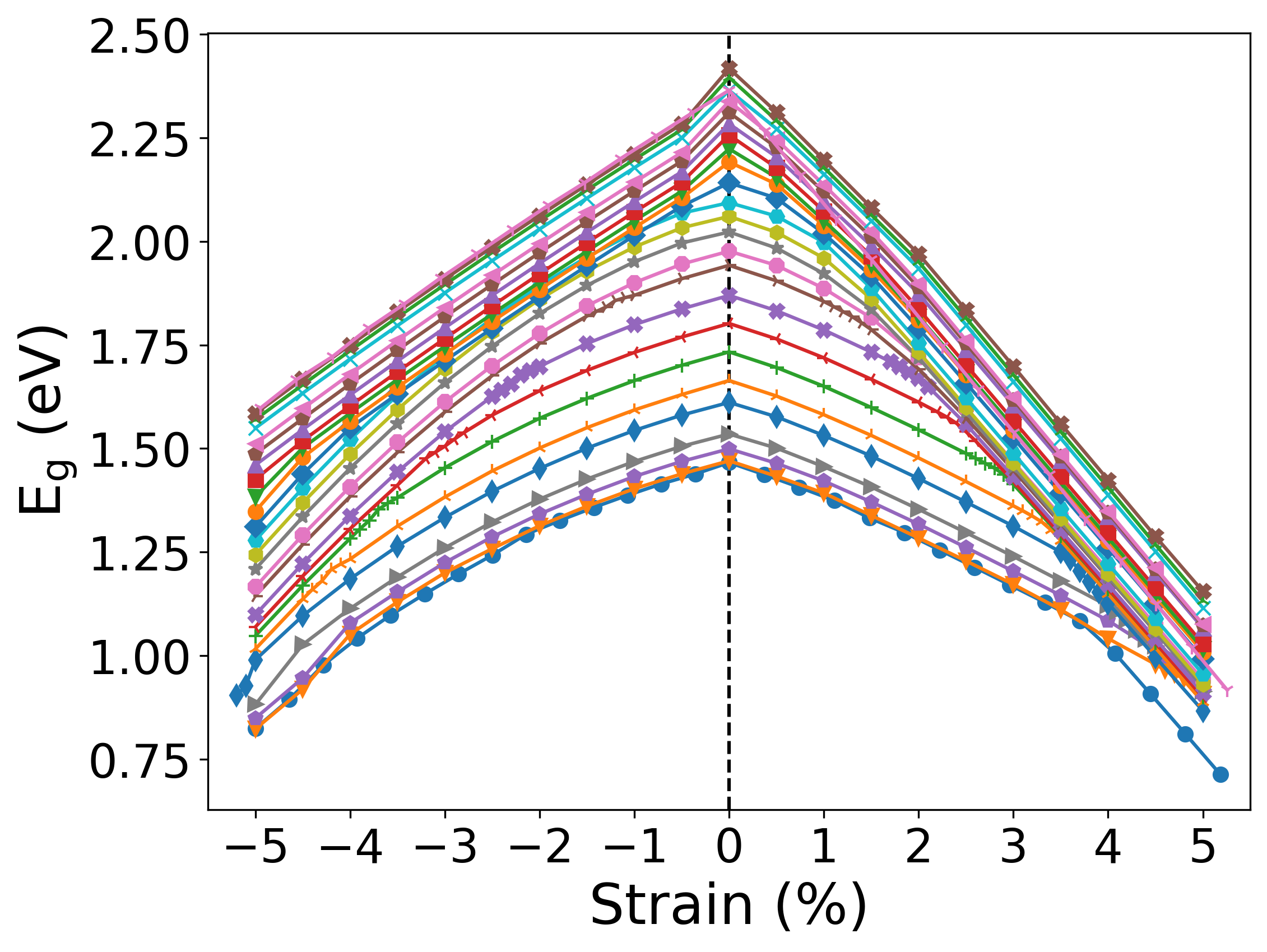}
    \hspace{0.2in}
    \includegraphics[width=3.4in]{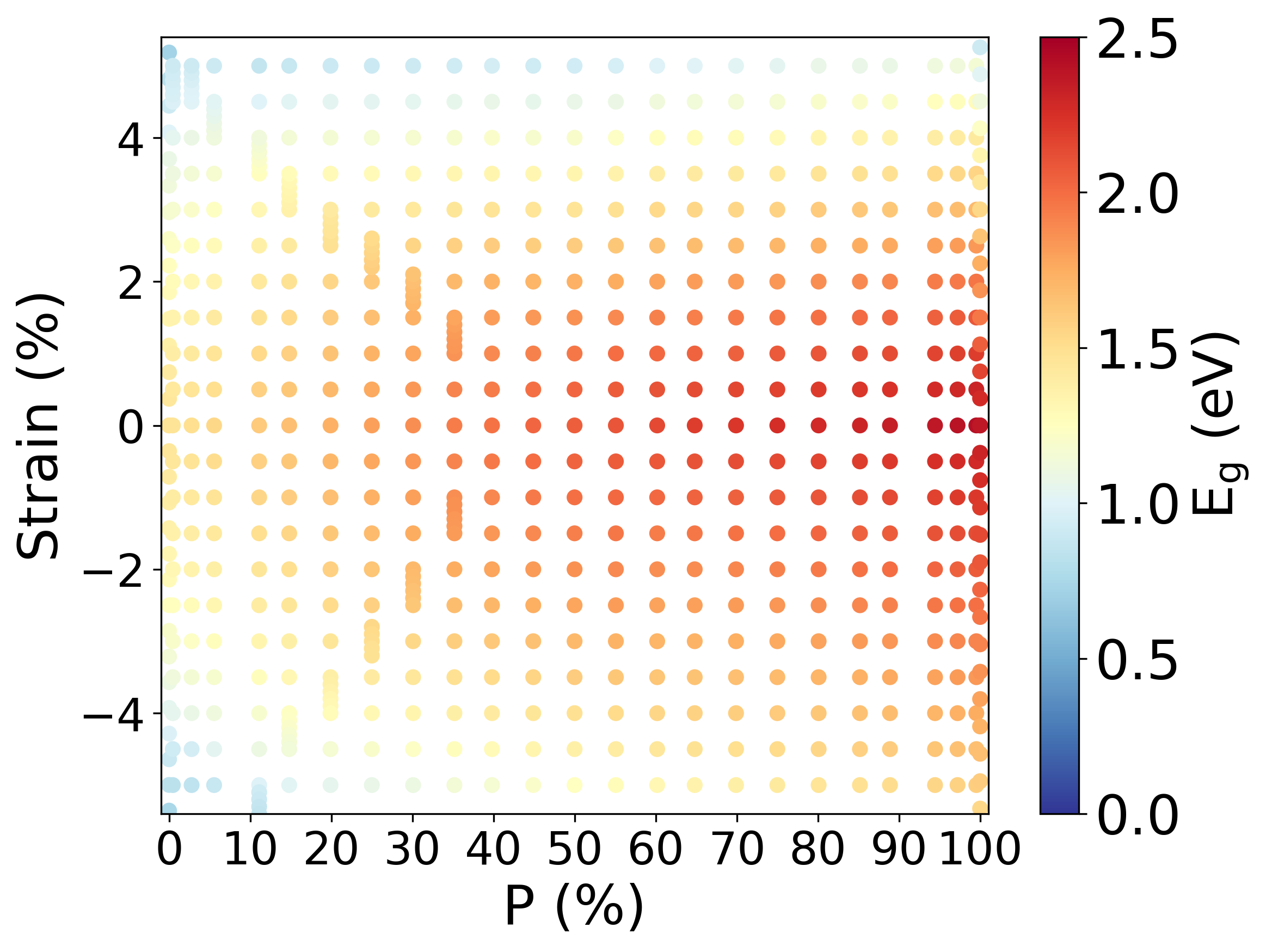}
    \label{fig:figS6b}
   } 
\end{figure}
\vspace*{\fill}
\clearpage
\vspace*{\fill}
\addtocounter{figure}{1}
\begin{figure}[h!]\ContinuedFloat
\centering
   \subfloat[Isotropic strain for GaAsN]{%
    \includegraphics[width=3.4in]{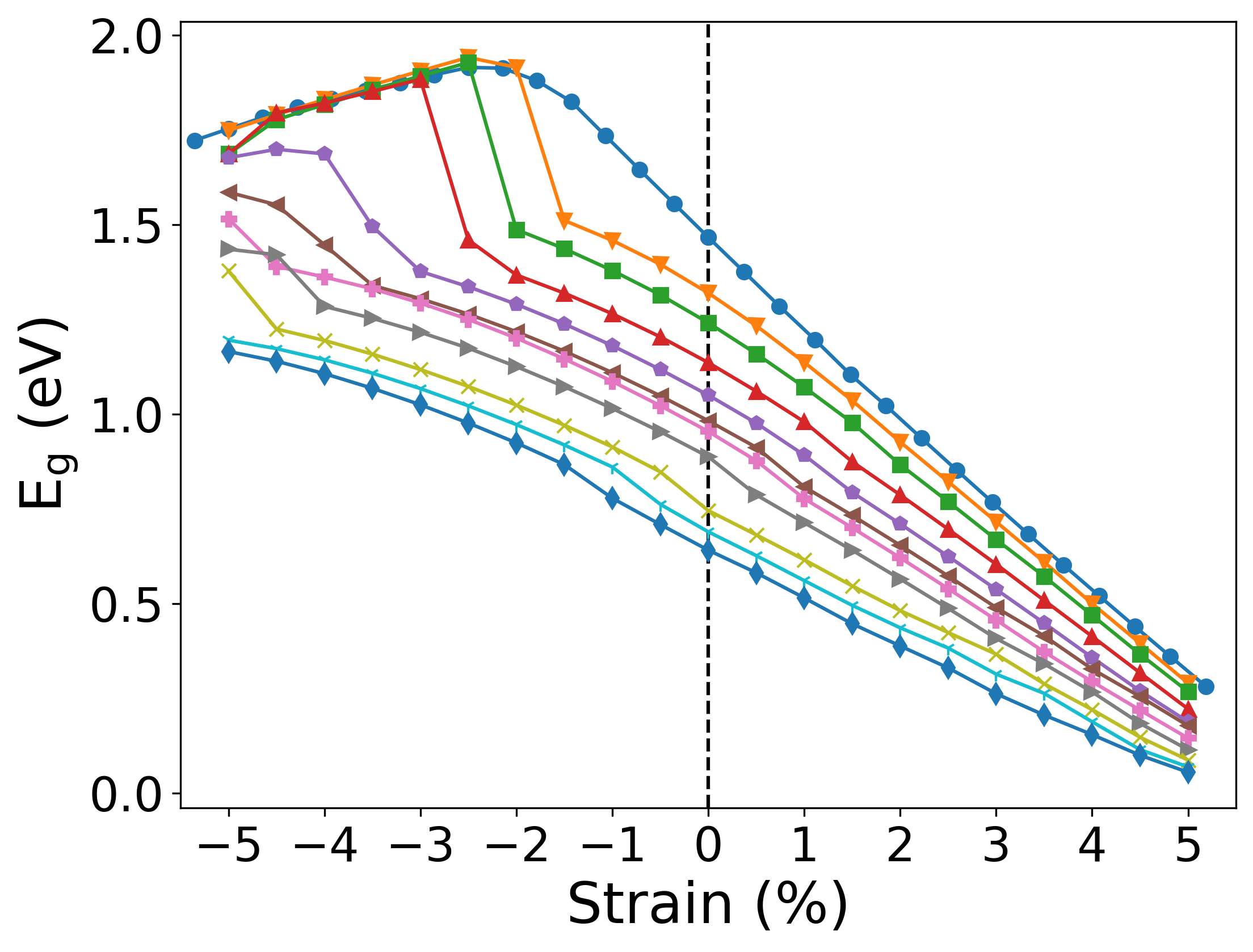}
    \hspace{0.2in}
    \includegraphics[width=3.4in]{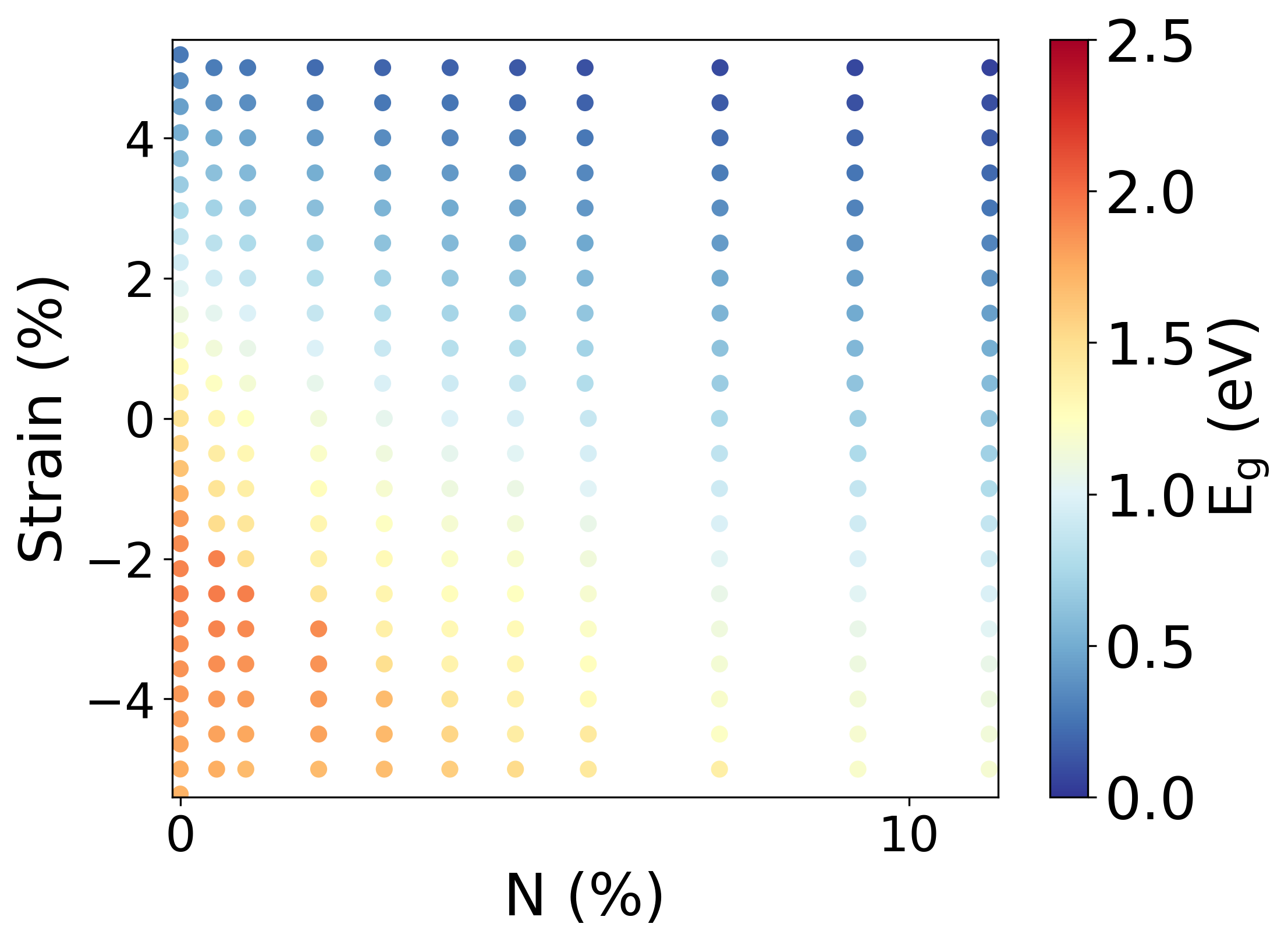}
    \label{fig:figS6c}
   } \\%
   \subfloat[Biaxial strain for GaAsN]{%
    \includegraphics[width=3.4in]{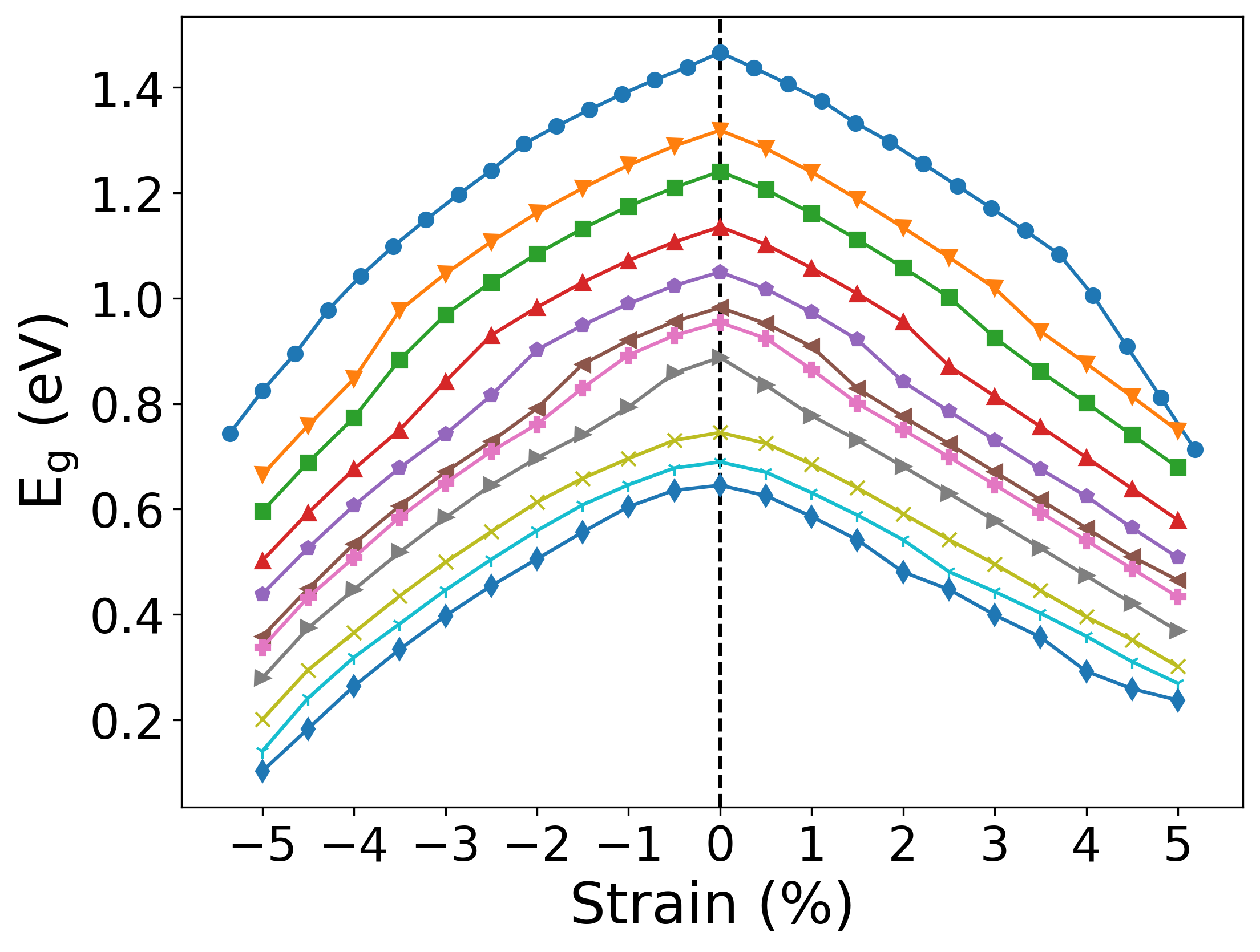}
    \hspace{0.2in}
    \includegraphics[width=3.4in]{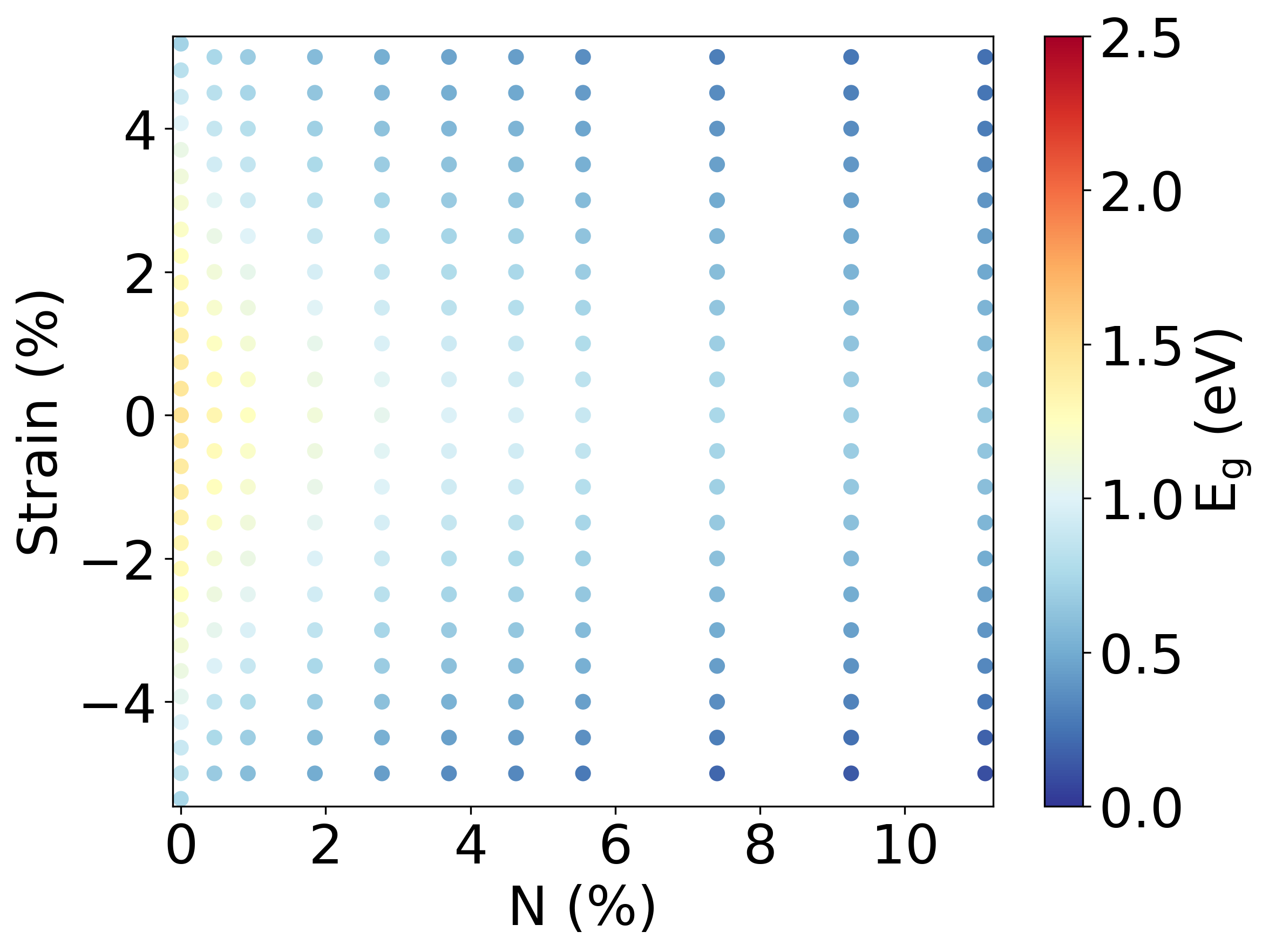}
    \label{fig:figS6d}
   } \\ %
   \subfloat[Biaxial strain for GaAsSb]{%
    \includegraphics[width=3.4in]{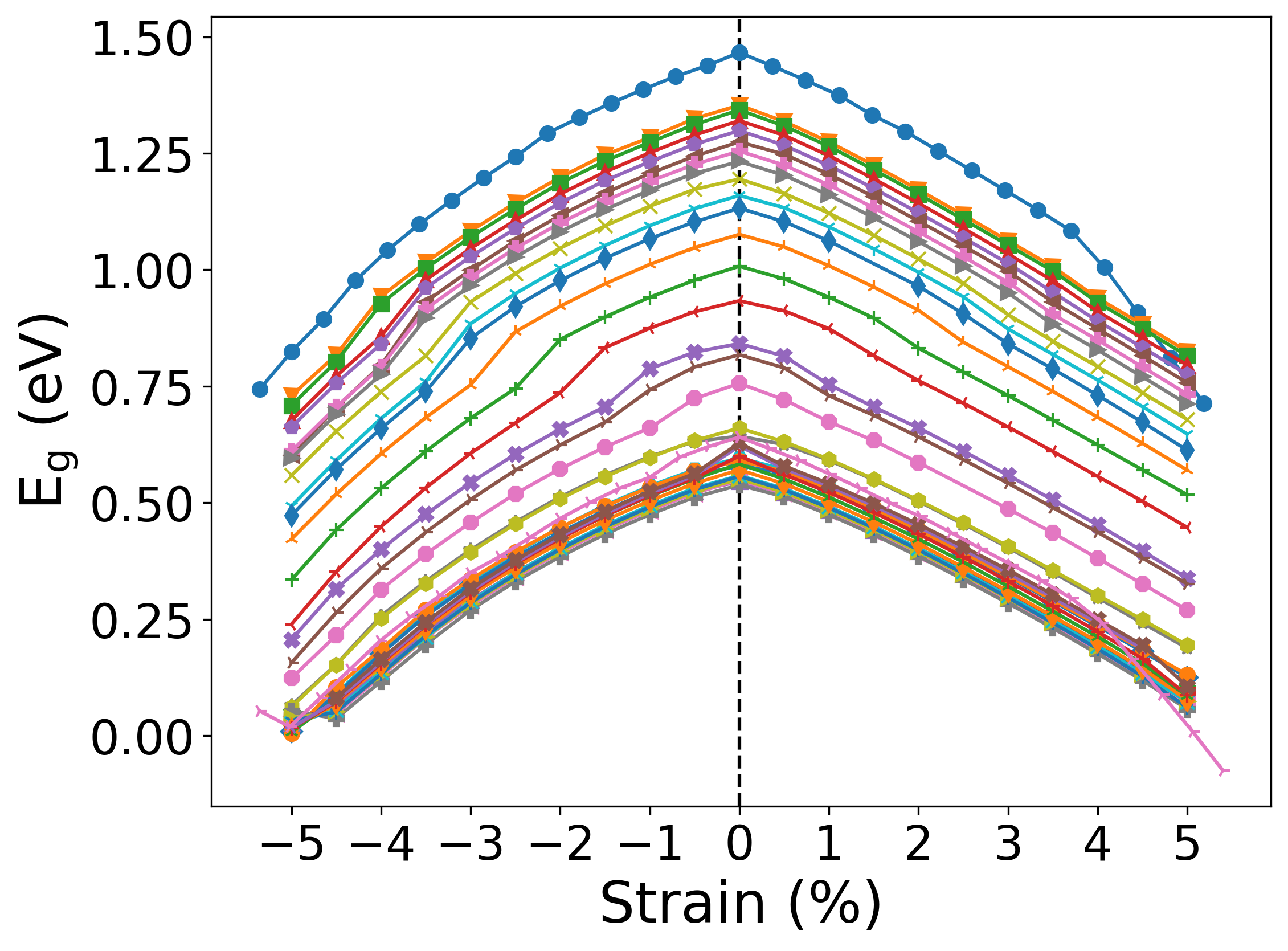}
    \hspace{0.2in}
    \includegraphics[width=3.4in]{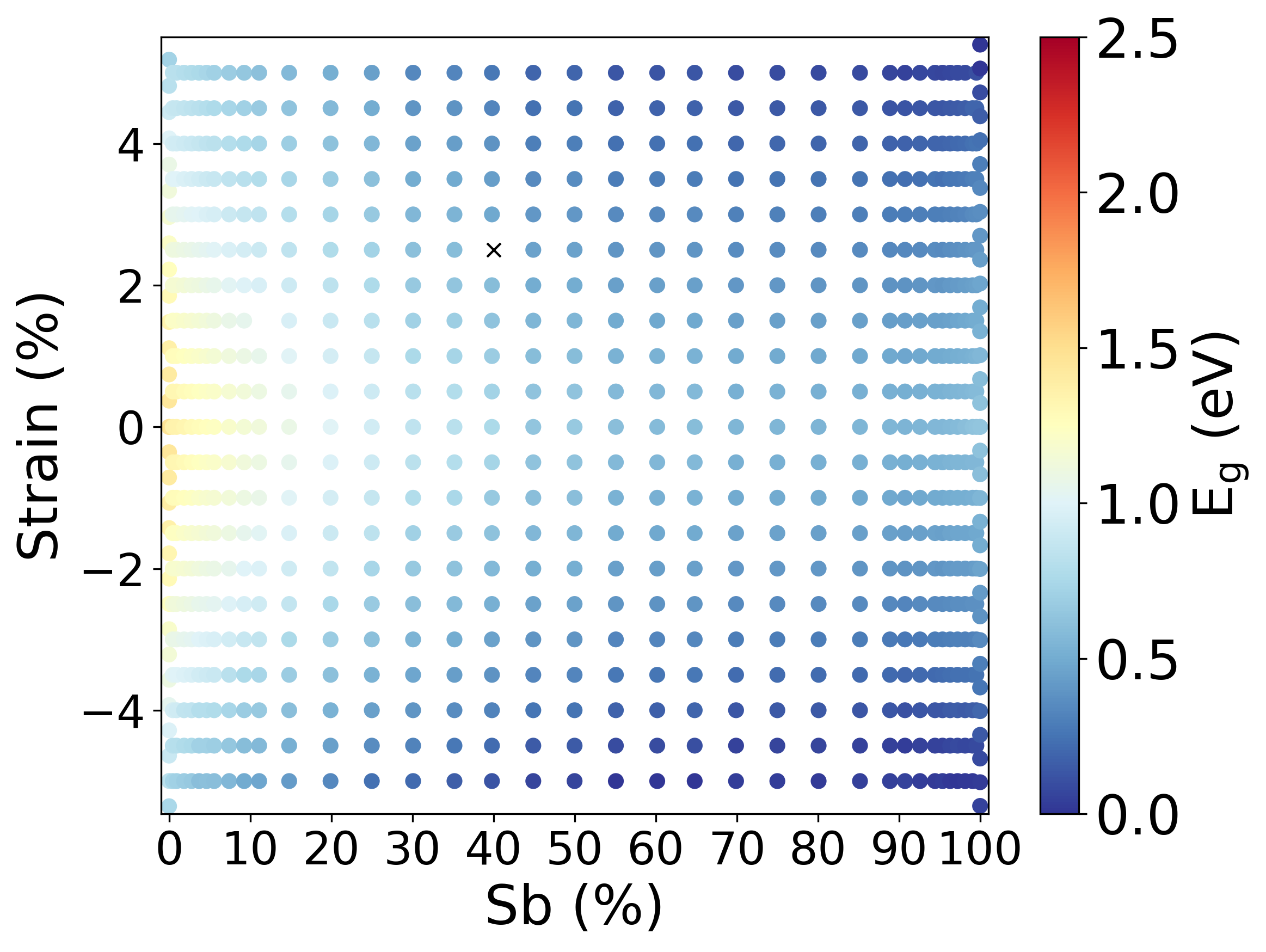}
    \label{fig:figS6e}
   } 
\end{figure}
\addtocounter{figure}{1}
\begin{figure}[h!]\ContinuedFloat
\centering
  \subfloat[Biaxial strain for GaAsBi]{%
    \includegraphics[width=3.4in]{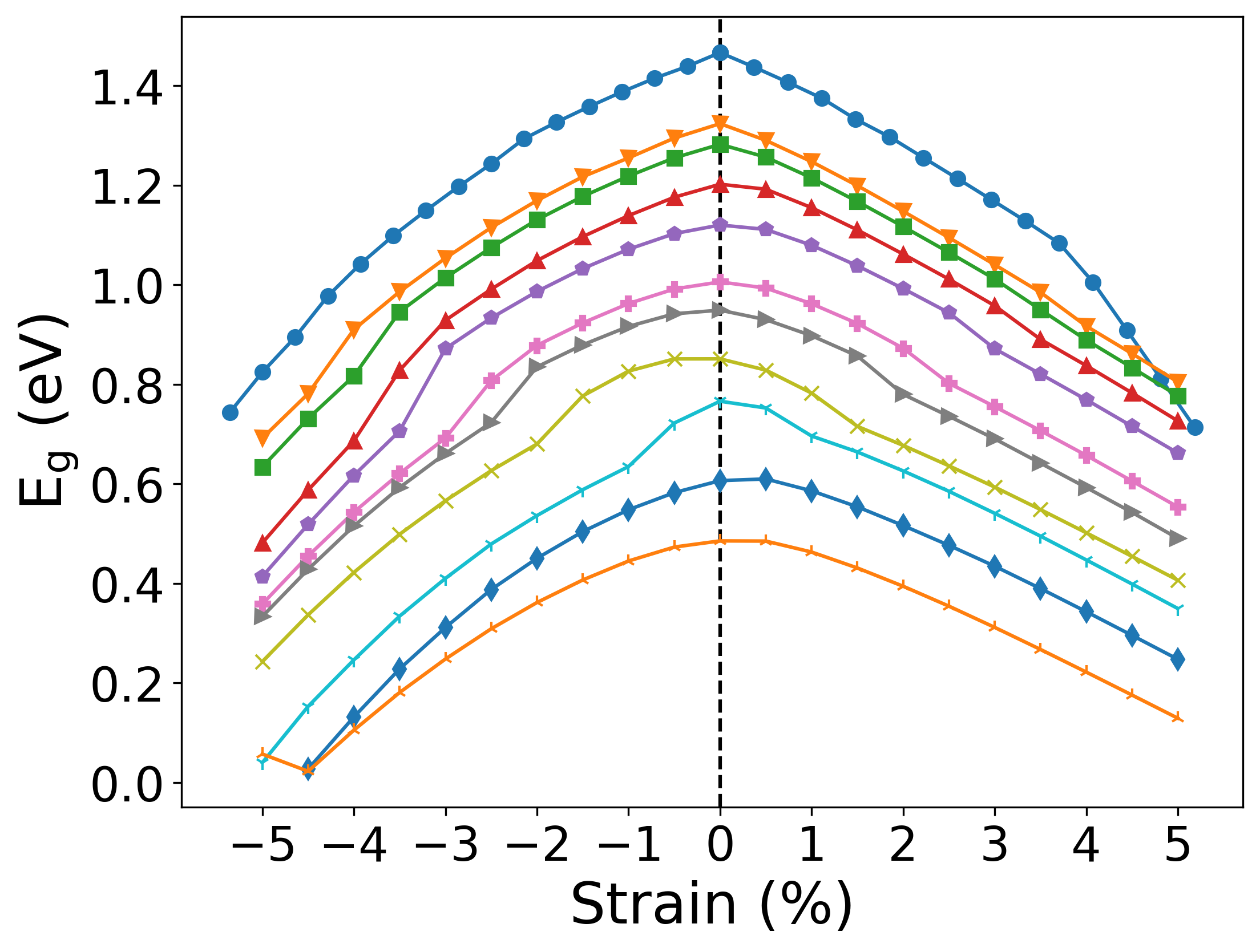}
    \hspace{0.2in}
    \includegraphics[width=3.4in]{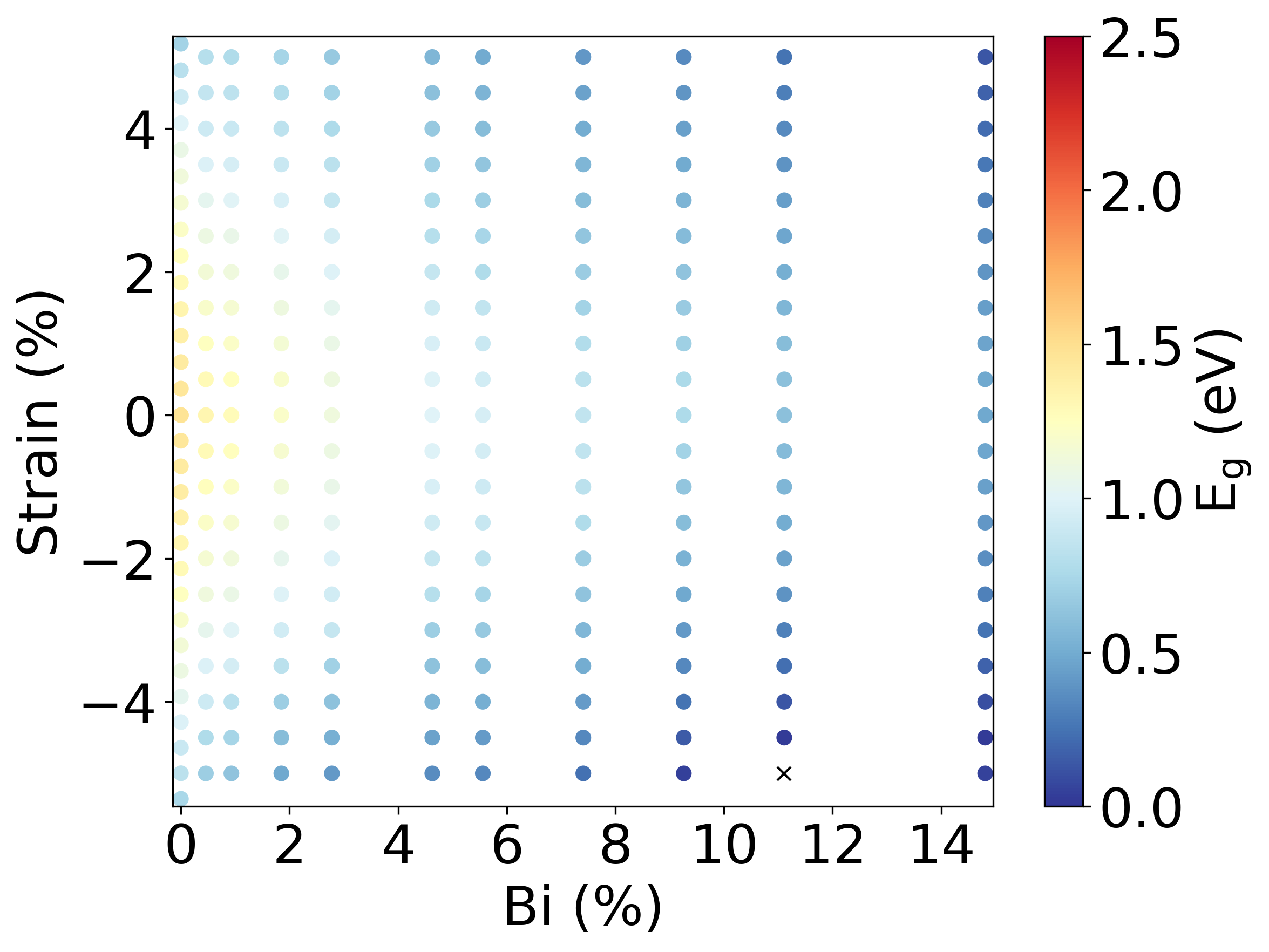}
    \label{fig:figS6f}
   } \\%
   \subfloat[Biaxial strain for GaPSb]{%
    \includegraphics[width=3.4in]{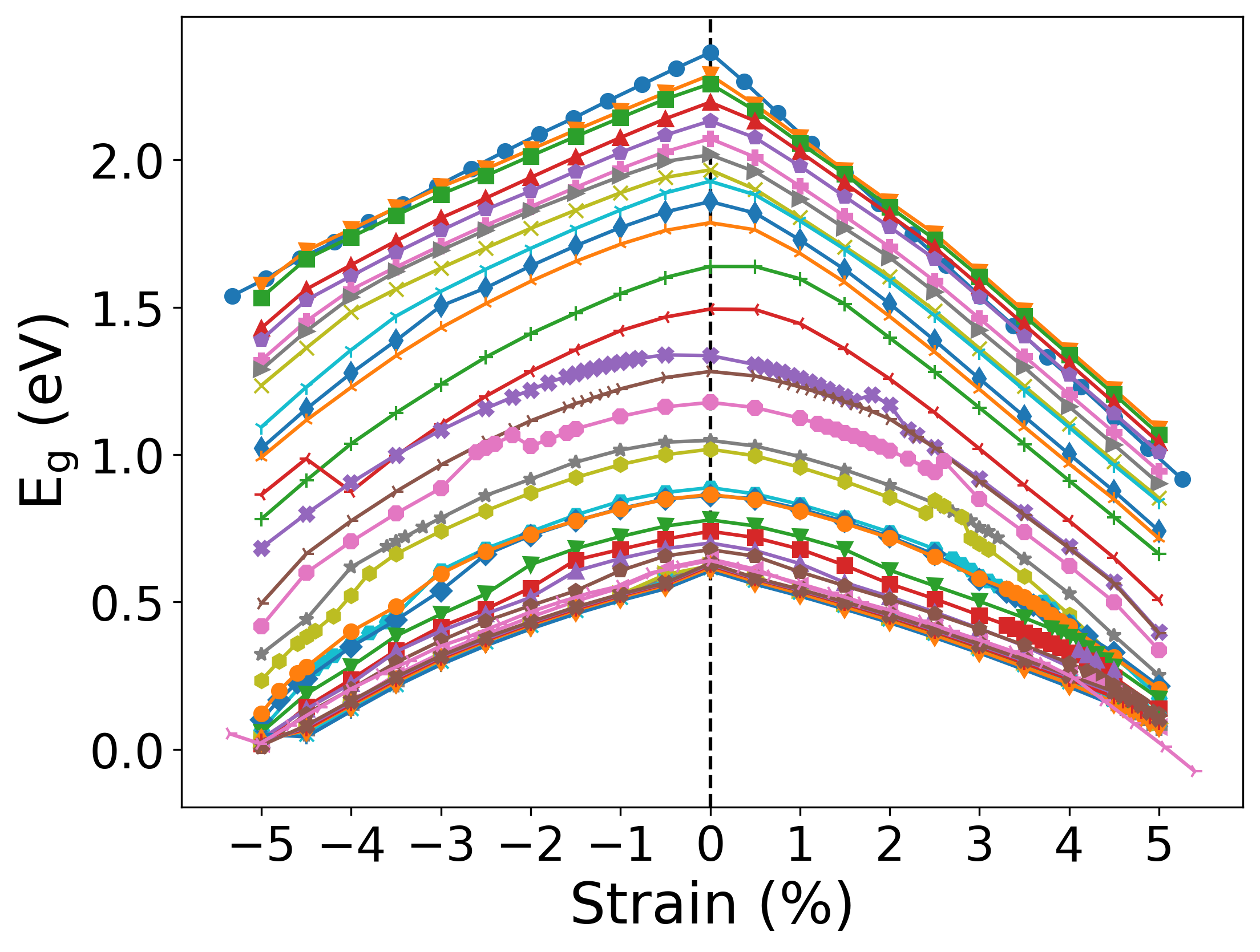}
    \hspace{0.2in}
    \includegraphics[width=3.4in]{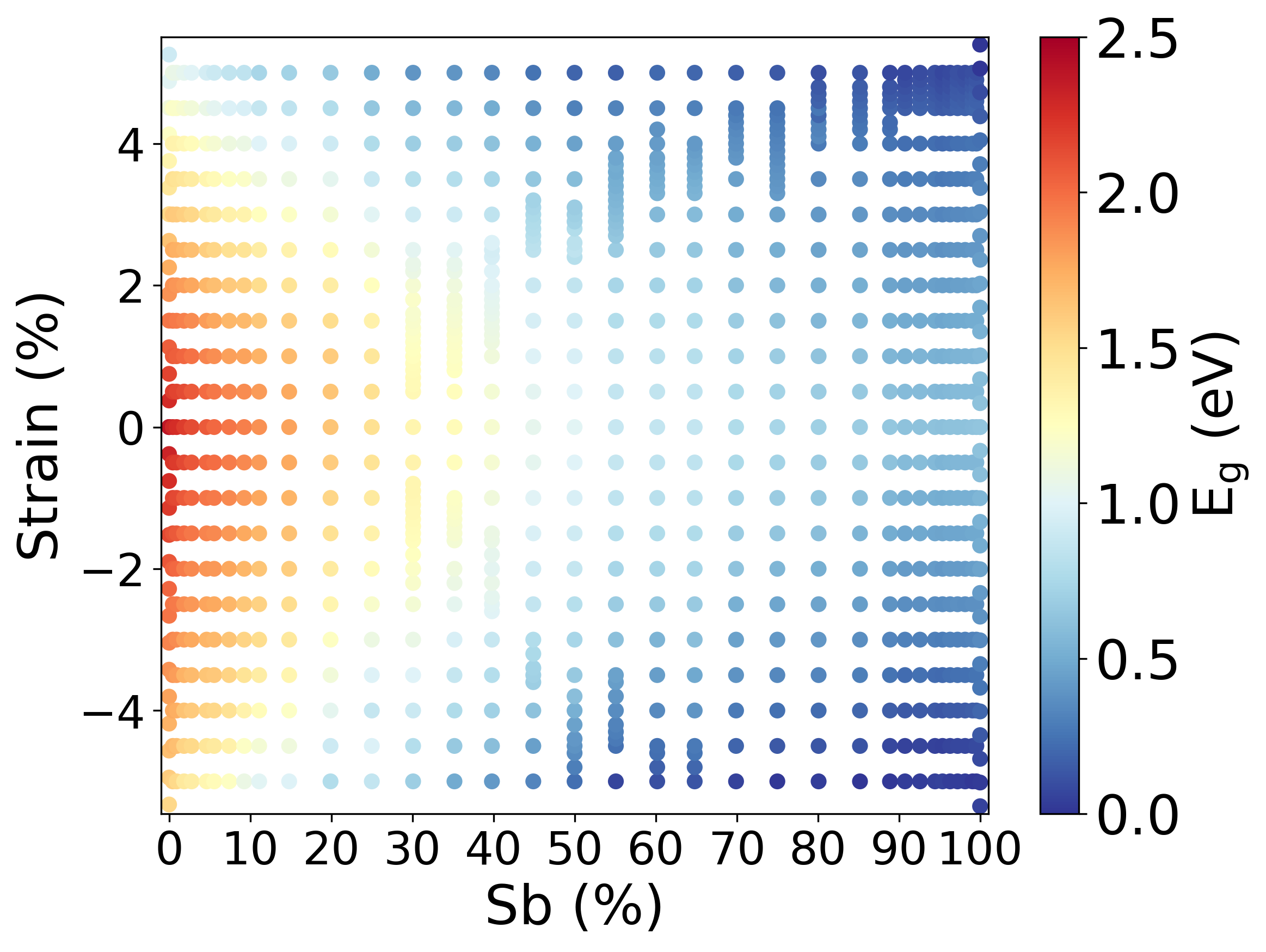}
    \label{fig:figS6g}
   } \\ %
   \subfloat[Biaxial strain for GaPBi]{%
    \includegraphics[width=3.4in]{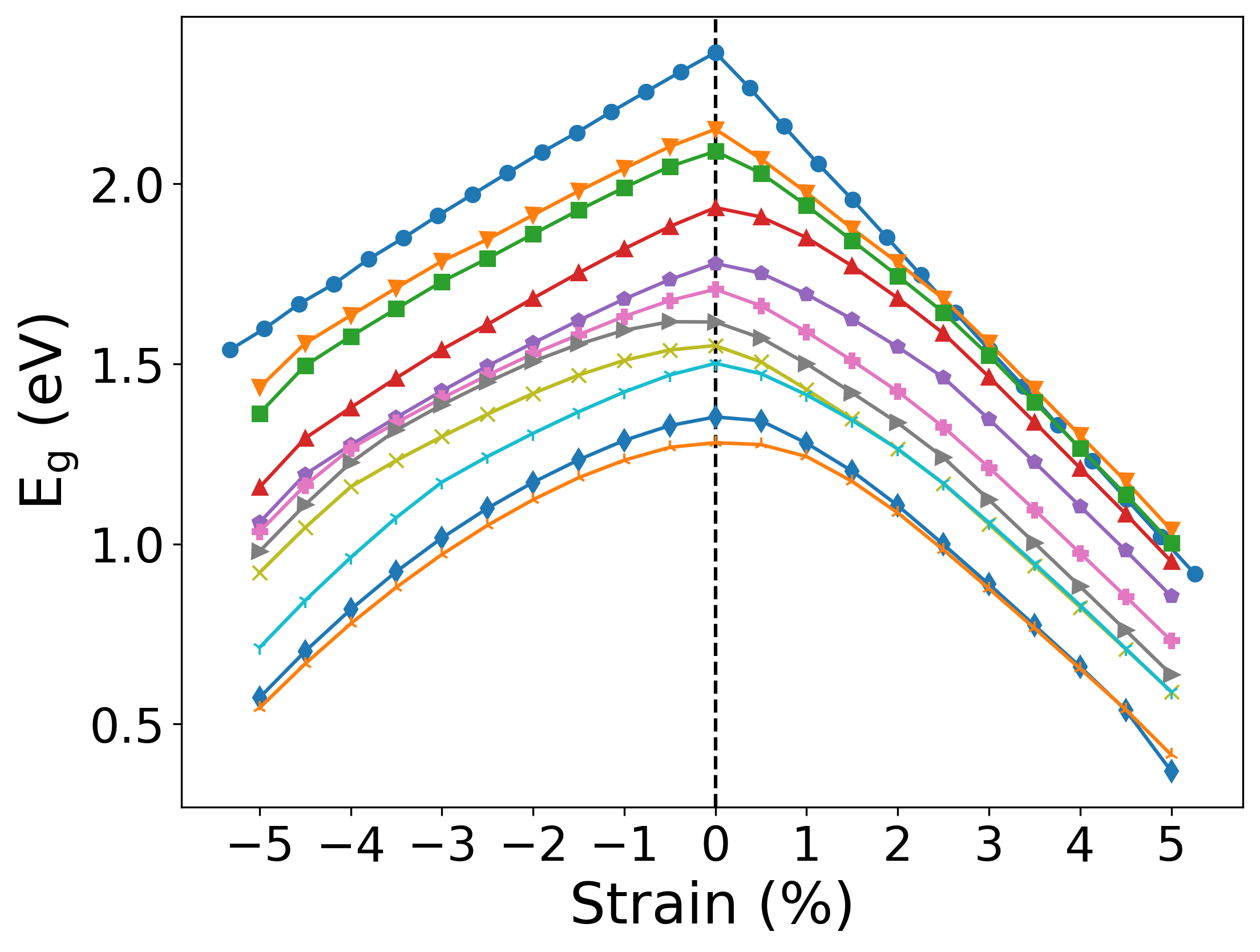}
    \hspace{0.2in}
    \includegraphics[width=3.4in]{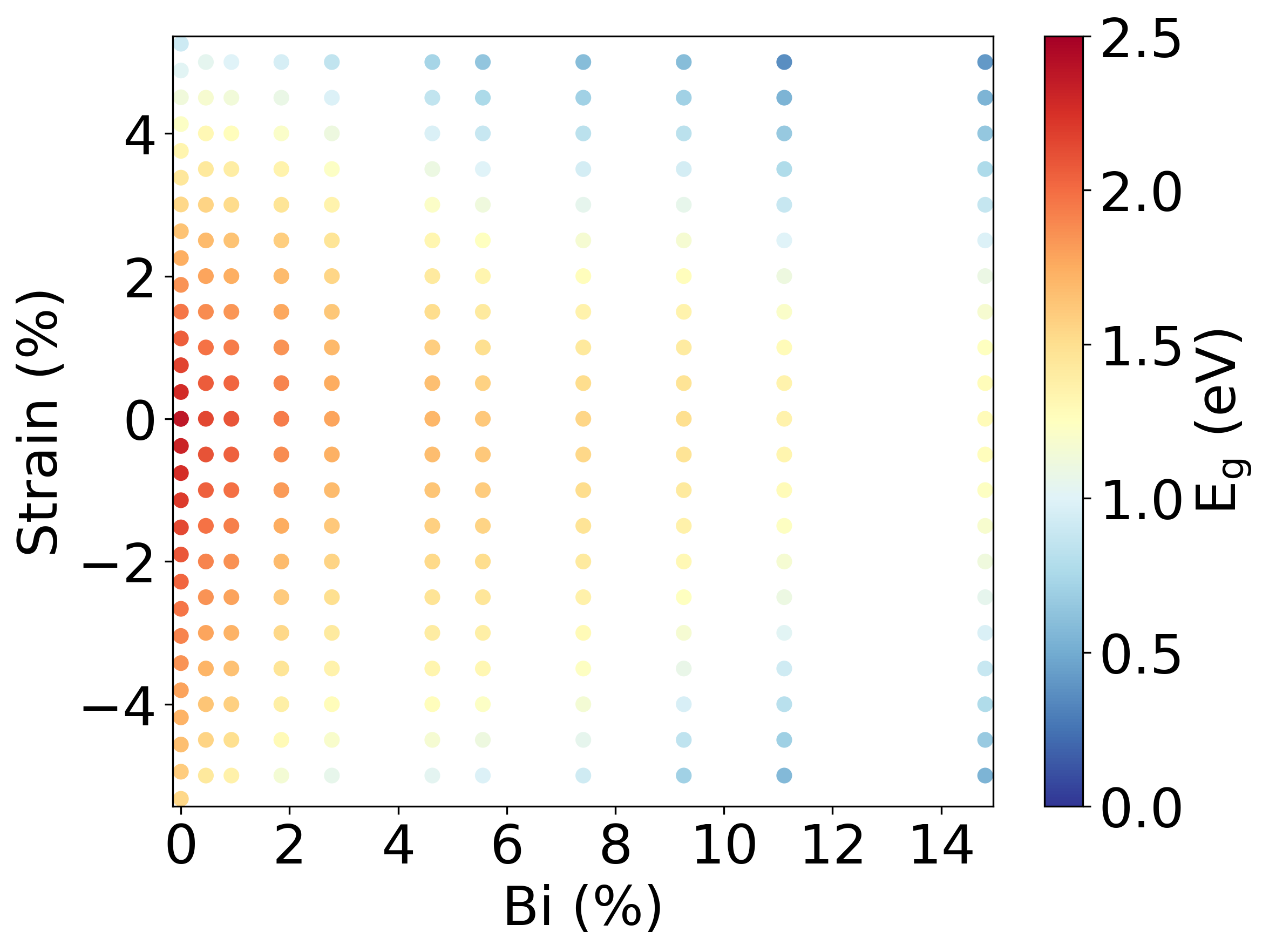}
    \label{fig:figS6h}
   } %
  
  \caption{Effect of strain on the bandgap of GaE$_{1-x}$Y$_x$ (E=As, P; Y=P, N, Sb, Bi). The left column shows the variation of bandgap magnitudes (E$_{\text{g}}$) as a function of strain at different compositions for the systems. The composition ($x\%$) legends are shown at the top. In the right column, we project the E$_{\text{g}}$ on the composition-strain space. The E$_{\text{g}}$ from these figures is interpolated to create Figs. 2--6 of the manuscript. We chose the same color bar scale limit for all the systems for consistency. The black crosses in (e) and (f) indicate the bandgap calculations that did not converge.}
  \label{fig:figS6}
\end{figure}

\clearpage

In Fig.~\ref{fig:figS6}, for the binary compounds (GaAs, GaP, and GaSb), we used the bandgap values from the `III-V binary semiconductors strain study' dataset of Ref.~\cite{Mondal2022Sup}. The strain axes were sampled primarily on the regular interval of 0.5\%. For GaAsP and GaPSb systems, sampling resolutions were increased close to the direct-indirect transition regions, as explained in Sec. III of the manuscript. For binaries, the intervals were not regular in the reference dataset. The composition axes were sampled in non-regular intervals. Because of the finite supercell size, the compositions could not be sampled in regular intervals. Note, however, the choice of compositions was arbitrary. Also, some of the DFT calculations at high strain could not be converged to the required accuracy and hence, are excluded. For GaAsN, under isotropic strain because of the effective conduction band redefinition (Fig.~\ref{fig:figS5}), as explained in Sec. III of the manuscript, the sudden rapid increase in bandgap values under compressive strain are noticeable (Fig.~\ref{fig:figS6c}). The sudden increase in band energies due to band redefinition is also visible for compressively strained GaPSb at 25\% Sb concentration (Fig.~\ref{fig:figS6g}).
\vspace*{\fill}
\section{\label{secSVI:interpolationdetails}Interpolation details}
\begin{figure}
\centering
  \subfloat[]{\label{fig:figS7a}\includegraphics[width=3.4in]{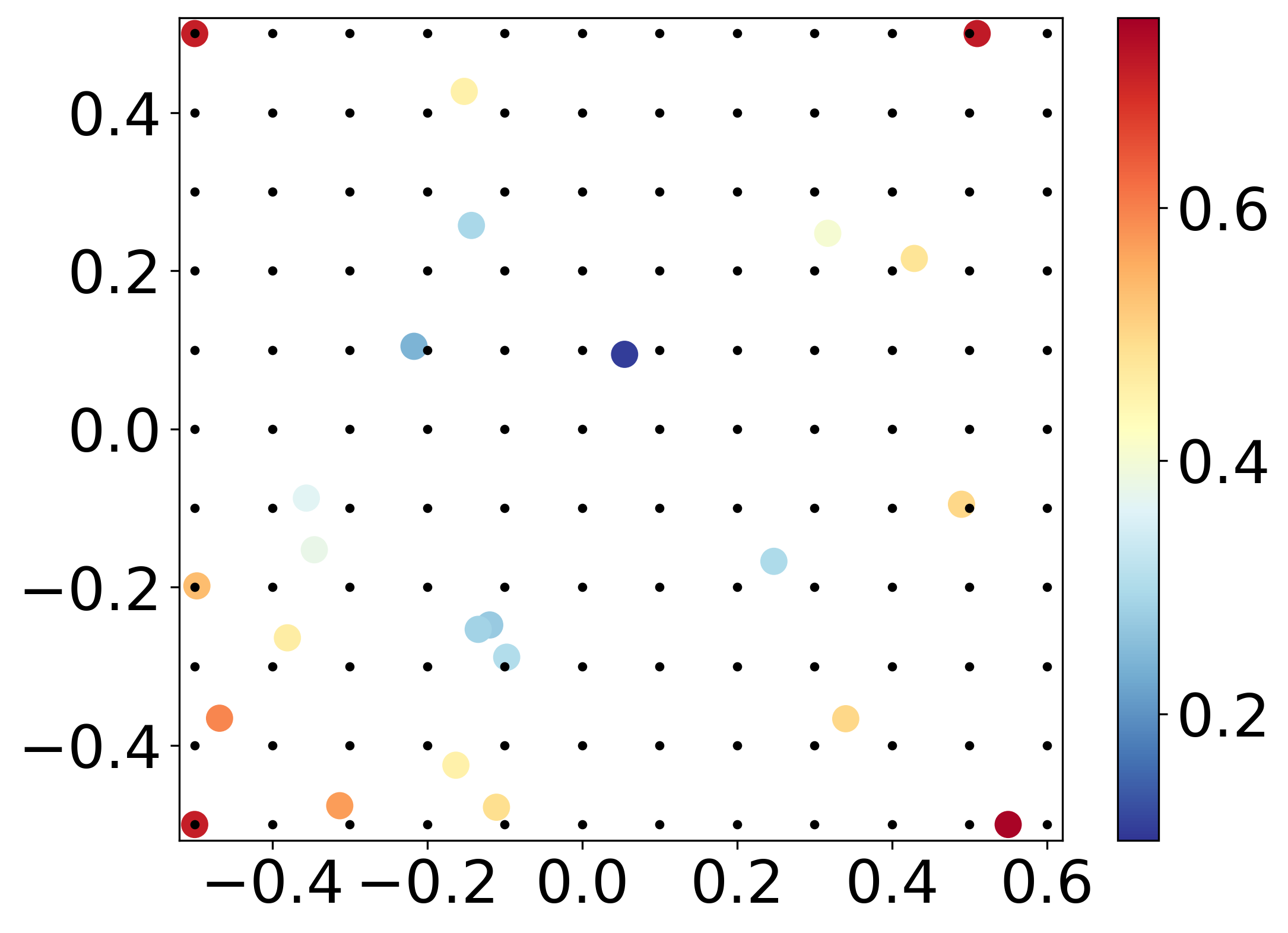}}
  \hspace{0.2in}
  \subfloat[]{\label{fig:figS7b}\includegraphics[width=3.4in]{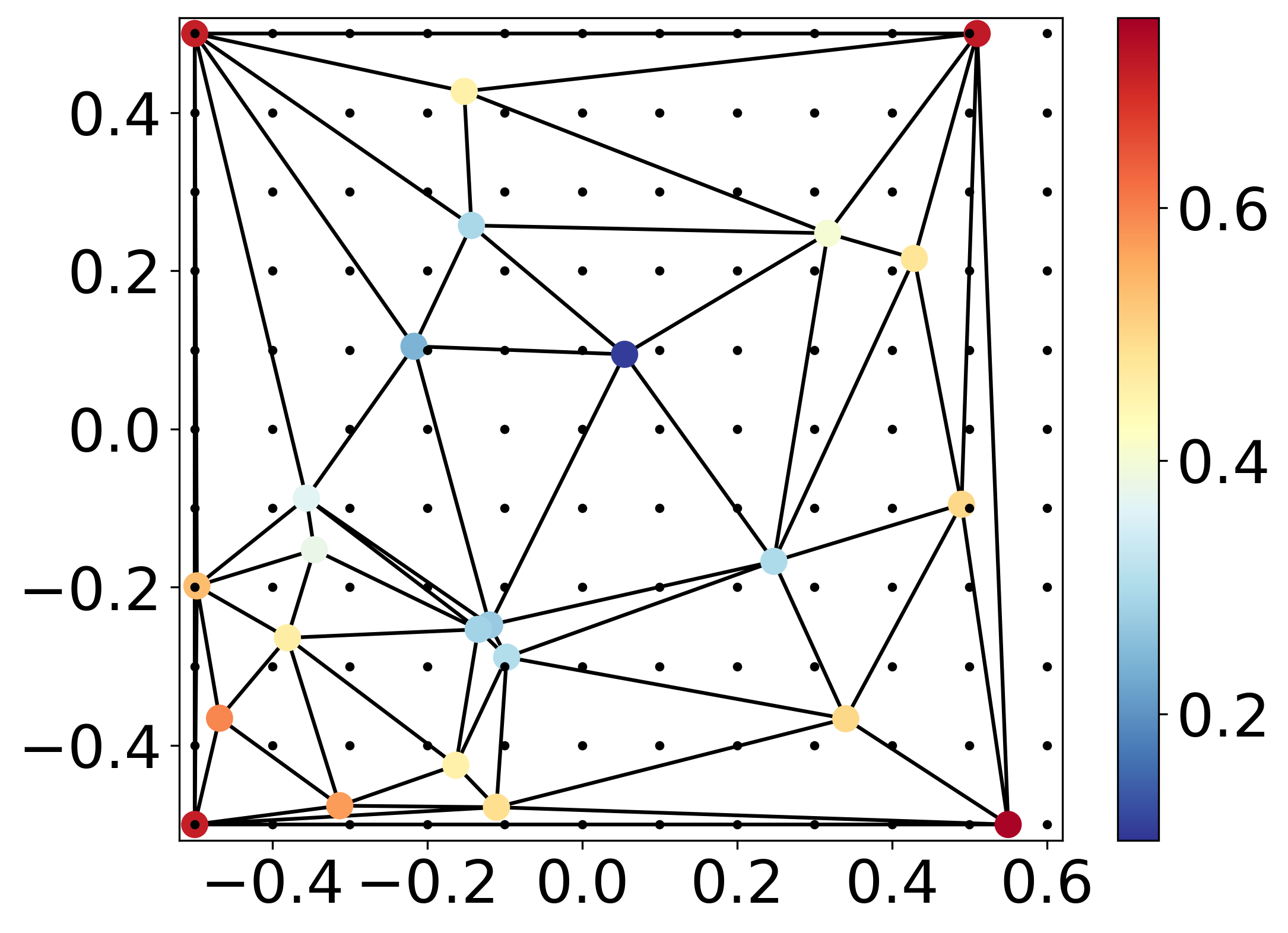}} \\
  \subfloat[]{\label{fig:figS7c}\includegraphics[width=3.4in]{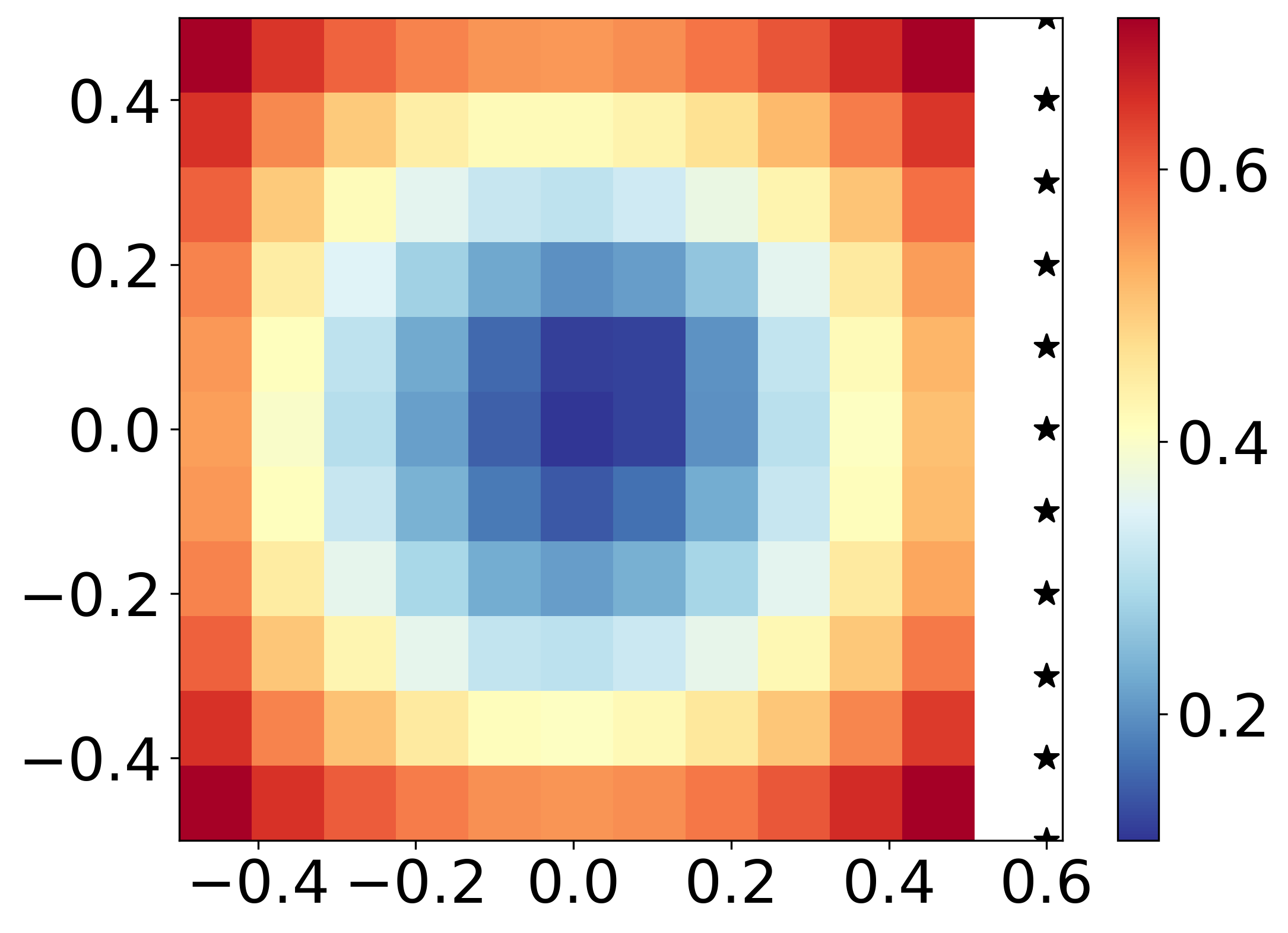}}
  \hspace{0.2in}
  \subfloat[]{\label{fig:figS7d}\includegraphics[width=3.4in]{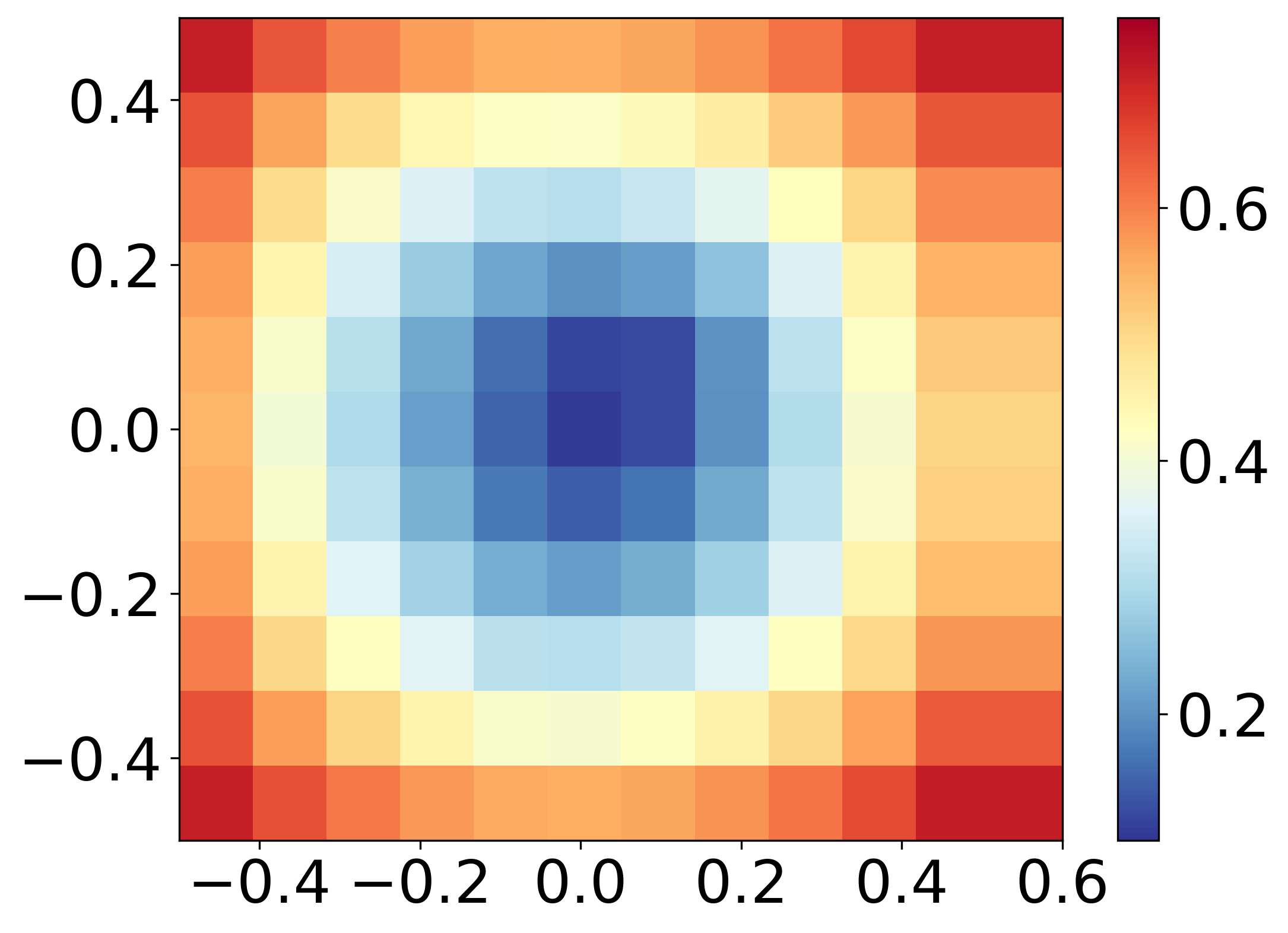}} \\
  \subfloat[]{\label{fig:figS7e}\includegraphics[width=3.4in]{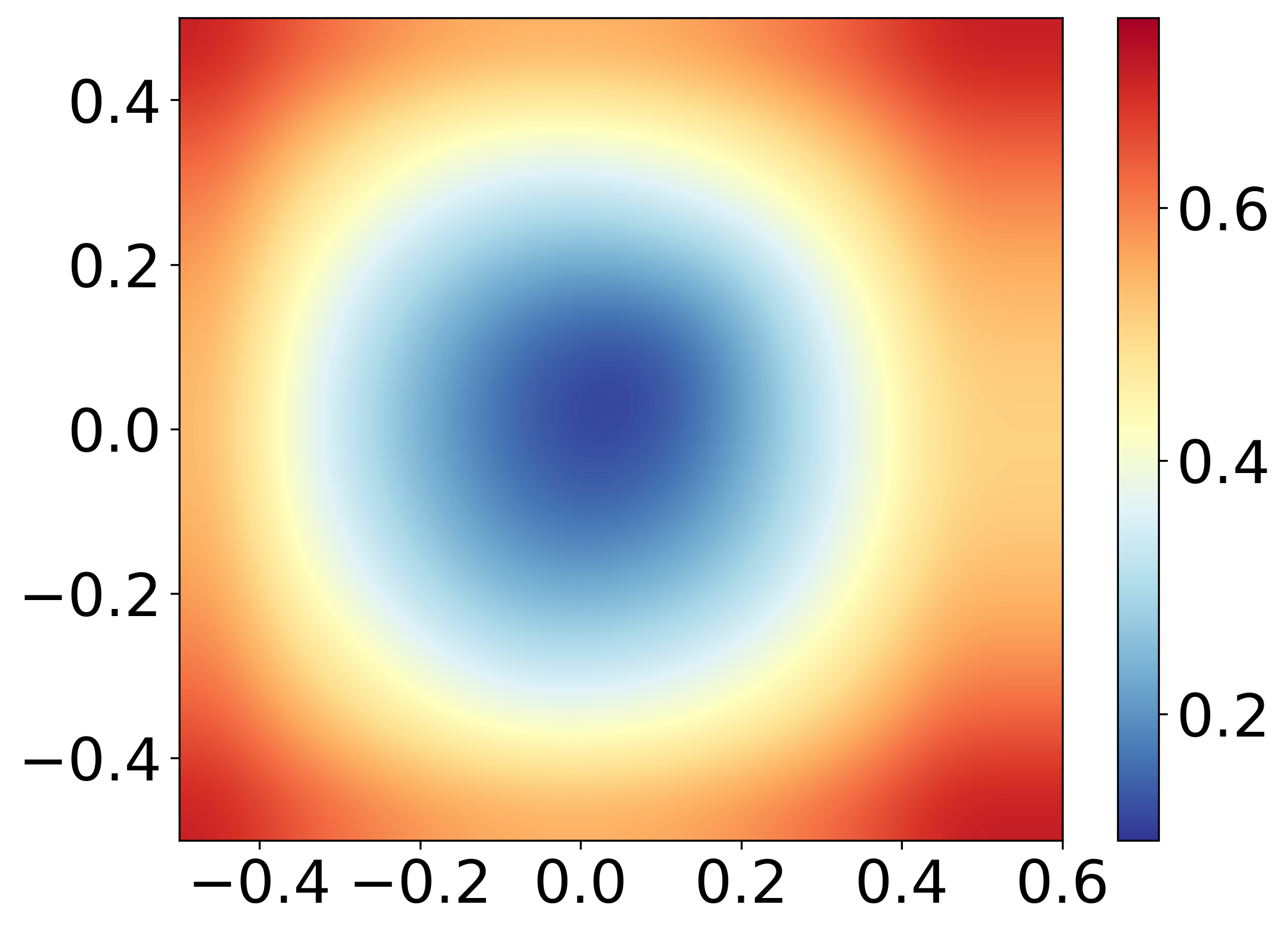}}
  \caption{The Clough-Tocher piecewise cubic, C1 smooth, curvature-minimizing 2D interpolation scheme. The color dots are the input points. The black dots indicate the interpolation grid. (a) Input data. (b) Delaunay triangulation, indicated by black triangles. (c) Nearest-neighbor interpolation points, indicated by the black stars. These points could not be triangulated. (d) Nearest-neighbor interpolation. (e) Bicubic interpolation over (d).}
  \label{fig:figS7}
\end{figure}

The E$_{\text{g}}$ from the right column figures of Fig.~\ref{fig:figS6} were interpolated over a fine grid to create Figs. 2--6 of the manuscript. We chose 0.1 grid resolution in both composition and strain for all the systems. We used Clough-Tocher piecewise cubic, C1 smooth, curvature-minimizing 2D interpolator from python \verb|scipy.interpolate| class \cite{SciPyNMeth2020}. The interpolation scheme follows:
\begin{itemize}
    \item[Step 1:] The input data are first triangulated with \verb|Qhull| \cite{Www.qhull.org} (Fig.~\ref{fig:figS7b}).
    \item[Step 2:] The interpolant then constructs a piecewise cubic interpolating B{\'e}zier polynomial on each triangle using the Clough-Tocher scheme \cite{Alfeld1984,Farin1986} (Fig.~\ref{fig:figS7c}). The interpolant is guaranteed to be continuously differentiable. The interpolant gradients are determined such that the interpolating surface curvature is approximately minimized \cite{Nielson1983,Renka1984}.
    \item[Step 3:] In the end, nearest-neighbor extrapolation is performed on the remaining points that can not be covered via \verb|Qhull| (Fig.~\ref{fig:figS7d}). Note that mostly the boundary points beyond the $x$ and $y$ limits of input points can not be triangulated under \verb|Qhull|. As long as those points are not far away from the available input points, the nearest-neighbor extrapolations to those points are acceptable.
    \item[Step 4:] Finally, bicubic interpolation, as implemented in python \verb|matplotlib.pyplot.imshow| class \cite{Hunter:2007,thomas_a_caswell_2022_7084615} is applied to smoothen the grid image (Fig.~\ref{fig:figS7e}).  
\end{itemize}
Note both the input data and interpolation grid are rescaled to unit square during interpolation to eliminate the effect of incommensurable units and/or large differences between the $x$ \& $y$ scales.
\vspace*{\fill}
\clearpage

\bibliography{supplement}